%% file: Streufert.tex
\pdfoutput=1

\providecommand{\ifmain}[1]{#1} 
 
\ifmain{ 
  \documentclass[12pt]{amsart} %
  \setlength{\topmargin}{0in}\setlength{\textheight}{8.6in} 
  \setlength{\oddsidemargin}{.25in}\setlength{\evensidemargin}{.25in}\setlength{\textwidth}{6in} 
  \newcommand{\tpath}{/Users/speter18/4t/4geek/tex} 

\input{Streufert-commands}
  \providecommand{\forpreamble}{} \forpreamble %
  \usepackage{hyperref} \hypersetup{colorlinks=true,linkcolor=black} 
  \begin{document} 
  } 
 
\ifmain{ 
  \numberwithin{lemma}{section} 
  \numberwithin{figure}{section} 
  \numberwithin{table}{section} 
  \newcommand{\plyR}[1]{\ensuremath{\mathit{#1}}}
  \newcommand{\plyr}[1]{\ensuremath{\scriptstyle\mathit{#1}}}
  \newcommand{\payoA}[1]{{\Small %
    $\mi{\mathit{#1}}\mo$}}
  \newcommand{\payoa}[1]{{\Tiny %
    $\mi{\mathit{#1}}\mo$}}
  \newcommand{\payob}[2]{{\Tiny %
    $\mi{\mathit{#1}}\\{\mathit{#2}}\mo$}}
  \newcommand{\payoc}[3]{{\Tiny %
    $\mi{\mathit{#1}}\\{\mathit{#2}}\\{\mathit{#3}}\mo$}}
  \newcommand{\rtau}{\overline{τ^{-1}}}
  }

\title{The Category G\MakeLowercase{m} for Extensive-Form Games}
\date{September 10, 2026. Results essentially identical to those in Sections 2, 3.1, and 4.1 of Streufert 2021.  Reworked, with cleaner definitions of out-tree and player transformation. {\em Keywords:} Game tree, Continuously labeled tree.  {\em Classifications}: MSC 91A70, JEL C73. {\em Contact information:} pstreuf@uwo.ca, 519-661-3500, Economics Department, Western University, London, Ontario, N6A 5C2, Canada.}  
\nichts{}

\thanks{The author is very grateful to the anonymous readers of Streufert 2020b, Streufert 2021, and another earlier version of this paper.}
\nichts{}
\nichts{}
\nichts{}

\maketitle
\iffinal{\vspace{-1mm}} %
\begin{centering}
Peter A. Streufert \\[-.5mm]
Economics Department \\[-.5mm]
Western University \\
\end{centering} 
\vspace{3mm} %

\begin{abstract} This paper introduces \ct{Gm}, which is a category for extensive-form games.  The category's objects are extensive-form games, each of which is regarded as a set of nodes which has been endowed with edges, information sets, actions, players, and utility functions.  Its morphisms are functions, from source nodes to target nodes, that respect this structure.  For instance, a game's information-set collection is newly regarded as a topological basis for the game's decision-node set, and then a morphism's continuity serves to respect the source game's information sets.  Given these definitions, a game isomorphism is characterized as a bijection whose restriction to decision nodes is a homeomorphism, whose induced player transformation is bijective, and whose induced run transformation is monotonic with respect to the total preorder determined by each player's utility function.  \end{abstract}

\vspace{11mm} %

\vspace{1mm}\section{Introduction}\label{A619}\showit
\markb{\sc \rf{A619}. Introduction}

\nssec{Overview}{F260}

An extensive-form game can be regarded as a set with structure.  More precisely, a game can be seen as a set of nodes which has been augmented with edges, information sets, actions, players, and utility functions.  This paper introduces game morphisms which map source nodes to target nodes in a way which respects this structure.  

There is quite a bit of structure.  First consider the edges.  In this dimension, game morphisms can be defined much like the graph morphisms of the category \ct{Grph} \nocite{Maclane98}(Mac Lane 1998, page~48).\nichts{}  Second consider the information sets.  This paper newly regards the collection of information sets as the basis of a topology over the decision nodes.  So in this dimension, game morphisms can be defined much like the continuous functions of the category \ct{Top} (Mac Lane 1998, page~12).  Beyond this, three layers of structure remain:\ actions, players, and utility functions.  Much of this paper is dedicated to developing game morphisms which respect these extra layers of structure.\nichts{}

All of this leads to a category for extensive-form games, which is called \ct{Gm}.   Proposition~\rf{A454} shows that \ct{Gm} is well-defined, Proposition~\rf{B178} characterizes its monomorphisms, and Proposition~\rf{A503}(\rf{F084}--\rf{F086}) characterizes its isomorphisms.  Further, Propositions 
\rf{B063},
\mbox{\rf{A481}(\rf{A483}--\rf{A490})}, 
\rf{A453}(\rf{F233},\rf{F234}),
\rf{E791}(b),
\rf{A454}(\rf{F237},\rf{F238}),
\rf{A455},
\rf{B185},
\rf{A491}(\rf{F079}--\rf{F194}), and
\rf{A503}(\rf{F088},\rf{F090})
list other appealing properties.  All suggest that \ct{Gm} is coherent, meaningful, and tractable.  

In some sense, Proposition~\rf{A503}(\rf{F084}--\rf{F086}) is the culmination of the paper.  It shows that a \ct{Gm} isomorphism is straightforwardly characterized as a bijection (from the source game's nodes to the target game's nodes) with three properties:\ [1] its restriction to decision nodes is a homeomorphism (with respect to the topologies generated by the games' information sets), [2] its induced player transformation is bijective, and [3] its induced run transformation is monotonic with respect to the total preorder determined by each player's utility function.

\nssec{Placement in Literature}{F277}

Categories have been used to study games in a variety of contexts.  First, there is a large literature on game semantics which uses specialized games to study logic and computation (some general references are \nocite{Hyland97}Hyland 1997, \nocite{Japaridze03}Japaridze 2003, and \nocite{AbramskyR24A}Abramsky and Reggio 2024).  This literature uses category theory to describe the algebra of composing games (see Hyland 1997, page 133).  In contrast, the present paper uses category theory to compare games.
\nichts{} 

In a manner broadly similar to game semantics, there is a recent computer-science literature which composes extensive-form games from game fragments called ``open games'' (some leading references are \nocite{GhaniHWZ18}Ghani, Hedges, Winschel, and Zahn 2018, \nocite{CapucGLN21v2}Capucci, Ghani, Ledent, and Nordvall Forsberg 2022, and \nocite{BoltHZ1923}Bolt, Hedges, and Zahn 2023).  These papers define open games as morphisms, which are to be composed together.  In contrast, the present paper defines games as objects, which are to be compared by morphisms. 
\nichts{}
\nichts{}  
\nichts{}

Accordingly, the remainder of this Section~\rf{F277} considers categories in which games are objects.  

\newcommand{\notesim}{\footnote{\label{F799}It is well understood that a simultaneous-move game can also be specified as an extensive-form game in which each player has one information set (see for example \nocite{OsborneR94}Osborne and Rubinstein 1994, page~202).  This literature review treats the two specifications as equivalent.  Only the latter is a \ct{Gm} game.}}
\nocite{Hedges18}Hedges 2018 defines morphisms for some open games.  Given that open games are themselves morphisms (see the second-previous paragraph), the resulting category for open games is a ``double category''.  This approach is very different from that of the present paper, which regards a game as a set of nodes with structure.  In addition, the category in Hedges 2018 is restricted to games of finite length, which are composed from perfect-information fragments and simultaneous-move{\notesim} fragments.  In contrast, the present paper considers games with up to countably infinite length, and with general information structure.

Elsewhere, categories are developed for simultaneous-move games.  These include \nocite{Lapit99}Lapitsky 1999, \nocite{Jimen14}Jim\'enez 2014, \nocite{ThomeV2325}Thom\'e and Viglizzo 2025, and \nocite{Kawamori26}Kawamori 2026.  A simultaneous-move game has a special information structure (see note~\rf{F799}) relative to the general information structure considered here.  

Also elsewhere, categories are developed for certain perfect-information games.\nichts{}  \nocite{HonseLR12}Honsell, Lenisa, and Redamalla 2012 builds a category for some perfect-information games which are defined coalgebraically.  \nocite{LarssonNS18}Larsson, Nowakowski, and Santos 2018 develops morphisms for certain perfect-information games on the basis of their plays.  \nocite{DuziST25}Duzi, Szeptycki, and Tholen 2025 provides a new proof of a topological equivalence via a category for certain perfect-information games.  Finally, \nocite{BasicEEPS24}Ba\v{s}i\'c, Ellis, Ernst, Popovi\'c, and Sieben 2024, and \nocite{BaltushkinES25A}Baltushkin, Ernst, and Sieben 2025 develop categories for some combinatoric games with perfect information.  Any perfect-information game has a special information structure relative to the general information structure of the games in the present paper.
\nichts{} 

The author's own line of research has always accommodated games with general information structure and up to countably infinite length.  The author's first approach was restricted to node-and-choice games (i.e., distinguished-action games), which require that each information set has its own actions (see Appendix~\rf{E615}).  Three papers resulted:\ \nocite{S1809ncp}Streufert 2018 (for ``preforms'' resembling the CLTs in the present paper), \nocite{S2001ncf}Streufert 2020a (for ``forms''), and finally \nocite{S2007ncgv2}Streufert 2020b (for games).  The author's second and current approach began with \nocite{S2105gmLocal}Streufert 2021, which removed the distinguished-action restriction.  The present paper is a revision of Streufert 2021 which more cleanly defines the concepts of out-tree and player transformation, and which takes its results from Streufert 2021, Sections 2, 3.1, and 4.1.  For the sake of brevity, the present revision omits Streufert 2021's results on subgames, equilibrium concepts, and subcategories.

Note that these omitted results in \nocite{S2105gmLocal}Streufert 2021 further suggest that the category \ct{Gm} is meaningful and tractable.  The results show that [1] the properties of perfect information and no-absentminded\-ness (i.e., linearity) are invariant to \ct{Gm} isomorphisms, [2] the solution concepts of Nash equilibrium and subgame-perfect equilibrium (Selten 1975) are also invariant to \ct{Gm} isomorphisms, [3] the full subcategory for distinguished-action games is essentially wide in \ct{Gm}, and [4] the full subcategory for choice-set games is essentially wide in the full subcategory for no-absentminded games.  (The latter two results are discussed by example in Appendix~\rf{F259}.)

\nssec{Motivation and related research}{F261}

Category theory has added clarity and perspective to many fields of mathematics.  So it seems valuable to do the same for the growing field of game theory.
\nichts{}
\nichts{}

The author's own motivation is rooted in the field of economics.  Specifically, the extensive-form games used by economists are formulated in many different styles (as demonstrated by the examples of Appendix~\rf{F258}).  This diversity of style arises not only from historical accident, but also from the different notational needs of different mathematical contexts.  Economists informally understand that [a] any ``substantial'' concept in one style should have the same meaning in any other style, and [b] any ``substantial'' result in one style should also hold in any other style.  This informal understanding might be formally developed.  In particular, when a concept or result is translated from one style to another, how would we define the sense in which the translation itself was correct or incorrect?  A good answer to this question would help us to identify, manipulate, and communicate the ``substance'' of game theory.  Category theory seems well-suited to this task.
\begin{picture}(0,0) %
  \put(-2.8,2.15){\color{white} \rule{20ex}{2.5ex}}
  \put(-2.3,2.35){{\sc\SMALL 1.\ Introduction}}
  \end{picture}

This is a large agenda, which necessarily extends well beyond the present paper.  The omitted results in Streufert 2021 (see third-previous paragraph) are a start.  Specifically, results [1] and [2] formalize the notion that perfect information, no-absentminded\-ness, Nash equilibrium, and subgame perfection are ``substantial'' in the sense of the previous paragraph.  Further, results [3] and [4] formally assess the ``substance'' of specifying games in the distinguished-action or action-set styles.  In particular, [3] shows the distinguished-action style is not restrictive, while [4] shows the action-set style implicitly imposes no-absentminded\-ness (see Appendix~\rf{F259} for further discussion).  Other concepts and styles remain to be explored.  Soon thereafter one can hope to formally translate concepts and results from one style to another, as the previous paragraph suggests.
\nichts{}

\nssec{Organization}{B031}
Sections~\rf{A620} and \rf{F219} use two steps to define the category \ct{Gm}.  Then Section~\rf{B023} characterizes monomorphisms and isomorphisms.  Appendix~\rf{F258} provides examples.  Appendices \rf{B025} and \rf{F223} support Sections~\rf{A620} and \rf{F219}, respectively.  Finally, Appendices \rf{B026} and \rf{F224} support Section~\rf{B023} regarding monomorphisms and isomorphisms, respectively.

\newcommand{\dve}{\dra\,\,}
\newcommand{\tableentities}{
\begin{table}[t]
{\small 
\begin{tabular}{cc} 

\multicolumn{2}{c}{Out-tree $(X,E)$ \hfill  (\rf{E776}) \rf{B032}} \\ \hline
$X$ & set of nodes $x$ \hfill  \rf{B032}\\
$E$ & set of (directed) edges $xy$ \hfill  \rf{B032}\\
$r$ & $\dve$ root node \hfill  \rf{B032}\\
$W$ & $\dve$ set of decision nodes $x$ \hfill  (\rf{E812}) \rf{B032}\\
$p$ & $\dve$ immediate predecessor function \hfill  \rf{B032}\\
$≼$ & $\dve$ weak tree order \hfill  \rf{B032}\\
$≺$ & $\dve$ strict tree order \hfill  \rf{B032}\\
$P$ & $\dve$ strict predecessor correspondence \hfill  \rf{B032}\\
$X⧷W$ & $\dve$ set of end nodes $x$ \hfill  \rf{B032}\\
$\ZZ$ & $\dve$ collection of runs $Z$ \hfill  (\rf{E813}) \rf{B032}\\
$\ZZf$ & $\dve$ collection of finite runs $Z$ \hfill  \rf{B032}\\
$\ZZi$ & $\dve$ collection of infinite runs $Z$ \hfill  \rf{B032}\\[3mm]

\multicolumn{2}{l}{CLT (Continuously labeled Tree) $Θ = (X,E,\HH,λ)$
  \hfill (\rf{C1})--(\rf{C4}) \rf{B033}} \\ \hline
$\HH$ & collection of information sets $H$ \hfill  \rf{B033}\\
$λ$ & labeling function (action-assigning function) \hfill  \rf{B033}\\
$A$ & $\dve$ set of actions $a$ \hfill  (\rf{F217}) \rf{B033}\\
$F$ & $\dve$ feasibility correspondence \hfill  (\rf{E779}) \rf{B033}\\
$n$ & $\dve$ next-node function \hfill  (\rf{E869}) \rf{B033}\\ [3mm]

\multicolumn{2}{l}{CLT morphism $θ = [Θ,Θ′,τ]$
  \hfill (\rf{cE}),(\rf{cH}),(\rf{cL}) \rf{E770}} \\ \hline
$τ$ & node transformation \hfill  \rf{E770}\\
$α$ & $\dve$ action transformation \hfill  (\rf{E782}) \rf{E770}\\
$ζ$ & $\dve$ run transformation (requires $θ$ is end-preserving) \hfill  (\rf{E786}) \rf{E771}\\ [3mm]

\multicolumn{2}{l}{Game $Γ = (X,E,\HH,λ,κ,U)$ \hfill (\rf{G1})--(\rf{G3}) \rf{B034} } \\ \hline
$κ$ & control-assigning function (player-assigning function) \hfill  \rf{B034}\\
$I$ & $\dve$ set of players $i$ \hfill  (\rf{F218}) \rf{B034}\\
$U$ & profile $⁅U_i⁆_{i∈I}$ listing utility function $U_i$ for each
      player $i$⋅ \hfill  \rf{B034}\\
$\lesssim_i$ & $\dve$ total preorder of $\ZZ$ implied by $U_i$ \hfill  (\rf{E784}) \rf{B034}\\[3mm]

\multicolumn{2}{l}{Game morphism $γ = [Γ,Γ′,τ]$
  \hfill (\rf{gZ}),(\rf{gK}),(\rf{gU}) \rf{E771} } \\ \hline
$ι$ & $\dve$ player transformation \hfill  (\rf{E785}) \rf{E771}\\[2mm]

\end{tabular} }
\caption{\small Terms defined in Sections \rf{A620} and \rf{F219}.  Out-trees, CLTs, CLT morphisms, games, and game morphisms are implicitly accompanied by their components and derivatives (\protect\rotatebox[origin=c]{180}{$\Lsh$}).  Definitions are in equations and sections on right side. } \label{A452}
\end{table}
}

\vspace{1mm}\section{Definition of \ct{CLT}}\label{A620}\showit
\markb{\sc \rf{A620}. Definition of \ct{CLT}}

\nssec{Initial remarks}{F241}

An extensive-form game is often defined in two steps:\ the first defines an ``extensive form'', and the second defines an ``extensive-form game''.  Similarly, this Section~\rf{A620} first defines the category \ct{CLT}, and then Section~\rf{F219} defines the category \ct{Gm}.  The acronym ``CLT'' stands for ``continuously labeled tree'' (Definition~\rf{E772}).  It is like an extensive form, but does not specify players (that is left for Section 3).

Together, Sections \rf{A620} and \rf{F219} specify (extensive-form) games as routinely as possible given the literature's wide variety of game specifications.  The particular specification adopted here belongs to the ``abstract'' style group of Section~\rf{E614}, resembles the ``KS game'' of Kline and Luckraz 2016 and Streufert 2019, and is virtually identical to the ``traditional game'' of \nocite{S2505-5fy}Streufert 2025.\nichts{}  In developing this game specification, the author's guiding principles were to avoid specialized assumptions, to use explicit notation, and to use standard mathematics whenever possible.  Nonetheless, almost every reader will find some aspect of this specification unfamiliar, and some of these difficulties are discussed in notes \rf{E814}, \rf{B216}, \rf{E815}, \rf{E882}, \rf{F800}, \rf{E816}, and \rf{E817}.  Excluded kinds of (extensive-form) games are discussed in note~\rf{E890}.

\tableentities

\nssec{Out-trees}{B032}
Before defining out-trees, define a {\em directed graph} to be a couple $(X,E)$ such that $E⋅⊆⋅X^2$.  The elements of $X$ are called {\em nodes}.  The elements of $E$ are called {\em edges}, and are denoted by $xy$ rather than $⁅x,y⁆$.  A {\em walk from $w$ to $y$ in a directed graph $(X,E)$} is a sequence $⁅x_0,x_1,x_2,...,x_m⁆$ such that \begin{gather}
\zz
x_0 = w,⋅x_m = y,⋅\text{and}⋅(∀ℓ{<}m)⋅x_ℓx_{ℓ+1}⋅∈⋅E. \label{E824}
\zz
\end{gather} Likewise, an {\em infinite walk from $w$ in a directed graph $(X,E)$} is a sequence $⁅x_0,x_1,x_2,...⁆$ such that $x_0 = w⋅\text{and}⋅(∀ℓ)⋅x_ℓx_{ℓ+1}⋅∈⋅E$.  Note that a walk in an arbitrary directed graph can visit a node more than once.  

\newcommand{\noteouttree}{\footnote{\label{F816}Similar definitions in the game-theory literature include \nocite{GaleS53}Gale and Stewart 1953; \nocite{Perea01}Perea 2001, page~10; Kline and Luckraz 2016, page 88; Streufert 2018; and Gonz\'alez-D\'iaz, Garc\'ia-Jurado, and Fiestras-Janeiro 2023, page 102.  Similar concepts in the graph-theory literature include \nocite{Tutte84}Tutte 1984, page~126; and \nocite{BangjensenG09}Bang-Jensen and Gutin 2009, page 21.  Several of these definitions apply only to finite trees, and several explicitly include the antisymmetry or irreflexivity of $E$.\nichts{}  The definition here applies equally well to trees of up to countably infinite length, and it implies both antisymmetry and irreflexivity.\nichts{}}} 
\newcommand{\notehazardtree}{\footnote{\label{E814}This specification of a tree may be unfamiliar for any of several reasons.  First, some game-tree specifications use undirected graphs rather directed graphs.  Second, some specifications use order theory rather than graph theory.  Third, some specifications use specialized notations which implicitly impose some of the properties of game trees.  For example, some game-tree properties are implicitly imposed if nodes are specified as sequences of past actions, or sets of past actions, or sets of outcomes.  Finally, it is common to specify smaller game trees by diagrams rather than by notation.}}
An {\em out-tree}{\noteouttree}$^,${\notehazardtree} is a directed graph $(X,E)$ such that \begin{gather}
\zz
(∃r∈X)(∀x∈X)⋅\text{there exists a unique walk from $r$ to $x$ in $(X,E)$}. \label{E776} 
\zz
\end{gather} (The term ``out-tree'' is used instead of its synonyms ``outwardly directed rooted tree'' and ``diverging arborescence'' only because it is shorter.)  It can be shown that each out-tree has exactly one $r$ that satisfies (\rf{E776}).  Call this the {\em root node} of the out-tree.  For example, $(⎨\f5⎬,∅)$ is a one-node out-tree because the one-node sequence $⁅\f5⁆$ is the unique walk from $\f5$ to itself in $(⎨\f5⎬,∅)$.  In this example, $\f5$ is the root node $r$.  Such one-node out-trees are the ``smallest'' of out-trees because $(∅,∅)$ is not an out-tree (it lacks the node $r$ required by (\rf{E776})).  In the opposite direction, the ``largest'' of out-trees can have uncountably infinite degree (note~\rf{E823}) and countably infinite length (there can be infinite walks of the sort defined below (\rf{E824})).\nichts{}

\newcommand{\notecorrespondence}{\footnote{\label{A629}To be clear, a {\em correspondence} $F{:}X⇉Y$ is taken to be a triple $[X,Y,F\gr]$ such that $F\gr⋅⊆⋅X×Y$, and a {\em function} $f{:}X→Y$ is taken to be a triple $[X,Y,f\gr]$ such that $f\gr⋅⊆⋅X×Y$ and $(∀x∈X)(∃!y∈Y)$ $(x,y)⋅∈⋅f\gr$ (the three components of a correspondence or function are its domain, codomain, and graph; the term ``correspondence'' is standard in economics but perhaps not elsewhere).\nichts{}}}
Consider an out-tree $(X,E)$.  Let \begin{gather}
\zz
W = π_1E, \label{E812}
\zz
\end{gather} where $π_1E$ is the projection of $E$ onto its first coordinate.  Call $W$ the {\em decision-node} set.  It can be shown that $E^{-1} = ⎨˙⁅y,x⁆˙|˙xy∈E˙⎬$ is the graph of a surjective function{\notecorrespondence} from the set $X⧷⎨r⎬$ of nonroot nodes to the decision-node set $W$.  Let $p$ denote this function, and call it the {\em immediate-predecessor function}.\footnote{\label{E823}Incidentally, the inverse image of $p$ at $x$ is $⎨˙y˙|˙x{=}p(y)˙⎬ = ⎨˙y˙|˙xy∈E˙⎬$.  This could be called the set of $x$'s ``immediate successors''.  A node can have uncountably many immediate successors.  In other words, an out-tree can have uncountable degree.} 
(If the out-tree $(X,E)$ is of the extreme form $(⎨x⎬,∅)$, then $W$, $X⧷⎨r⎬$, and $p$ are all empty.)

Further, let $≼$ be the binary relation on $X$ defined by $x⋅≼⋅y$ iff there is a walk from $x$ to $y$.  Call $≼$ the {\em weak tree order}.  Note $≼$ is reflexive because of the one-node walk from any node to itself.  Next, let $≺$ be the binary relation on $X$ defined by $x⋅≺⋅y$ iff $x⋅≼⋅y$ and $x⋅≠⋅y$.   Call $≺$ the {\em strict tree order}.\nichts{}  It can be shown that $≼$ is a partial order on $X$ and that $≺$ is its asymmetric part.  Further, define the correspondence (note~\rf{A629}) $P{:}X⇉W$ by $P(y) = ⎨\,x∈X\,|\,x≺y\,⎬$.  Call $P$ the {\em strict predecessor correspondence}.  It is easily shown that $(∀y∈X⧷⎨r⎬)$ $p(y)⋅∈⋅P(y)$.  

It can be shown that each walk in the out-tree $(X,E)$ visits each of its nodes exactly once,\footnote{This implies that each walk in $(X,E)$ is a ``path'', where a ``path'' is a special kind of walk which visits each of its nodes exactly once.  This paper does not use the term ``path''.}  and that each walk in $(X,E)$ is characterized by the set of nodes that it visits.\nichts{}  Because of this, each walk in $(X,E)$ will be identified with the set of nodes that it visits.  For example, for each node $y⋅∈⋅X$, the set $P(y)⋃⎨y⎬$ is the walk from the root node $r$ to the node $y$ (expressed as a set rather than a sequence). 

Further, call $X⧷W$ the set of {\em end nodes}, and let $\ZZf$ be the collection of all walks from $r$ to an end node (each walk is expressed as a set $Z⋅⊆⋅X$).\nichts{}  Further, let $\ZZi$ be the collection of\nichts{\tiny (the node sets of)} all infinite walks from $r$ (as defined below (\rf{E824}), but each walk is expressed as a set $Z⋅⊆⋅X$).  Finally, let  \begin{gather}
\zz
\ZZ = \ZZf⋃\ZZi, \label{E813} 
\zz
\end{gather} and call $\ZZ$ the collection of {\em runs} (elsewhere ``plays'').  Note that $\ZZ$ is nonempty (because if the out-tree $(X,E)$ is of the extreme form $(⎨x⎬,∅)$, then the end-node set is $X⧷W = ⎨x⎬$ and the only run in $\ZZ$ is $Z = ⎨x⎬$).  Possibly $\ZZ = \ZZf$, possibly $\ZZ = \ZZi$, and possibly both $\ZZf$ and $\ZZi$ are nonempty.

\begin{figure}[h] \newcommand{\hgth}{4.60}
\begin{picture}(0,\hgth) \myoutergrid{\hgth} \renewcommand{\sellg}[1]{#1}
  \put( 0.00, 2.15){\makebox(0,0){\scalebox{0.88}{ \begin{pspicture} \end{pspicture} }}} \end{picture}
\caption{\small An out-tree $(X,E)$ with a partition $\HH$ of its decision-node set $W⋅⊆⋅X$.  The directions of the edges in $E$ are shown by underlining the tree's root node $r$.} \label{E822} \end{figure}

\nssec{Continuously labeled trees (CLTs)}{B033}
Consider an out-tree $(X,E)$.  Let $\HH$ be a partition of the decision-node set~$W$, and call its elements {\em information sets}.  For example, consider Figure~\rf{E822}.  There the decision-node set is $W = ⎨\f0,\f1,\f3,\f4⎬$, and the shaded regions depict that $W$ is partitioned by the information sets belonging to the collection $\HH = ⎨⎨\f0⎬,⎨\f1⎬,⎨\f3˛\f4⎬⎬$.

One can regard $\HH$ as the basis of a topology for $W$.  Call this topology the {\em information topology} for $W$.  Thus a set of decision nodes is open in the information topology for $W$ iff it is the union of a collection of information sets from $\HH$.  This paper uses the information topology for $W$ and the discrete topology everywhere else.

\newcommand{\noteweirdA}{\footnote{\label{B216}It may seem strange to derive $A$.  Alternatively, one could firstly let $A$ be a set called the set of ``actions'', and secondly let $λ$ be a surjective function from the edge set $E$ to the action set $A$.  This alternative specification is avoided because it would necessitate adding $A$ to the tuple $(X,E,\HH,λ)$ specifying a CLT (Definition~\rf{E772}).  Instead, (\rf{F217}) derives $A$ from $λ$.  To be absolutely clear, recall from note~\rf{A629} that a function is regarded as a triple.  So the assumption that $λ = [λ^{\mathsf{dom}},λ^{\mathsf{cod}},λ\gr]$ is a surjective function from $E$ implies that  $λ^{\mathsf{dom}} = E$ and $λ^{\mathsf{cod}} = \bar{λ}(E)$.  In this context, (\rf{F217}) defines $A$ to be $λ^{\mathsf{cod}} = \bar{λ}(E)$.  Accordingly, Table~\rf{A452} lists $λ$ as a component of a CLT, and $A$ as a derivative of a CLT.}}
\newcommand{\notehazardlabel}{\footnote{\label{E815}This specification of a labeling function may be unfamiliar for any of several reasons.  First, there is a bijection between the edge set $E$ and the successor-node set $X⧷⎨r⎬$.  Because of this, a labeling function $λ$ on $E$ could be alternatively specified by a function on $X⧷⎨r⎬$.  Second, some specifications use specialized notations which implicitly assign actions to edges or successor nodes.  For example, if nodes are specified as sequences of past actions, then the action assigned to a successor node is simply the node's most recent past action.  Third, a locally injective (\rf{E778}) labeling function $λ$ can be expressed as a map from decision-node/feasible-action couples to nodes, as suggested by the map $n$ in note~\rf{E873}.  Fourth, it is common to assign actions to edges in a tree diagram rather than through formal notation.}}
\pagebreak 
Next, let $λ$ be a surjective function from the edge set $E$ to a discrete (i.e.\ unstructured) codomain.  Call $λ$ the {\em labeling function} (elsewhere an ``action-assigning function''), and say that $λ(xy)$ {\em labels} the edge $xy$.  Then let $A$ be the codomain of $λ$.{\noteweirdA}  In other words (since $λ$ is surjective), let \begin{gather}
\zz
A = \bar{λ}(E), \label{F217}
\zz
\end{gather} where $\bar{λ}(E) = ⎨⋅λ(xy)⋅|⋅xy∈E⋅⎬$ denotes the image of the set $E$.  Although $A$ could be called the set of ``labels'', $A$ will instead be called the set of {\em actions}.{\notehazardlabel}  

It will be assumed that $λ$ is {\em locally injective} in the sense that, for any two edges of the form $xy_1⋅∈E$ and $xy_2⋅∈⋅E$, \begin{gather}
\zz
y_1⋅≠⋅y_2⋅\text{implies}⋅λ(xy_1)⋅≠⋅λ(xy_2). \label{E778} 
\zz
\end{gather} This implies that the edges leaving each decision node $x⋅∈⋅W$ are injectively labeled.  For example, the labeling function $λ$ shown in Figure~\rf{E820} is locally injective even though it is not injective.  Local injectivity is equivalent to the concept of ``determinism'' in Blackburn, de Rijke, and Venema 2001, page~3.

\begin{figure}[t] \newcommand{\hgth}{4.60}
\begin{picture}(0,\hgth) \myoutergrid{\hgth} \renewcommand{\selle}[1]{#1}
  \put( 0.00, 2.15){\makebox(0,0){\scalebox{0.88}{ \begin{pspicture} \end{pspicture} }}} \end{picture}
\caption{\small A continuously labeled tree (CLT).  A CLT consists of an out-tree $(X,E)$, a partition $\HH$ of the tree's decision-node set $W⋅⊆⋅X$, and a labeling (action-assigning) function~$λ$ defined on the tree's edge set $E$.} \label{E820} \end{figure}

\newcommand{\notefeas}{\footnote{\label{E882}This derivation of the feasibility correspondence may be unfamiliar because the literature occasionally has the feasible set as a function of the information set rather than the decision node.  In this alternative, a well-defined feasibility correspondence implicitly imposes continuity (\rf{E780}).}}
From $λ$ derive the correspondence $F{:}W⇉A$ by\begin{gather}
\zz
(∀x∈W)⋅F(x) = ⎨\,a∈A\,|\,(∃y)\,λ(xy){=}a\,⎬. \label{E779}
\zz
\end{gather} Thus $F(x)$ is the set of actions that label the edges leaving $x$.  Call $F$ the {\em feasibility correspondence}, and call $F(x)$ the set of actions that are {\em feasible} at $x$.{\notefeas}  For example, $F(\f3) = ⎨\fb,\fc⎬$ in Figure~\rf{E820}.

\newcommand{\noteFgraph}{\footnote{$F\gr$ is the graph of the correspondence $F$, as in note \rf{A629}.  Notice that $(x,a)⋅∈⋅F\gr$ iff $a⋅∈⋅F(x)$.}} 
\newcommand{\noteln}{\footnote{\label{E873}Casually, $λ(x∙)$ and $n(x,∙)$ are local inverses at each decision node $x$.  More specifically, $λ(x∙)$ maps immediate successors $y$ to feasible actions $a$, and conversely, $n(x,∙)$ maps feasible actions $a$ to immediate successors $y$.  Details are in Lemma~\rf{A816}(\rf{A822}).}}
For future reference, we define a function $n$ which takes each decision node $x⋅∈⋅W$ and each feasible action $a⋅∈⋅F(x)$ to the node $n(x,a)$ at the end of the edge from~$x$ that is labeled by $a$.  Casually, $n(x,a)$ is the result of choosing $a$ at $x$.  Formally, assume $λ$ is locally injective and define{\noteFgraph} $n{:}F\gr→X⧷⎨r⎬$ implicitly by{\noteln} \begin{gather}
\zz
(∀(x,a)∈F\gr)⋅λ(˙x˙n(x˛a)˙) = a. \label{E869} 
\zz
\end{gather} Thus $n(x˛a)$ is defined to be the unique $y_*$ such that $λ(xy_*) = a$.  Call $n$ the {\em next-node function} (it is well-defined by Lemma \rf{A816}(\rf{A817})).  For example, $n(\f3,\fc) = \f6$ in Figure~\rf{E820}.

\newcommand{\notelhc}{\footnote{This is equivalent to assuming that $F{:}W⇉A$ is lower semicontinuous in the sense of Berge 1963,\nocite{Berge6397} page 109, when its domain is endowed with the information topology, and when its codomain is endowed with the discrete topology.}}
\nichts{}
\nichts{}
From a different perspective, $F = ⁅F(x)⁆_{x∈W}$ can be regarded as a set-valued function from $W$ to $\PP(A)$.  It will be assumed that $F$ is continuous as a function from~$W$, endowed with the information topology, to $\PP(A)$, endowed with the discrete topology.{\notelhc}  This is equivalent to \begin{gather}
\zz
(∀H∈\HH,x_1∈H,x_2∈H)⋅F(x_1) = F(x_2), \label{E780}
\zz
\end{gather} which requires that two nodes in one information set have the same feasible set.  For example in Figure~\rf{E820}, nodes $\f3$ and $\f4$ share an information set, and in accord with~(\rf{E780}), the two nodes share the feasible set $F(\f3) = F(\f4) = ⎨\fb,\fc⎬$. 

\begin{ndef}\label{E772} A {\bf continuously labeled tree (CLT)} is a tuple $Θ =$ $(X,E,\HH,λ)$ such that\begin{gather}
\zz
(X,E)⋅\text{is an out-tree (\rf{E776})}, \tag{C1}\label{C1} \\
\HH⋅\text{is a partition of $W$}, \tag{C2}\label{C2} \\
λ⋅\text{is a locally injective (\rf{E778}) surjective function from $E$},⋅\text{and} \tag{C3}\label{C3}\\
⁅F(x)⁆_{x∈W}⋅\text{is continuous (\rf{E780}) from}⋅W \tag{C4}\label{C4} 
\zz
\end{gather} (where $E$ determines $W$ by (\rf{E812}) and $λ$ determines $F$ by (\rf{E779})). \end{ndef}

Note that a tree diagram, with underlined root node, shaded information sets, and labeled edges, can unambiguously specify all the components of a (small) CLT.  For example, Figure \rf{E820} defines a CLT.  In an entirely different vein, note that the ``smallest'' CLTs have a one-node out-tree $(X,E) = (⎨x⎬,∅)$, an empty partition $\HH$, and an empty labeling function $λ$.  Here the decision-node set $W$ is empty.

\nssec{CLT morphisms}{E770} 

This Section~\rf{E770} will define CLT morphisms.  This first paragraph fixes some general notation.  For $f{:}X→Y$, define $\bar{f}{:}\PP(X)→\PP(Y)$ by $\bar{f}(A) = ⎨\,f(x)\,|\,x∈A\,⎬$.\nichts{}  Also, for $f{:}X→Y$, $A\,⊆\,X$, and $B⋅⊆⋅Y$, let $f|_{A,B}$ be the function with domain $A$, codomain $B$, and the graph of $f|_A$.  This construction is well-defined iff $\bar{f}(A)⋅⊆⋅B$.

\newcommand{\notegrph}{\nichts{}}
Consider a source CLT $Θ$ and a target CLT $Θ′$.  A {\em node transformation} is a function of the form $τ{:}X→X′$.  Definition~\rf{E773} will assume that $τ$ {\em preserves edges} in the sense that \begin{gather}
\zz
(∀xy∈E)⋅τ(x)τ(y)⋅∈⋅E′. \label{E870} 
\zz
\end{gather}  Edge preservation implies that $[(X,E),(X′,E′),τ]$ is a graph morphism,{\notegrph} and easily leads to the several consequences in Proposition~\rf{A481}(\rf{A483}--\rf{A482}).  Among these consequences is $\dtau(W)⋅⊆⋅W′$, which implies that $τ|_{W,W′}$ is well-defined (recall the general notation of the previous paragraph).

Definition~\rf{E773} will also assume that $τ|_{W,W′}$ is continuous.  As would be expected from general topology, this means that the inverse image of every target information set is the union of a collection of source information sets.  However, Proposition~\rf{B063} provides another perspective by taking advantage of the fact that $W$ and $W′$ are endowed with partition topologies.  The proposition shows that continuity is equivalent to the image of each source information set being included in a target information set.  In other words, continuity says that source information sets cannot be ``split''.\nichts{}  

\begin{prop}\label{B063} Suppose $Θ$ and $Θ′$ are CLTs, $τ{:}X→X′$, and $\dtau(W)⋅⊆⋅W′$.  Then $τ|_{W,W′}$ is continuous iff $(∀H∈\HH)(∃H′∈\HH′)$ $\dtau(H)⋅⊆⋅H′$.  (Proof~\rf{B063p}.) \end{prop}

For example, consider Figure~\rf{A467}.  There $τ|_{W,W′} = \id_W$ is discontinuous because the inverse image $⎨\,x∈W\,|\,τ(x)∈⎨\f4⎬\,⎬ = ⎨\f4⎬$ of the target information set $⎨\f4⎬$ fails to be open in the source information topology.\nichts{}  Equivalently, $τ|_{W,W′} = \id_W$ is discontinuous because the image $\dtau(⎨\f3,\f4⎬) = ⎨\f3,\f4⎬$ of the source information set $⎨\f3,\f4⎬$ is not included in a target information set.  In other words, $τ|_{W,W′} = \id_W$ is discontinuous because it ``splits'' the source information set $⎨\f3,\f4⎬$.   In contrast, consider Figure~\rf{A803} or Figure~\rf{A767}.  In each of those cases, $τ|_{W,W′} = \id_W$ is continuous since no source information sets are split.

\begin{figure}[t] \newcommand{\hgth}{4.60}
\begin{picture}(0,\hgth) \myoutergrid{\hgth} \renewcommand{\sellb}[1]{#1}
  \put( 0.00, 2.15){\makebox(0,0){\scalebox{0.88}{ \begin{pspicture} \end{pspicture} }}} \end{picture}
\caption{\small The CLT $Θ$ (from Figure~\rf{E820}) and the CLT $Θ′$ have different information sets.  The tuple $[Θ,Θ′,\id_X]$ is not a morphism because $τ|_{W,W′} = \id_W$ is not continuous.} \label{A467} \end{figure}

To interpret continuity in the context of game theory, we only need two observations: [1] that large information sets correspond to less information, and [2] that continuity prevents information sets from being split.  Thus, continuity corresponds to preserving a lack of information. 

\pagebreak
Next, from the tuple $[Θ,Θ′,τ]$, derive $α = ⁅α_x{:}F(x)→F′(τ(x))⁆_{x∈W}$ at each decision node $x⋅∈⋅W$ by \begin{gather}
\zz
(∀a∈F(x))⋅α_x(a) = λ′(\,τ(x)\,τ(n(x{,}a))\,) \label{E782}
\zz
\end{gather} (where the source labeling function $λ$ determines the source next-node function $n$ by (\rf{E869})).  Call $α = ⁅α_x⁆_{x∈W}$ the tuple's {\em action transformation}.  To unpack its definition, first note that at each $x⋅∈⋅W$, the statement $α_x{:}F(x)→F′(τ(x))$ means that the function $α_x$ sends source actions feasible at $x$ to target actions feasible at the image of $x$.  Then consider a source action $a$ which is feasible at $x$.  By (\rf{E782}), this $a$ is sent to the target action that labels the target edge from the image of $x$ to the image of the next source node determined by $x$ and $a$.\nichts{} 

\begin{prop}\label{E871} Suppose $Θ$ and $Θ′$ are CLTs, and $τ{:}X→X′$ preserves edges (\rf{E870}).  Then the action transformation $α = ⁅α_x⁆_{x∈W}$ (\rf{E782}) is well-defined. (Proof~\rf{E871p}.) \end{prop}

\newcommand{\notepfalpha}{\footnote{\label{F708}Take a source information set $H⋅∈⋅\HH$ and two decision nodes $x_1$ and $x_2$ in $H$.  It suffices to show that the domains and codomains of $α_{x_1}{:}F(x_1)→F′(τ(x_1))$ and $α_{x_2}{:}F(x_2)→F′(τ(x_2))$ are equal.  Axiom (\rf{C4}) for $Θ$ implies $F(x_1) = F(x_2)$.  Further, Proposition~\rf{B063} and the assumed continuity of $τ|_{W,W′}$ implies there is $H′⋅∈⋅\HH′$ such that $⎨τ(x_1),τ(x_2)⎬$ $⊆$ $H′$.  Thus (\rf{C4}) for $Θ′$ implies $F′(τ(x_1)) = F′(τ(x_2))$.}}
As a whole, $α = ⁅α_x⁆_{x∈W}$ is a function-valued function with domain $W$.  Definition~\rf{E773} will assume that $α$ is continuous as a function from $W$, endowed with the information topology, to the set of functions, endowed with the discrete topology.  By topology, this is equivalent to assuming that  \begin{gather}
\zz
(∀H∈\HH,x_1∈H,x_2∈H)⋅α_{x_1} = α_{x_2}. \label{E781} 
\zz
\end{gather} Further, it can be shown{\notepfalpha} that if $τ|_{W,W′}$ is continuous, then (\rf{E781}) is equivalent to \begin{gather}
\zz
(∀H∈\HH,x_1∈H,x_2∈H)⋅(a∈F(x_1))⋅α_{x_1}(a) = α_{x_2}(a). \label{E874} 
\zz
\end{gather} In this precise sense, the continuity of $α$ is equivalent to requiring that two nodes in the same information set transform actions in the same way.

For example, consider Figure~\rf{A803}.  Here \begin{gather}
\zz
α_{\f3}(\fb) = λ′(\,τ(\f3)\,τ(n(\f3,\fb))\,)  = λ′(\,τ(\f3)\,τ(\f5)\,)  = λ′(\,\f3\,\f5\,) = \fe⋅\text{and} \nt
α_{\f4}(\fb) = λ′(\,τ(\f4)\,τ(n(\f4,\fb))\,) = λ′(\,τ(\f4)\,τ(\f7)\,) = λ′(\,\f4\,\f7\,) = \ff.
\notag
\zz
\end{gather} Since these differ, $α_{\f3}⋅≠⋅ α_{\f4}$.  Thus (\rf{E781}) fails since $\f3$ and $\f4$ belong to the same source information set $H = ⎨\f3,\f4⎬⋅⊆⋅W$.  Hence $α$ is discontinuous from $W$.  In contrast, Figure~\rf{A468} provides an example of a continuous $α$.  There $α_{\f3}(\fb) = α_{\f4}(\fb) = \fe$ and $α_{\f3}(\fc) = α_{\f4}(\fc) = \ff$.  Thus by (\rf{E874}), $α_{\f3} = α_{\f4}$.  So by (\rf{E781}) and the fact that $H = ⎨\f3,\f4⎬⋅⊆⋅W$ is the only nonsingleton source information set, $α$ is continuous from $W$. 

\begin{figure}[t] \newcommand{\hgth}{4.60}
\begin{picture}(0,\hgth) \myoutergrid{\hgth} \renewcommand{\sellc}[1]{#1}
  \put( 0.00, 2.15){\makebox(0,0){\scalebox{0.88}{ \begin{pspicture} \end{pspicture} }}} \end{picture}
\caption{\small The CLT $Θ$ (from Figure~\rf{E820}) and the CLT $Θ′$ have different actions.  The tuple $[Θ,Θ′,\id_X]$ is not a morphism because its $α = ⁅α_x⁆_{x∈W}$ is not continuous (\rf{E781}) from $W$.} \label{A803} \end{figure}
 
\begin{figure}[t] \newcommand{\hgth}{4.60}
\begin{picture}(0,\hgth) \myoutergrid{\hgth} \renewcommand{\sella}[1]{#1}
  \put( 0.00, 2.15){\makebox(0,0){\scalebox{0.88}{ \begin{pspicture} \end{pspicture} }}} \end{picture}
\caption{\small The CLT $Θ$ (from Figure~\rf{E820}) and the CLT $Θ′$ have different actions.  The tuple $[Θ,Θ′,\id_X]$ is a morphism.} \label{A468} \end{figure}

Note that the continuity of $α$ from $W$ (\rf{E781}) allows ``local'' action changes in the sense that actions can be changed differently at different information sets.  For example, consider Figure~\rf{A468}, and note $α_{\f0}(\fb) = \fb$ and $α_{\f3}(\fb) = \fe$.  This is consistent with the continuity of $α$ (\rf{E781}) because $\f0$ and $\f3$ are in different information sets.

\pagebreak

\newcommand{\yyyclt}{\vspace{1.8ex}}
\newcommand{\yyyez}{\vspace{.2ex}}

\begin{ndef}\label{E773} A {\bf CLT morphism} is a tuple $θ = [Θ,Θ′,τ]$ such that $Θ$ and $Θ′$ are CLTs (Definition~\rf{E772}), $τ{:}X→X′$, \vspace{1mm} %
\begin{gather}
\zz
τ⋅\text{is edge-preserving (\rf{E870})}, \tag{cE}\label{cE} \\ 
τ|_{W,W′}⋅\text{is continuous},⋅\text{and} \tag{cH}\label{cH} \\ 
⁅α_x⁆_{x∈W}⋅\text{is continuous (\rf{E781}) from}⋅W \tag{cL}\label{cL} \\[-5.5mm] %
\notag %
\zz
\end{gather} (where $[Θ,Θ′,τ]$ determines its action transformation $⁅α_x⁆_{x∈W}$ by (\rf{E782})). \end{ndef}

\yyyclt

Mnemonically, (\rf{cE}) abbreviates ``edge preservation'', (\rf{cH}) ``information-set preservation'', and (\rf{cL}) ``label preservation'' (the leading ``c'' in (\rf{cE}), (\rf{cH}), and (\rf{cL}) signifies ``CLT'').  Note that the source root $r$ may or may not be taken to the target root $r′$. 

\yyyclt

\newcommand{\notefbar}{\footnote{\label{A639}Concerning the symbol $\dtau$, recall that Section~\rf{E770}'s first paragraph defined $\bar{f}(A)$ to be $⎨\,f(x)\,|\,x∈A\,⎬$ for arbitrary $f{:}X→Y$ and $A⋅⊆⋅X$.}}
\begin{prop}\label{A481} \mbox{Suppose $[Θ,Θ′,τ]$ is a morphism. Then the following hold.} \begin{tlist}
\yyyez\yl{A483} $(∀x∈X,y∈X)⋅x⋅≼⋅y⋅⇒⋅τ(x)⋅≼′⋅τ(y)$.
\yyyez\yl{A484} $(∀x∈X,y∈X)⋅x⋅≺⋅y⋅⇒⋅τ(x)⋅≺′⋅τ(y)$.
\yyyez\yl{A486} $(∀y∈X)⋅\dtau(P(y))⋅⊆⋅P′(τ(y))$.\notefbar
\yyyez\yl{A485} $(∀y∈X⧷⎨r⎬)⋅τ(p(y)) = p′(τ(y))$.
\yyyez\yl{A482} $\dtau(W)⋅⊆⋅W′$.$^{\rf{A639}}$  
\yyyez\yl{A804} $(∀xy∈E)⋅α_x(λ(xy)) = λ′(τ(x)τ(y))$.
\yyyez\yl{A490} $(∀x∈W,a∈F(x))⋅τ(n(x,a)) = n′(τ(x),α_x(a))$. (Proof~\rf{A481p}.)
\end{tlist} \end{prop}

\vspace{6.5mm} %
\nssec{The category \ct{CLT}}{F221}

\vspace{2.5mm} %
This paragraph defines the category \ct{CLT}, which is called the {\em category of CLTs (continuously labeled trees)}.  Let an object be a CLT $Θ$ (Definition~\rf{E772}).  Let an arrow be a CLT morphism $[Θ,Θ′,τ]$ (Definition~\rf{E773}).   Let source, target, identity, and composition be 
\vspace{2mm} %
\begin{gather}
\zz
[Θ,Θ′,τ]^\src = Θ,⋅[Θ,Θ′,τ]^\trg = Θ′,⋅\id_Θ = [Θ,Θ,\id_X],⋅\text{and} \label{E790} \\
[Θ′,Θ″,τ′]○[Θ,Θ′,τ] = [Θ,Θ″,τ′○τ] \notag \\[-4.5mm] %
\notag %
\zz
\end{gather} 
(where $\id_X$ is an identity in \ct{Set}).

\yyyclt

\begin{prop}\label{A453} ⋅ \begin{tlist}
\yyyez\yl{F232} \ct{CLT} is well-defined. 
\yyyez\yl{F233} If $Θ$ is a CLT, then the action transformation of $\id_Θ$ is $⁅\id_{F(x)}⁆_{x∈W}$.
\yyyez\yl{F234} If $[Θ,Θ′,τ]$ and $[Θ′,Θ″,τ′]$ are morphisms, then the action transformation of $[Θ′,Θ″,τ′]○[Θ,Θ′,τ]$ is $⁅α′_{τ(x)}○α_x⁆_{x∈W}$. (Proof~\rf{A453p}.) \end{tlist} \end{prop}

\pagebreak

\section{Definition of \ct{Gm}}\label{F219}\showit
\markb{\sc \rf{F219}. Definition of \ct{Gm}}

\nssec{Games}{B034}
\newcommand{\notehazardcontrol}{\footnote{\label{E816}This specification of a control-assigning function may be slightly unfamiliar.  First, the term ``control-assigning'' is introduced here since it is less ambiguous than ``player-assigning'' or ``move-assigning''.  Second, some other specifications freshly partition the decision-node set, and then assign each player to exactly one of the new partition's cells.  This is effectively a map from decision nodes to players.  Third, a few other specifications map information sets to players.  This implies a map from decision nodes to players and implicitly imposes continuity (\rf{E783}).}}
\newcommand{\noteweirdI}{\footnote{\label{F800}It may seem strange to derive $I$.  Alternatively, one could firstly let $I$ be a set called the set of ``players'', and secondly let $κ$ be a surjective function from the decision-node set $W$ to the player set $I$.  This alternative specification is avoided because it would necessitate adding $I$ to the tuple $(X,E,\HH,λ,κ,u)$ specifying a game (Definition~\rf{E775}).  Instead, (\rf{F218}) derives $I$ from $κ$.  (This is very similar to the derivation of $A$ from $λ$ in equation~(\rf{F217}) and note~\rf{B216}.)  To be absolutely clear, recall from note~\rf{A629} that a function is regarded as a triple.  So the assumption that $κ = [κ^{\mathsf{dom}},κ^{\mathsf{cod}},κ\gr]$ is a surjective function from $W$ implies that  $κ^{\mathsf{dom}} = W$ and $κ^{\mathsf{cod}} = \bar{κ}(I)$.  In this context, (\rf{F218}) defines $I$ to be $κ^{\mathsf{cod}} = \bar{κ}(W)$.  Accordingly, Table~\rf{A452} lists $κ$ as a component of a game, and $I$ as a derivative of a game.}}
To define a game,%
\begin{picture}(0,0) %
  \put( 2.0,2.15){\color{white} \rule{25ex}{2.5ex}}
  \put( 2.33,2.18){{\sc\SMALL 3.\ Definition of \ct{Gm}}}
  \end{picture}
begin with a CLT $Θ = (X,E,\HH,λ)$.  Then let $κ$ be a surjective function from the decision-node set $W$ to a discrete (i.e.\ unstructured) codomain.  Let~$I$ be the codomain of $κ$.{\noteweirdI} In other words (since $κ$ is surjective), let\begin{gather}
\zz
I = \bar{κ}(W). \label{F218}
\zz
\end{gather}   Call $κ$ the {\em control-assigning function} (elsewhere the ``player-assigning function'' or ``move-assigning function''), call a member of $I$ a {\em player}, and say that player $κ(x)$ {\em controls} node $x$.{\notehazardcontrol}  

Definition~\rf{E775} will assume that $κ$ is continuous. This means that, for each $i⋅∈⋅I$, the inverse image $⎨\,x∈W\,|\,κ(x){=}i\,⎬$ is open in the information topology.  This is equivalent to saying that, for each $i⋅∈⋅I$, $⎨\,x∈W\,|\,κ(x){=}i\,⎬$ is the union of a collection of information sets.  Equivalently,  \begin{gather}
\zz
(∀H∈\HH,x_1∈H,x_2∈H)⋅κ(x_1) = κ(x_2), \label{E783} 
\zz
\end{gather} which requires that two nodes in one information set are assigned to the same player.

\newcommand{\notehazardutil}{\footnote{\label{E817}This specification of a utility function may be slightly unfamiliar.  First, in finite-horizon games, it is common to assign utility numbers to end nodes rather than to runs.  In this special case, there is a bijection between the end-node set $X⧷W$ and the run collection $\ZZ$ (here $\ZZ$ equals the collection $\ZZf$ of finite runs, as defined prior to (\rf{E813})). Second, infinite utility numbers are included because, in economics, many popular utility functions generate ${-}∞$ utility when some consumption level is zero.  Such a utility function is often part of a consumer dynamic optimization problem, and such a problem can be specified as a one-player game.  Third, preferences are occasionally specified by a total preorder like (\rf{E784}) rather than by a utility function.\nichts{}}}
\newcommand{\noterep}{\footnote{To be clear, a total preorder is a total and transitive binary relation.  In economics, totality is called ``completeness'' and a total preorder is called an ``ordering''.}}
Next, for each $i⋅∈⋅I$, consider a function $U_i{:}\ZZ→\eR$, where $\eR$ denotes the set of extended real numbers, that is, $Ṛ⋃⎨{-}∞,∞⎬$.{\notehazardutil}  Call $U_i$ the {\em utility function} of player~$i$ (elsewhere the ``payoff function'' of player $i$).  From the function $U_i{:}\ZZ→\eR$, derive the total preorder{\noterep} $\lesssim_i$ on $\ZZ$ defined by \begin{gather}
\zz
(∀Z_1∈\ZZ,Z_2∈\ZZ)⋅Z_1 \lesssim_i Z_2⋅\text{iff}⋅U_i(Z_1)⋅≤⋅U_i(Z_2). \label{E784}
\zz
\end{gather} In economics, the function $U_i$ is said to ``represent'' the total preorder $\lesssim_i$, and conversely, $\lesssim_i$ is said to be the ``ordinal content'' of $U_i$.\nichts{}

\begin{ndef}\label{E775} An {\bf (extensive-form) game} is a tuple $Γ = (X,E,\HH,λ,κ,U)$ such that \begin{gather}
\zz
(X,E,\HH,λ)⋅\text{is a CLT (Definition~\rf{E772})}, \tag{G1}\label{G1}\\
κ⋅\text{is a continuous (\rf{E783}) surjective function from $W$},⋅\text{and} \tag{G2}\label{G2}\\
U⋅\text{is of the form}⋅⁅U_i{:}\ZZ→\eR⁆_{i∈I} \tag{G3}\label{G3}
\zz
\end{gather} (where $(X,E)$ determines $W$ and $\ZZ$ by (\rf{E812}) and (\rf{E813}), and $I$ is $κ$'s codomain (\rf{F218})). \end{ndef}

Figure \rf{E613} in Appendix~\rf{F258} provides six examples.  Note that a CLT diagram, with controlling players and utility vectors, can unambiguously specify almost all the components of a (small) game.  The only thing that may need to be specified externally is the order in which the utility vectors list the players.  In an entirely different vein, note that the ``smallest'' games have a one-node out-tree $(⎨x⎬,∅)$, an empty partition $\HH$, an empty labeling function $λ$, an empty control-assigning function $κ$, and an empty utility-function vector $U$.  Here the player set $I$ is empty. 

\vspace{3ex} %

\nssec{Game morphisms}{E771}

A game morphism $[Γ,Γ′,τ]$ (Definition~\rf{E774}) will be built on top of a CLT morphism $[Θ,Θ′,τ]$ (Definition~\rf{E773}).  The essential difference is that a game morphism preserves not only edges, information sets, and labels, but also [1] the control that players exercise at decision nodes, and [2] the ordinal content (\rf{E784}) of the utility numbers that players assign to runs.

Consider two games $Γ$ and $Γ′$ together with a node transformation $τ{:}X→X′$.  The triple $[Γ,Γ′,τ]$ is said to {\em preserve control} iff\footnote{In (\rf{E788})'s concluding equation, each side is well-defined because $(∀w∈W)$ $τ(w)⋅∈⋅W′$ by Proposition~\rf{A481}(\rf{A482}).} \begin{gather}
\zz
(∀x_1∈W,x_2∈W)⋅κ(x_1) = κ(x_2)⋅\text{implies}⋅κ′(τ(x_1)) = κ′(τ(x_2)). \label{E788}
\zz
\end{gather} This requires that any two source nodes controlled by one player must be sent to target nodes controlled by one player.  Figure~\rf{A805} shows an example which violates this assumption.  

\begin{figure}[h] \newcommand{\hgth}{2.50} %
\begin{picture}(0,\hgth) \myoutergrid{\hgth} \renewcommand{\sellc}[1]{#1}
  \put( 0.00, 1.00){\makebox(0,0){\scalebox{0.90}{ \begin{pspicture} \end{pspicture} }}} \end{picture}
\caption{\small The games $Γ$ and $Γ′$ have different players.  The tuple $[Γ,Γ′,\id_{X}]$ is not a game morphism since it is not control-preserving~(\rf{E788}).} \label{A805} \end{figure}

\pagebreak
From a control-preserving tuple $[Γ,Γ′,τ]$, derive the function $ι{:}I→I′$ with graph  \begin{gather}
\zz
ι\gr = ⎨˙(κ(x),κ′(τ(x)))˙|˙x∈W˙⎬. \label{E785} 
\zz
\end{gather} Call $ι$ the {\em player transformation} of $[Γ,Γ′,τ]$.  Equation (\rf{E785}) says that $ι$ maps the player controlling decision node $x$ to the player controlling the image of $x$.  Proposition~\rf{E791}(a) shows that $ι$ is well-defined.  Further, Proposition~\rf{E791}(b) implies that $(∀x∈W)$ $ι(κ(x)) = κ′(τ(x))$, which says (like (\rf{E785}) did) that for each decision node $x⋅∈⋅W$, the source player controlling $x$ is sent by $ι$ to the target player controlling the image of $x$. 

\begin{prop}\label{E791} Suppose that $Γ$ and $Γ′$ are games, that $[Θ,Θ′,τ]$ is a CLT morphism, and that $[Γ,Γ′,τ]$ is a control-preserving (\rf{E788}) tuple.  Then (a) the player transformation $ι$ (\rf{E785}) is well-defined, and (b) $ι○κ = κ′○τ|_{W,W′}$. (Proof~\rf{E791p}.) \end{prop}

Before considering utility functions, recall (from near (\rf{E813})) that $X⧷W$ is the set of end nodes, and say that a CLT morphism $[Θ,Θ′,τ]$ {\em preserves ends} iff \begin{gather}
\zz
\dtau(X⧷W)⋅⊆⋅X′⧷W′. \label{E787}
\zz
\end{gather} Figure~\rf{A460} shows a morphism which does not preserve ends, because the source end node~$\f2$ is mapped to the target decision node $\f{12}$.  Accordingly, the run $⎨\f0,\f2⎬$ is mapped to the non-run walk $⎨\f{10},\f{12}⎬$.  Arguably, the source utility assigned to the source run $⎨\f0,\f2⎬$ should imply nothing about the target utilities assigned to the target runs $⎨\f{10},\f{12},\f{13}⎬$ and $⎨\f{10},\f{12},\f{14}⎬$ because the connection between that source run and those two target runs is so nebulous.  For this reason, \ct{Gm} game morphisms will only be defined over the CLT morphisms that preserve ends.%
\footnote{Streufert 2020b (Section 3.2) shows how to define a more general notion of game morphism that is defined over all CLT morphisms.  This generalization weakens monotonicity ((\rf{gU}) in Definition~\rf{E774}) by requiring that the condition holds over just the set of source runs whose ends the morphism preserves (in Figure~\rf{A460}, this set is just the singleton $⎨⎨\f0,\f1⎬⎬$; in Streufert 2020b (equation~[g4], page~15), this set is denoted $\ZZ^θ⋅⊆⋅\ZZ$, where $θ$ is the morphism).  This extra generality does not seem to be worth the extra work because [a] partial utility preservation is not that interesting and [b] end-preservation (\rf{E787}) is not that restrictive.  Regarding~[b], every CLT isomorphism is end-preserving (Streufert 2020b, Proposition~2.3(c)), and every CLT morphism from a CLT with only infinite plays is end-preserving (simply because there are no ends to be preserved).  Further, any inclusions like Figure~\rf{A461} are also end-preserving (such inclusions arise naturally in the context of subgame perfection (\nocite{Selten75}Selten 1975)).}

\begin{figure}[h] \newcommand{\hgth}{3.60}
\begin{picture}(0,\hgth) \myoutergrid{\hgth} \renewcommand{\sella}[1]{#1}
  \put( 0.00, 1.80){\makebox(0,0){\scalebox{0.90}{ \begin{pspicture} \end{pspicture} }}} \end{picture}
\caption{\small The CLT morphism $[Θ,Θ^*,τ]$ defined by $τ(\mathsf{x}){=}\mathsf{x{+}10}$. This morphism is not end-preserving (\rf{E787}).} \label{A460} \end{figure}

In contrast, Figure~\rf{A461} shows an end-preserving morphism.  Here it seems reasonable that a player's preference between the source runs $⎨\f0,\f1⎬$ and $⎨\f0,\f2⎬$ could be preserved in a preference between the target runs $⎨\f{50},\f{10},\f{11}⎬$ and $⎨\f{50},\f{10},\f{12}⎬$.  Note that the images $⎨\f{10},\f{11}⎬$ and $⎨\f{10},\f{12}⎬$ of the two source runs are preceded by the target node $\f{50}$, and that $\f{50}$ is not the image of any source node. 

\begin{figure}[h] \newcommand{\hgth}{4.00}
\begin{picture}(0,\hgth) \myoutergrid{\hgth} \renewcommand{\sellb}[1]{#1}
  \put( 0.00, 2.00){\makebox(0,0){\scalebox{0.90}{ \begin{pspicture} \end{pspicture} }}} \end{picture}
\caption{\small The CLT morphism $[Θ,Θ′,τ]$ defined by $τ(\mathsf{x}){=}\mathsf{x{+}10}$.  This morphism is end-preserving (\rf{E787}).} \label{A461} \end{figure}
 
Accordingly, consider an end-preserving CLT morphism $[Θ,Θ′,τ]$.  Then derive the function $ζ{:}\ZZ→\ZZ′$ by\nichts{} \begin{gather}
\zz
ζ(Z) = P′○τ(r)⋃\dtau(Z). \label{E786}
\zz
\end{gather} Call $ζ$ the {\em run transformation}.  Proposition~\rf{A458} shows that $ζ$ is well-defined.  For example, in Figure~\rf{A461},\begin{gather}
\zz
ζ(⎨\f0,\f2⎬) = P′○τ(r)\,⋃\,\dtau(⎨\f0,\f2⎬) = ⎨\f{50}⎬\,⋃\,⎨\f{10},\f{12}⎬= ⎨\f{50},\f{10},\f{12}⎬, \notag
\zz
\end{gather} where $P′○τ(r) = P′○τ(\f0) = P′(\f{10}) = ⎨\f{50}⎬$.  

\enlargethispage*{\baselineskip}

In English, $ζ$ maps a source run $Z$ to its image $\bar{τ}(Z)$ and then prepends $\bar{τ}(Z)$ with the walk from $r′$ to $τ(r)$.  This verbal characterization is accurate because the walk from $r′$ to $τ(r)$ is $P′○τ(r)⋃⎨τ(r)⎬$, and because $τ(r)⋅∈⋅\bar{τ}(Z)$ since $r⋅∈⋅Z$ because $Z$ is a run (in the example, $τ(r) = τ(\f0) = \f{10}$).

\begin{prop}\label{A458} If a CLT morphism $[Θ,Θ′,τ]$ is end-preserving (\rf{E787}), then its run transformation $ζ$ (\rf{E786}) is well-defined. (Proof~\rf{A458p}.) \end{prop}

\begin{ndef}\label{E774} A {\bf game morphism} is a tuple $γ = [Γ,Γ′,τ]$ such that $Γ$ and $Γ′$ are games (Definition~\rf{E775}),  \begin{gather}
\zz
[Θ,Θ′,τ]⋅\text{is an end-preserving (\rf{E787}) CLT morphism (Definition~\rf{E773})}, \tag{gZ}\label{gZ}\\ 
[Γ,Γ′,τ]⋅\text{is control-preserving (\rf{E788}),}⋅\text{and} \tag{gK}\label{gK}\\
(∀i∈I)⋅ζ⋅\text{is monotone from}⋅(\ZZ,\lesssim_i)⋅\text{to}⋅(\ZZ′,\lesssim′_{ι(i)}) \tag{gU}\label{gU} 
\zz
\end{gather} (where $[Γ,Γ′,τ]$ determines $ι$ and $ζ$ by (\rf{E785}) and (\rf{E786})). \end{ndef}

To summarize briefly, (\rf{gZ}) explicitly preserves ends, and implicitly preserves edges, information sets and labels via the Definition \rf{E773} of a CLT morphism.  Then (\rf{gK}) preserves control.  Finally, the definition (\rf{E784}) of $\lesssim_i$ implies that (\rf{gU}) is equivalent to \begin{gather}
\zz
(∀i∈I,Z_1∈\ZZ,Z_2∈\ZZ)⋅U_i(Z_1)⋅≤⋅U_i(Z_2)⋅\text{implies}⋅U_{ι(i)}(ζ(Z_1))⋅≤⋅U_{ι(i)}(ζ(Z_2)). \label{E881}
\zz
\end{gather} In this sense, (\rf{gU}) preserves the ordinal content of the utility functions.  

\pagebreak

\nssec{The category \ct{Gm}}{F222}
This paragraph defines%
\begin{picture}(0,0) %
  \put( 1.21,1.50){\color{white} \rule{25ex}{2.5ex}}
  \put( 1.43,1.54){{\sc\SMALL 3.\ Definition of \ct{Gm}}}
  \end{picture}
the category \ct{Gm}, which is called the {\em category of exten\-sive-form games}.  Let an object be a game $Γ$ (Definition~\rf{E775}).  Let an arrow be a game morphism $[Γ,Γ′,τ]$ (Definition~\rf{E775}).  Let source, target, identity, and composition be\begin{gather}
\zz
[Γ,Γ′,τ]^\src = Γ,⋅[Γ,Γ′,τ]^\trg = Γ′,⋅\id_Γ = [Γ,Γ,\id_X],⋅\text{and} \label{E789} \\
[Γ′,Γ″,τ′]○[Γ,Γ′,τ] = [Γ,Γ″,τ′○τ]. \notag
\zz
\end{gather}  

\begin{prop}\label{A454} ⋅ \begin{tlist}
\yl{F236} \ct{Gm} is well-defined.  
\yl{F237} Suppose $Γ$ is a game.  Then the player transformation of $\id_Γ$ is $\id_I$, and its run transformation is $\id_{\ZZ}$.
\yl{F238} Suppose $[Γ,Γ′,τ]$ and $[Γ′,Γ″,τ′]$ are morphisms.  Then the player transformation of $[Γ′,Γ″,τ′]○[Γ,Γ′,τ]$ is $ι′○ι$, and its run transformation is $ζ′○ζ$.  (Proof~\rf{A454p}.) \end{tlist} \end{prop}

\begin{prop}\label{A455} Define $\FB$ from \ct{Gm} to \ct{CLT} by\begin{gather}
\zz
\FO⋅{:}⋅(X,E,\HH,λ,κ,U) \mapsto (X,E,\HH,λ)⋅\text{and} \nt 
\FA⋅{:}⋅[Γ,Γ′,τ] \mapsto [\FO(Γ),\FO(Γ′),τ]. \notag
\zz
\end{gather} Then $\FB$ is a well-defined functor. (Proof~\rf{A455p}.) \end{prop}

The ``forgetful'' functor of Proposition~\rf{A455} allows Section~\rf{A620}'s results for CLTs to be applied to games in a natural but rigorous way.  For example, Propositions \rf{A455} and \rf{A453}(\rf{F234}) imply that if $γ$ and $γ′$ are two game morphisms, then the action transformation of $γ′○γ$ is $⁅α′_{τ(x)}○α(x)⁆_{x∈W}$.

\vspace{1mm}\section{Special Morphisms}\label{B023}\showit
\markb{\sc \rf{B023}. Special Morphisms}

\nssec{Monomorphisms}{B035}

\begin{prop}\label{B185} In \ct{CLT}, a morphism $[Θ,Θ′,τ]$ is monic iff $τ$ is injective. (Proof~\rf{B185p}.) \end{prop} 

Proposition~\rf{B185} characterizes a monomorphism in \ct{CLT} by the injectivity of $τ$.  The injectivity of $τ$ implies the injectivity of each action transformation $α_x$ (Lemma~\rf{A713}).  Figure~\rf{B029} provides an example of monomorphism in \ct{CLT} (here $\inc_{X,X′}$ is the inclusion function from $X$ to $X′$).  Another example is Figure~\rf{A767}.

\begin{figure}[h] \newcommand{\hgth}{2.90}
\begin{picture}(0,\hgth) \myoutergrid{\hgth} \renewcommand{\sellc}[1]{#1}
  \put(-0.40, 1.10){\makebox(0,0){\scalebox{0.90}{ \begin{pspicture} \end{pspicture} }}} \end{picture}
\caption{\small $[Θ,Θ′,\inc_{X,X′}]$ is a monomorphism in \ct{CLT}.} \label{B029} \end{figure}

\pagebreak

\begin{prop}\label{B178} In \ct{Gm}, a morphism $[Γ,Γ′,τ]$ is monic iff its run transformation $ζ$ is injective. (Proof~\rf{B178p}.) \end{prop} 

\newcommand{\yyymonic}{\vspace{3mm}}

\yyymonic
Proposition~\rf{B178} characterizes a monomorphism in \ct{Gm} by the injectivity of its run transformation $ζ$.  By $ζ$'s definition (\rf{E786}), the injectivity of $ζ$ is weaker than the injectivity of $τ$ in Proposition~\rf{B185}, and further, the next paragraph will show by example that the injectivity of $ζ$ is strictly weaker.  The salient difference between Propositions \rf{B185} and \rf{B178} is that game morphisms are built on CLT morphisms that are end-preserving (Definition~\rf{E774} condition (\rf{gZ})).%
\footnote{Accordingly, consider the \ct{CLT} subcategory having only end-preserving morphisms.  Here I would conjecture that monomorphisms are characterized by injective run transformations.}
\nichts{} 

\newcommand{\noteforget}{\footnote{This is consistent with Proposition \rf{A455}'s forgetful functor because a functor need not take a monomorphism to a monomorphism.\nichts{}}}
\yyymonic
For an example, consider Figure~\rf{B159}.  Here $ζ$ is injective, since it maps the two source runs to the two target runs.  Yet $τ$ is not injective since $τ(\ex{41}) = τ(\ex{42}) = \ex{40}$.  Thus Propositions~\rf{B185} and~\rf{B178} imply $[Γ,Γ′,τ]$ is monic but $[Θ,Θ′,τ]$ is not.{\noteforget}  To develop intuition for $[Γ,Γ′,τ]$ being monic, let $γ$ denote $[Γ,Γ′,τ]$ and consider the problem of specifying distinct $γ^1 = [Γ^*,Γ,τ^1]$ and $γ^2 = [Γ^*,Γ,τ^2]$ such that $γ○γ^1 = γ○γ^2$.  The non-injectivity of $τ$ at $\f{41}$ and $\f{42}$ suggests specifying an $x^*⋅∈⋅X^*$ such that $τ^1(x^*) = \f{41}$ and $τ^2(x^*) = \f{42}$.  But since $γ^1$ and $γ^2$ must be built on end-preserving CLT morphisms, and since $\f{41}$ and $\f{42}$ are not end nodes in $Γ$, $x^*$ cannot be an end node in $Γ^*$.  This suggests specifying a successor $y^*$ of $x^*$.  But then (\rf{cE}) for $γ^1$ implies $τ^1(y^*) = \f{81}$, and (\rf{cE}) for $γ^2$ implies $τ^2(y^*) = \f{82}$.  These equations and the definition of $τ$ imply $τ○τ^1(y^*) = \f{81}$ and $τ○τ^2(y^*) = \f{82}$, which contradict $γ○γ^1 = γ○γ^2$.\nichts{}

\begin{figure}[h] \newcommand{\hgth}{3.90} %
\begin{picture}(0,\hgth) \myoutergrid{\hgth} \renewcommand{\selle}[1]{#1}
  \put(-1.50, 1.60){\makebox(0,0){\scalebox{0.90}{ \begin{pspicture} \end{pspicture} }}} \end{picture} %
\caption{\small The \ct{Gm} monomorphism $[Γ,Γ′,τ]$ defined by $τ(x) = \ex{40}$ for $x⋅∈⋅⎨\ex{41},\ex{42}⎬$ and $τ(x) = x$ otherwise.  Its underlying $[Θ,Θ′,τ]$ is not a monomorphism in \ct{CLT}.} \label{B159} \end{figure}

\yyymonic
Monomorphisms in \ct{Gm} can also have non-injective action and player transformations.  Figure~\rf{B159}'s example has a non-injective action transformation because the definition of $α$ (\rf{E782}) implies both $α_{\f0}(\fb) = \ff$ and $α_{\f0}(\fd) = \ff$.\nichts{}  
Meanwhile, Figure~\rf{A748}'s example has a non-injective player transformation because the definition of $ι$ (\rf{E785}) implies both $ι(\plyR{P1}) = \plyR{P1}$ and $ι(\plyR{P2}) = \plyR{P1}$.\nichts{}\toenote{Extra. Another interesting game monomorphism is Figure~\rf{A785}.}

\begin{figure}[h] \newcommand{\hgth}{2.00}
\begin{picture}(0,\hgth) \myoutergrid{\hgth} \renewcommand{\sellb}[1]{#1}
  \put( 0.00, 1.00){\makebox(0,0){\scalebox{0.90}{ \begin{pspicture} \end{pspicture} }}} \end{picture}
\caption{\small The \ct{Gm} monomorphism $[Γ,Γ′,\id_{X}]$.  Its player transformation~$ι$ takes both $\plyR{P1}$ and $\plyR{P2}$ to~$\plyR{P1}$.} \label{A748} \end{figure}

\pagebreak ⋅ %
\nssec{Isomorphisms}{B036}
Proposition~\rf{A491}(\rf{F080}) uses the term ``homeomorphism'' from topology.  In general, recall that a function from one topological space to another is a homeomorphism iff the function is bijective and both the function and its inverse are continuous.  Proposition~\rf{A491}(\rf{F080}) applies this concept to $τ|_{W,W′}$.  To see the details, first note the proposition assumes $[Θ,Θ′,τ]$ is a morphism, which by Definition~\rf{E773} implies $τ|_{W,W′}$ is continuous.  Second note the proposition's part (\rf{F079}) requires $τ$ is bijective, which by Lemma~\rf{A537}(\rf{A501x}) implies $τ|_{W,W′}$ is bijective.  Thus in this context, $τ|_{W,W′}$ is a homeomorphism iff $(τ|_{W,W′})^{-1}$ is continuous.  

For an example, consider Figure~\rf{A767}.  Here $τ|_{W,W′} = \id_W$ is not a homeomorphism because $(τ|_{W,W′})^{-1} = \id_{W′}$ is not continuous.  To see this, note that the inverse splits the information set $⎨\f3,\f4⎬$, and that the splitting of information sets characterizes discontinuity by Proposition~\rf{B063}.

For a symmetric perspective which directly concerns the splitting of information sets, assume that $Θ$ and $Θ′$ are CLTs and that $τ{:}X→X′$ is bijective.  Then $τ|_{W,W′}$ is a homeomorphism iff the set-to-set function $\dtau|_{\HH,\HH′} = (\HH⋅∋⋅H \mapsto \dtau(H)⋅∈⋅\HH′)$ is a bijection with inverse $\overline{τ^{-1}}|_{\HH′,\HH} = (\HH⋅∋⋅\overline{τ^{-1}}(H′)⋅\mapsfrom⋅H′⋅∈⋅\HH′)$ (Lemma~\rf{F226}(\rf{F229}$⇔$\rf{F231})).  Thus a homeomorphism does not split information sets in either direction.\footnote{Incidentally, Lemma~\rf{F226}(\rf{F229}$⇔$\rf{F813}) shows $τ|_{W,W′}$ is a homeomorphism iff $(∀H∈\HH)$ $\dtau(H)⋅∈⋅\HH′$.  This seemingly weak condition is equivalent to $\dtau|_{\HH,\HH′}$ being well-defined.}

\begin{figure}[h] \newcommand{\hgth}{4.90} %
\begin{picture}(0,\hgth) \myoutergrid{\hgth} \renewcommand{\selld}[1]{#1}
  \put( 0.00, 2.20){\makebox(0,0){\scalebox{0.90}{ \begin{pspicture} \end{pspicture} }}} \end{picture}
\caption{\small The morphism $[Θ,Θ′,\id_X]$ is not an isomorphism because $τ|_{W,W′} = \id_W$ is not a homeomorphism.} \label{A767} \end{figure}

\newcommand{\noteWW}{\footnote{\label{F242}Condition (\rf{F079}) follows from condition (\rf{F080}) when neither $Θ$ nor $Θ′$ has an end node, that is, when $X⧷W = ∅$ and $X′⧷W′ = ∅$.  This happens when both $Θ$ and $Θ′$ have only infinite runs.}}

\begin{prop}\label{A491} In \ct{CLT}, a morphism $[Θ,Θ′,τ]$ is an isomorphism iff {\noteWW}
\begin{picture}(0,0) %
  \put(-.5,0.04){\color{white} \rule{2ex}{2ex}}
  \end{picture}
\subi \begin{gather} \zz
\label{F079} τ⋅\text{is a bijection$^{\text{\rf{F242}}}$ and} \tag{a} \\ 
\label{F080} τ|_{W,W′}⋅\text{is a homeomorphism.} \tag{b}   
\zz
\end{gather}\subo Further, if $[Θ,Θ′,τ]$ is an isomorphism, the following hold. \begin{tlist} \setcounter{tlistno}{2}
\yyyez\yl{A546} $E⋅∋⋅xy \mapsto τ(x)τ(y)⋅∈⋅E′$ is a bijection. 
\yyyez\yl{A502} $τ|_{X⧷W,˙X′⧷W′}$ is a bijection.
\yyyez\yl{A492} $(∀x∈X,y∈X)⋅x⋅≼⋅y⋅⟺⋅τ(x)⋅≼′⋅τ(y)$. 
\yyyez\yl{A493} $(∀x∈X,y∈X)⋅x⋅≺⋅y⋅⟺⋅τ(x)⋅≺′⋅τ(y)$. 
\yyyez\yl{A533} $τ(r) = r′$.
\yyyez\yl{A494} $(∀y∈X)⋅\dtau(P(y))⋅=⋅P′(τ(y))$.
\yyyez\yl{F225} $\dtau|_{\HH,\HH′}$ is a bijection.  Its inverse is $\overline{τ^{-1}}|_{\HH′,\HH}$.
\yyyez\yl{F081} In the action transformation $⁅α_x⁆_{x∈W}$ of $[Θ,Θ′,τ]$, each $α_x$ is a bijection. 
\yyyez\yl{F087} $[Θ,Θ′,τ]$ is end-preserving.
\yyyez\yl{F186} The run transformation $ζ$ of $[Θ,Θ′,τ]$ is the bijection $\dtau|_{\ZZ,\ZZ′}$.\  Its inverse is $\overline{τ^{-1}}|_{\ZZ′,\ZZ}$.{\yyyez}
\yl{F082} The inverse of $[Θ,Θ′,τ]$ is $[Θ′,Θ,τ^{-1}]$. 
\yyyez\yl{F083} The action transformation of $[Θ′,Θ,τ^{-1}]$ is $⁅α^{-1}_{τ^{-1}(x′)}⁆_{x′∈W′}$. 
\yyyez\yl{F192} $[Θ′,Θ,τ^{-1}]$ is end-preserving.
\yyyez\yl{F194} The run transformation of $[Θ′,Θ,τ^{-1}]$ is $ζ^{-1}$.  (Proof \rf{A491p}.)
\end{tlist}\end{prop}
 
\vspace{5mm} %
To explore Proposition~\rf{A491}(\rf{F079}--\rf{F080})'s characterization, this paragraph relaxes (\rf{F079}) and (\rf{F080}) independently.  In Figure~\rf{B029}, $τ$ is not bijective in violation of (\rf{F079}), even though $τ|_{W,W′} = \id_{⎨\f0⎬}$ is a homeomorphism in accord with (\rf{F080}).  Here the tuple $[Θ′,Θ,τ^{-1}]$ cannot be constructed because $τ^{-1}$ does not exist.  In Figure~\rf{A767}, $τ$ is bijective but $τ|_{W,W′}$ is not a homeomorphism.  Here the tuple $[Θ′,Θ,τ^{-1}] = [Θ′,Θ,\id_{X′}]$ is not a morphism because $\id_{X′}$ is not continuous in violation of (\rf{cH}) in Definition~\rf{E773}.

\newcommand{\notez}{\footnote{\label{F196}By the definitions (\rf{E784}) of $⁅\lesssim_i⁆_{i∈I}$ and $⁅\lesssim′_{i′}⁆_{i′∈I′}$, the monotonicity of $ζ$ and $ζ′$ is equivalent to $(∀i∈I,Z_1∈\ZZ,Z_2∈\ZZ)$ $U_i(Z_1)⋅≤⋅U_i(Z_2)⋅⇔⋅U′_{ι(i)}○ζ(Z_1)⋅≤⋅U′_{ι(i)}○ζ(Z_2)$.  Also, Proposition~\rf{A503}'s condition~(\rf{F084}) and Proposition~\rf{A491}(\rf{F186}) imply that $ζ$ and $\dtau|_{\ZZ,\ZZ′}$ are interchangeable.}}
\vspace{8mm} %
Proposition~\rf{A503}'s condition (\rf{F086}) uses the term ``preorder isomorphism''.  This is the concept of order isomorphism from order theory, extended from orders to preorders (Davey and Priestley 2002, page~3; Viglizzo 2023). \nocite{DaveyP02} \nocite{Vigli23} In particular, for each player $i⋅∈⋅I$, $ζ$ is an preorder isomorphism from $(\ZZ,\lesssim_i)$ to $(\ZZ′,\lesssim′_{ι(i)})$ iff $ζ{:}\ZZ→\ZZ′$ is a bijection and both $ζ$ and $ζ^{-1}$ are monotone.   Note that Proposition~\rf{A503}'s condition (\rf{F084})  and Proposition~\rf{A491}(\rf{F186}) imply that $ζ{:}\ZZ→\ZZ′$ is a bijection.  Further, Proposition~\rf{A503}'s assumption that $[Γ,Γ′,τ]$ is a morphism implies $ζ$ is monotone.  Thus Proposition~\rf{A503}'s condition~(\rf{F086}) adds only that $ζ^{-1}$ is monotone.{\notez}

\pagebreak
\newcommand{\noteinj}{\footnote{\label{F195}Technically, condition (\rf{F085}) can be can be weakened to the injectivity of $ι{:}I→I′$ because condition~(\rf{F084}) implies the surjectivity of $ι$.  To see this, in steps, the image $\bar{ι}(I)$ by $I$'s definition (\rf{F218}) is equal to $\bar{ι}(κ(W))$, which by Proposition~\rf{E791}(b) is equal to $\bar{κ}′(\bar{τ}(W))$, which by condition~(\rf{F084}) and Proposition~\rf{A491}(\rf{F080}) is equal to $\bar{κ}′(W′)$, which by the definition of $I′$ (\rf{F218}) is equal to~$I′$.}}  
\begin{prop}\label{A503} In \ct{Gm}, a morphism $[Γ,Γ′,τ]$ is an isomorphism iff {\noteinj} 
\begin{picture}(0,0) %
  \put(-.5,0.04){\color{white} \rule{2ex}{2ex}}
  \end{picture}
\subi \begin{gather}
\zz
\label{F084} [Θ,Θ′,τ]⋅\text{is an isomorphism in \ct{CLT}}, \tag{a} \\
\label{F085} ι⋅\text{is a bijection,$^{\text{\normalfont\rf{F195}}}$ and} \tag{b} \\
\label{F086} (∀i∈I)⋅ζ⋅\text{is a preorder isomorphism from}⋅(\ZZ,\lesssim_i)⋅\text{to}⋅(\ZZ′,\lesssim′_{ι(i)}). \tag{c}
\zz
\end{gather}\subo Further, if $[Γ,Γ′,τ]$ is an isomorphism, the following hold. \begin{tlist} \setcounter{tlistno}{3} 
\yyyez\yl{F088} The inverse of $[Γ,Γ′,τ]$ is $[Γ′,Γ,τ^{-1}]$. 
\yyyez\yl{F090} The player transformation of $[Γ′,Γ,τ^{-1}]$ is $ι^{-1}$.
(Proof~\rf{A503p}.)
\end{tlist} \end{prop}

\vspace{3mm} %
Since there is a forgetful functor from \ct{Gm} to \ct{CLT} (Proposition~\rf{A455}), any \ct{Gm} isomorphism has an underlying \ct{CLT} isomorphism in accord with Proposition~\rf{A503}'s condition (\rf{F084}).  To explore the remainder of Proposition~\rf{A503}(\rf{F084}--\rf{F086})'s characterization, the remainder of this paragraph will relax conditions (\rf{F085}) and (\rf{F086}) independently.  Figure~\rf{A748} satisfies (\rf{F084}) and (\rf{F086}) but not~(\rf{F085}).  Here the ``inverse'' tuple $[Γ′,Γ,τ^{-1}] = [Γ′,Γ,\id_{X′}]$ is not a morphism because it violates (\rf{gK}) of Definition~\rf{E774}---specifically, nodes $\f0$ and $\f1$ are controlled by the same player in $Γ′$ but $τ^{-1}(\f0)$ and $τ^{-1}(\f1)$ are controlled by different players in $Γ$.  Finally, Figure~\rf{A785} satisfies (\rf{F084}) and (\rf{F085}) but not (\rf{F086}).  Here the ``inverse'' tuple $[Γ′,Γ,τ^{-1}] = [Γ′,Γ,\id_{X′}]$ is not a morphism because it violates (\rf{gU}) of Definition~\rf{E774}---specifically, $⎨0˛2⎬˙\lesssim′_{\plyr{P1}}˙⎨0,1⎬$ but not $ζ^{-1}(⎨0˛2⎬)˙\lesssim_{\plyr{P1}}˙ζ^{-1}(⎨0˛1⎬)$.  

\begin{figure}[h] \newcommand{\hgth}{3.50} %
\begin{picture}(0,\hgth) \myoutergrid{\hgth} \renewcommand{\selld}[1]{#1}
  \put(-0.60, 1.50){\makebox(0,0){\scalebox{0.90}{ \begin{pspicture} \end{pspicture} }}} \end{picture}
\caption{\small The morphism $[Γ,Γ′,\id_X]$ is not an isomorphism because of the utility functions.} \label{A785} \end{figure}

\vspace{3mm} %
Incidentally, all one-node CLTs are isomorphic to one another.  To see this via Proposition~\rf{A491}(\rf{F079}--\rf{F080})'s characterization, consider two one-node CLTs $Θ$ and $Θ′$.  Here the only conceivable transformation $τ$ maps the one node of $Θ$ to the one node of $Θ′$, and the homeomorphism $τ|_{W,W′}$ is the empty function.  Further, all one-node games are isomorphic to one another.  To see this via Proposition~\rf{A503}(\rf{F084}--\rf{F086})'s characterization, recall that all one-node CLTs are isomorphic, and note that each one-node game has no players.

\sloppy

\appendix

\pagebreak
\section{Examples}\label{F258}
\markb{\sc Appendix \rf{F258}. Examples}

\nssec{General orientation}{F262}

Readers less unacquainted with game theory might begin with Figure~\rf{E613}(c). The nodes are numbers, the root node is underlined, the edges are shown by line segments, the information sets are shown by shaded regions, and the actions are letters.  The players are $\mathit{P1, P2}$ and $\mathit{P3}$, and the utility (payoff) profiles are beneath the five terminal nodes.\footnote{This game is from Selten 1975, Figure~1, and is commonly known as the ``horse'' game simply because it looks like a stick-drawing of a horse.  For a story for the game, suppose a student (called player $\mathit{P1}$) must decide between the bad action of not doing her homework (called $\fb$) and the correct action of doing her homework (called $\fc$).  Knowing that the homework has been finished (node $\f1$), a dog (player $\mathit{P2}$) must decide between the dumb action of eating the homework ($\fd$) and the good action of taking a nap ($\fg$).  Finally, without knowing whether the student chose bad (node $\f3$) or the student chose correct and the dog chose dumb (node $\f4$), the teacher (player $\mathit{P3}$) must decide between excusing the student ($\fe$) and failing the student ($\ff$).  The student most prefers being excused without doing the homework (run $⎨\f0{,}\f3{,}\f5⎬$), and least prefers failing after doing the homework (run $⎨\f0{,}\f1{,}\f4{,}\f8⎬$).  The dog likes eating homework (runs $⎨\f0{,}\f1{,}\f4{,}\f7⎬$ and $⎨\f0{,}\f1{,}\f4{,}\f8⎬$).  The teacher does not want to excuse a badly behaving student (run $⎨\f0{,}\f3{,}\f5⎬$) or to fail a correctly behaving student (run $⎨\f0{,}\f1{,}\f4{,}\f8⎬$).}

\begin{figure}[p] \newcommand{\hgth}{19.80}
\begin{picture}(0,\hgth) \myoutergrid{\hgth} 
  \put(-0.60,10.00){\makebox(0,0){\scalebox{0.86}{ \begin{pspicture} \end{pspicture} }}} \end{picture}
\caption{\small Examples of different specification styles.  Roughly, generic styles (c,d) are in top row, backward-looking styles (a,b) are lower left, and forward-looking styles (e,f) are lower right.  In these examples, players and utilities do not change.} \label{E613} \end{figure} 

It may be helpful to see a brief history of extensive-form game specifications.  Broadly speaking, there are three kinds of nodes.  (i)~As in Figure~\rf{E613}(f), von Neumann and Morgenstern 1944 (Section~10)\nichts{} specified each node as a set of outcomes, and used set inclusion to arrange these nodes in a tree.\nocite{NeumaM44} (ii) As in Figure~\rf{E613}(c,d,e), \nocite{Kuhn5397}Kuhn 1953 specified each node as an abstract entity without any internal structure, and then separately specified the edges of the tree connecting the nodes.  (iii) As in Figure~\rf{E613}(b), Harris 1985 and Osborne and Rubinstein 1994 (page 200) specified each node as a sequence of past actions, and then used concatenation to arrange these sequences in a tree.  That created the first tractable specification of infinite-horizon (that is, countably-infinite-length) extensive-form games.  More recently, Al\'os-Ferrer and Ritzberger 2016 returned to the outcome-set style of von Neumann and Morgenstern 1944 and extended it to allow for an infinite horizon.\nocite{AlosfR16}  Finally, Streufert 2019 showed how the sequences of Harris 1985 and Osborne and Rubinstein 1994 can be changed into sets, as in Figure~\rf{E613}(a).\nocite{five-1903} \nocite{Harri85}

\nssec{Style groups}{E614} 
The examples of Figure~\rf{E613} represent five groups of specification styles.  These five groups are discussed in the following five paragraphs.  Most of these groups contain many different styles from different academic fields.  Within this diverse literature, nodes are sometimes called ``vertices'', ``states'', ``histories'', or ``positions'', and actions are sometimes called ``labels'', ``alternatives'', ``choices'', or ``programs''. 

\newcommand{\noteffy}{\nichts{}}
\newcommand{\notelts}{\footnote{Some labelled transition systems and process graphs have recursive transitions.  These do not support extensive-form games because extensive-form games require trees.  Similarly, stochastic games, such as those in Mertens 2002,\nocite{Merte02} are not extensive-form games because they have recursive transitions.}}
{\em Abstract group (e.g., Figure~\rf{E613}(c,d))}.  The styles in this group place no restrictions on the notation for nodes and actions.  Examples from economics include the style in \nocite{KrepsW82}Kreps and Wilson 1982; the style in Myerson 1991, Section 2.1; and the style in Mas-Colell, Whinston, and Green 1995, page~227.\nocite{Myers91} \nocite{MascoWG95} Examples from elsewhere include the ``labeled transition system'' style in Blackburn, de Rijke, and Venema 2001, page 3;\nocite{BlackRV01} the style in Shoham and Leyton-Brown 2009, page 125;\nocite{ShohaL09} and the ``epistemic process graph'' style in van Benthem 2014, page 70.{\notelts}  A final example is the ``\ct{Gm}'' style in this paper, which is virtually identical to the ``traditional style'' in Streufert 2025.{\noteffy}  Since this group specifies nodes and actions abstractly, it can be understood to subsume the other four groups.  Relatedly, this paper's \ct{Gm} games can be understood to include most extensive-form games.%
\footnote{\label{E890}\ct{Gm} games include most but not all extensive-form games. 
[a]~The many differences between members of the abstract group (Section~\rf{E614} and Figure~\rf{E613}(c,d)) are being suppressed.  Details are in notes \rf{E814}, \rf{E815}, \rf{E882}, \rf{E816}, and~\rf{E817}.\nichts{}
[b] Although a \ct{Gm} game can specify a simultaneous move via multiple information sets (as in Osborne and Rubinstein 1994, page 202), it cannot specify multiple players moving at the same information set (as in Al\'os-Ferrer and Ritzberger 2016, page~140, first full paragraph).  This limitation [b] was briefly discussed in note~\rf{F799}.\nichts{} 
[c] Although countably infinite runs are allowed, a \ct{Gm} game is discrete in the sense that each node has a finite number of predecessors.  This excludes games in continuous time (\nocite{DocknJLS00}Dockner, J{\o}rgensen, Long, and Sorger 2000) and games which resume after a countably infinite number of moves (these and yet more general games are explored in Al\'os-Ferrer and Ritzberger 2016, Chapter~5).  Only this limitation [c] is ``substantial'' in the sense of Section~\rf{F261}.
[d] There are some games which are not directly comparable to \ct{Gm} games because they are not built on trees.  These include games constructed from open games (as discussed in the second paragraph of Section~\rf{F277}), games constructed using coalgebra (as briefly mentioned later in Section~\rf{F277}), and games built on 5-ary relations known as ``pentaforms'' (Streufert 2025).
(The author intends to build a category for pentaform games which is isomorphic to the category \ct{Gm}.)
\nichts{}
}
\nichts{} 

{\em Sequence group (e.g., Figure~\rf{E613}(b))}.  The styles in this group specify actions abstractly and then specify each node as the sequence of past actions leading up to it.  This group is arguably the most popular, and is ideally suited for repeated games and for stationary economic models in general.  Specific examples from economics include the styles in Harris 1985\nocite{Harri85} and Osborne and Rubinstein 1994, page 200.\nocite{OsborneR94}  Examples from elsewhere include the ``protocol'' style in Parikh and Ramanujam 1985;\nocite{PariRama85} the ``history-based multi-agent structure'' style in Pacuit 2007;\nocite{Pacu07} the ``sequence-form representation'' style in Shoham and Leyton-Brown 2009, page 129; the style in Mann, Sandu, and Sevenster 2011, Chapter 2;\nocite{MannSS11} the ``logical game'' style in Hodges 2013, Section 2;\nocite{Hodge13} and the ``epistemic forest model'' style in van Benthem 2014, page 130.\nocite{Benth14}   

{\em Action-set group (e.g., Figure~\rf{E613}(a))}.  The styles in this group specify actions abstractly and then specify each node as the set (rather than the sequence) of past actions leading up to it.  This small group includes the styles of Streufert 2015 and 2019.\nocite{A2-1508}  Its technical advantage is that a set of actions is a simpler mathematical object than a sequence of actions.  These action-set nodes are ``dual'' to the outcome-set nodes of the outcome-set group in the sense that action sets look backward while outcome sets look forward.

{\em Node-set group (e.g., Figure~\rf{E613}(e))}.  The styles in this group specify nodes abstractly and then specify each action as the set of nodes to which the action leads (or similarly, as the set of edges that end at the nodes to which the action leads).  This group includes the style of Kuhn 1953, the style of Selten 1975,\nocite{Selten75} the ``simple'' style of Al\'os-Ferrer and Ritzberger 2016, Section 6.3, and the style of Gonz\'alez-D\'iaz, Garc\'ia-Jurado, and Fiestras-Janeiro 2023, Section 3.1.\nocite{GonzalezGF23}

{\em Outcome-set group (e.g., Figure~\rf{E613}(f))}.  The styles in this group specify both nodes and actions as sets of outcomes.  This group includes the style in von Neumann and Morgenstern 1944, Section 10, as well as the style in Al\'os-Ferrer and Ritzberger 2016, Section 6.2.  Note that outcome-set nodes shrink as one goes forward, away from the root node.  In this sense, the node collection of an outcome-set game closely resembles the filtration of a stochastic process, and stochastic processes are often used by economists to study stochastic dynamic issues.

\nssec{Distinguished Actions}{E615} 
The five-group catalog of Section~\rf{E614} can be complemented with a partial second ``dimension''.  In Figure~\rf{E613}(d), the actions $\f{left}$ and $\f{right}$ are shared by the three information sets.  Figure~\rf{E613}(c) is different.  There, the actions of the three information sets are distinguished from one another.  Distinguished actions allow one to economize on notation and terminology.  For example, a strategy in a general game is a function taking each decision node to the action to be played at that node.  Alternatively, a strategy in a game with distinguished actions can be defined as merely the set of actions to be played.  An early user of distinguished actions was Kreps and Wilson 1982 (page~867).  Distinguished actions were also assumed throughout the ``node-and-choice''  categories of Streufert 2018, 2020a, and 2020b (distinguished actions are sometimes called ``choices'').

Games in the abstract and sequence style groups of Section~\rf{E614} may or may not have distinguished actions.  However, games in the action-set, node-set, and outcome-set groups must have distinguished actions because of specialized group properties (see Streufert 2019, Section~7, in which non-distinguished actions are called ``shared alternatives'').

\nssec{Equivalence between style groups}{F259}
One might conjecture that the five style groups are somehow equivalent, and indeed, this conjecture has been partially confirmed by the literature.  Broadly speaking, Al\'os-Ferrer and Ritzberger 2016 showed that each game in a certain node-set style could be mapped to a game in a certain outcome-set style, and vice versa.  Similarly, Kline and Luckraz 2016 showed that each game in a certain abstract style could be mapped to a game in a certain sequence style, and vice versa.\nocite{KlineL16} Similarly, Streufert 2019 connected an action-set style with a sequence style, and an abstract style with a node-set style.  Thus representative styles from all five style groups have been linked together.

However, the situation is nuanced.  [a] The results in the previous paragraph concern only one representative style from each style group, and do not formally define the groups themselves.  [b] Distinguished actions are required by some groups but not others (as noted in Section~\rf{E615}).  [c] Only the abstract and sequence groups are capable of expressing absentmindedness \nocite{PiccioneR97}(Piccione and Rubinstein 1997), which is also known as nonlinearity (Isbell 1957). \nocite{Isbel57}  Thus the action-set, node-set, and outcome-set groups are less general in a substantial sense (details in Streufert 2019, Section~7).  [d] The previous paragraph's literature only establishes ad hoc functions between different styles, and so the meaning of each cross-style function rests solely on its own intuitive appeal.  It is hoped that these ad hoc functions can be unified, justified, and extended by the isomorphisms of \ct{Gm}.  Indeed, Streufert 2021 (Section~5) has begun to do so, as discussed at the end of Sections~\rf{F277} and \rf{F261} in connection with ``results [3] and [4]''.

\vspace{1mm}\section{For Definition of \ct{CLT}}\label{B025}
\markb{\sc Appendix \rf{B025}. For Definition of \ct{CLT}}

\begin{lemma}\label{A816} Suppose $(X,E)$ is an out-tree (\rf{E776}), and $λ{:}E→A$ is surjective and locally injective (\rf{E778}).  Derive the feasibility correspondence $F$ (\rf{E779}) and the next-node function~$n$~(\rf{E869}).  Then the following hold. \begin{tlist}
\yl{B056} $⨆_{x∈W}F(x) = A$.
\yl{A817} The function $n$ is well-defined.
\yl{A822} Fix a decision node $x_o⋅∈⋅W$.  Then \begin{gather}
\zz
F(x_o)⋅∋⋅a \mapsto n(x_o,a)⋅∈⋅⎨y_{\sss+}|x_oy_{\sss+}∈E⎬\notag
\zz
\end{gather} is a bijection.  Its inverse is $F(x_o)⋅∋⋅λ(x_oy)$ $\mapsfrom$ $y⋅∈⋅⎨y_{\sss+}|x_o y_{\sss+}∈E⎬$.\footnote{Intuitively, $n(x_o,∙)$ (mapping action to next node) and $λ(x_o˙∙)$ (mapping next node to action) are ``local inverses'' at $x_o$.}  
\end{tlist}\end{lemma}

\begin{pf} {\em (\rf{B056})}. For the forward inclusion $⨆_{x∈W}F(x)⋅⊆⋅A$, note that each $F(x)⋅⊆⋅A$ by the definition (\rf{E779}) of $F$.  For the reverse inclusion, take $a⋅∈⋅A$.  Then the definition (\rf{F217}) of $A$ implies there is \lic{B057} $xy⋅∈⋅E$ such that \li{B058} $λ(xy) = a$.  Then \rf{B057} and the definition (\rf{E812}) of $W$ imply $x⋅∈⋅W$.  Also \rf{B058} and the definition (\rf{E779}) of $F$ imply $a⋅∈⋅F(x)$.

{\em (\rf{A817})}.  By the definition (\rf{E869}) of $n$, it suffices to show that for each $(x,a)⋅∈⋅F\gr$, there is a unique $y⋅∈⋅X⧷⎨r⎬$ such that $λ(xy) = a$.  Toward that end, take $(x,a)⋅∈⋅F\gr$.  Then by the definition of $F$, there is a node $y⋅∈⋅X$ such that $λ(xy) = a$.  This $y$ is unique, for if another $y_*$ satisfies $λ(xy_*) = a$, then the local injectivity (\rf{E778}) of $λ$ implies $y = y_*$. 

Finally, $y⋅≠⋅r$ holds by two observations.  First, $y⋅∈⋅π_2E$ because $xy⋅∈⋅E$.  Second, $r⋅∉⋅π_2E$ because if a node $x_*⋅∈⋅X$ satisfies $x_*r⋅∈⋅E$, then the definition (\rf{E776}) of an out-tree implies the existence of a walk from $r$ to $x_*$, which by $x_*r⋅∈⋅E$ implies the existence of a nontrivial walk from $r$ to $r$, which violates (\rf{E776}).

{\em (\rf{A822})}.  To see that the forward function is well-defined, consider a feasible action $a⋅∈⋅F(x_o)$.  Then part~(\rf{A817}) implies $n(x_o,a)$ is well-defined, and $n$'s definition (\rf{E869}) implies $x_o˙n(x_0,a)⋅∈⋅E$.  To see that the reverse function is well-defined, consider a node $y$ such that $x_oy⋅∈⋅E$.  Then $λ(x_oy)$ is well-defined (since the domain of $λ$ is $E$), and $F$'s definition (\rf{E779}) implies $λ(x_oy)⋅∈⋅F(x_o)$.  Thus it suffices to show \begin{gather}
\zz
(∀a∈F(x_o))⋅λ(x_o˙n(x_o,a)) = a⋅\text{and} \label{F819} \\
(∀y∈⎨y_{\sss+}|x_o y_{\sss+}∈E⎬)⋅n(x_o,λ(x_o y)) = y.  \label{F820}
\zz
\end{gather} For (\rf{F819}), take a feasible action $a⋅∈⋅F(x_o)$.  Then $λ(x_on(x_o,a)) = a$ holds since $n(x_o,a)$ is defined by (\rf{E869}) to be the unique $y_*$ that satisfies $λ(x_oy_*) = a$.   For (\rf{F820}), take a node $y$ that immediately follows $x_o$ in the sense that $x_o y⋅∈⋅E$.  Consider the action $λ(x_oy)$.  The definition of $F(x_o)$ implies $λ(x_oy)⋅∈⋅F(x_o)$, which implies $(x_o,λ(x_oy))⋅∈⋅F\gr$, which implies $(x_o,λ(x_oy))$ is in the domain of $n$.  So we may define $y_*$ by \ilc{F817} $y_* = n(x_o,λ(x_oy))$.  Then the definition (\rf{E869}) of $n$ implies that $y_*$ satisfies $λ(x_oy_*) = λ(x_oy)$ [to see this, just regard $λ(x_oy)$ as some constant action].  Thus local injectivity (\rf{E778}) implies $y_* = y$, which by \rf{F817} implies $n(x_o,λ(x_oy)) = y$. \end{pf}

\begin{npf}[{{\bf for Proposition~\rf{B063}}}]\label{B063p} By general topology, $τ|_{W,W′}$ is continuous iff \begin{gather}
\zz
(∀x∈W,H′∈\HH')⋅τ(x)⋅∈⋅H′⋅\text{implies}⋅(∃H∈\HH)⋅x⋅∈⋅H⋅\text{and}⋅\dtau(H)⋅⊆⋅H′. \label{B064}
\zz
\end{gather} Thus it suffices to show that (\rf{B064}) is equivalent to \begin{gather}
\zz
(∀H∈\HH)(∃H′∈\HH′)⋅\dtau(H)⋅⊆⋅H′. \label{B065}
\zz
\end{gather} For the following argument, note that $\HH$ partitions $W$ by (\rf{C2}) for $Θ$, and that $\HH′$ partitions $W′$ by (\rf{C2}) for $Θ′$.

For the forward direction, assume (\rf{B064}) and take \lic{B066} $H⋅∈⋅\HH$.  Then $\HH$ partitioning $W$ implies there is \il{B082} $x⋅∈⋅W$ such that \il{B067} $x⋅∈⋅H$.  Further, the proposition's assumption that $\dtau(W)⋅⊆⋅W′$, together with $\HH′$ partitioning $W′$, imply there is \il{B068} $H′⋅∈⋅\HH′$ such that \il{B069} $τ(x)⋅∈⋅H′$.  Note \rf{B082}, \rf{B068}, \rf{B069}, and (\rf{B064}) imply there is \il{B070} $H_o⋅∈⋅\HH$ such that \il{B071} $x⋅∈⋅H_o$ and \il{B072} $\dtau(H_o)⋅⊆⋅H′$.  Further, \rf{B066} and \rf{B067}, and \rf{B070} and \rf{B071}, and $\HH$ partitioning $W$ imply $H = H_o$.  Thus \rf{B072} implies $\dtau(H)⋅⊆⋅H′$. 

For the reverse direction, assume (\rf{B065}), and take \lic{B073} $x⋅∈⋅W$ and \il{B076} $H′⋅∈⋅\HH′$ such that \il{B077} $τ(x)⋅∈⋅H′$.  Note \rf{B073} and $\HH$ partitioning $W$ imply there is \il{B074} $H⋅∈⋅\HH$ such that \il{B075} $x⋅∈⋅H$.  So it suffices to show $\dtau(H)⋅⊆⋅H′$.  Note \rf{B074} and (\rf{B065}) imply there is \il{B079} $H′_*⋅∈⋅\HH′$ such that \il{B080} $\dtau(H)⋅⊆⋅H′_*$.  Easily, \rf{B075} implies $τ(x)⋅∈⋅\dtau(H)$, which by \rf{B080} implies \il{B081} $τ(x)⋅∈⋅H′_*$.  Further, \rf{B076} and \rf{B077}, and \rf{B079} and \rf{B081}, and $\HH′$ partitioning $W′$ imply $H′ = H′_*$.  Thus \rf{B080} implies $\dtau(H)⋅⊆⋅H′$. \end{npf}

\begin{npf}[{{\bf for Proposition~\rf{E871}}}]\label{E871p} Consider any decision node $x⋅∈⋅W$ and any feasible action $a⋅∈⋅F(x)$.  Since $n$ is well-defined by Lemma~\rf{A816}(\rf{A817}), the definition of $n$ (\rf{E869}) implies $x˙n(x˛a)⋅∈⋅E$, which by edge-preservation (\rf{E870}) implies $τ(x)˙τ(n(x˛a))⋅∈⋅E′$, which by (\rf{C3}) implies $λ′(˙τ(x)˙τ(n(x˛a))˙)$ exists.  Thus since \begin{gather}
\zz
λ′(\,τ(x)\,τ(n(x{,}a))\,) = α_x(a) \label{E866}
\zz
\end{gather} by $α_x(a)$'s definition (\rf{E782}), the value $α_x(a)$ is well-defined.  So it remains to show that the value $α_x(a)$ is in the codomain $F′(τ(x))$.  By the definition (\rf{E779}) of $F′$ at $τ(x)⋅∈⋅W′$, \begin{gather}
\zz
F′(τ(x)) = ⎨⋅a′∈A′⋅|⋅(∃y′)⋅λ′(˙τ(x)˙y′˙){=}a′˙⎬. \notag 
\zz
\end{gather} By (\rf{E866}), the right-hand side's equality holds at $y′ = τ(n(x˛a))$ and $a′ = α_x(a)$.  So $α_x(a)$ belongs to the set on the right-hand side. \end{npf}

\begin{npf}[{{\bf for Proposition \rf{A481}}}]\label{A481p} {\em (\rf{A483})--(\rf{A482})}. Edge preservation (\rf{cE}) implies that the tuple $[(X,E),(X′,E′),τ]$ is a graph morphism, which implies properties (\rf{A483})--(\rf{A482}).

{\em (\rf{A804})}.  Take an edge $xy⋅∈⋅E$.  In steps, the definition (\rf{E779}) of $F$ implies $λ(xy)⋅∈⋅F(x)$, which by the definition (\rf{E782}) of $α_x$ implies \begin{gather}
\zz
α_x(λ(xy)) = λ′(\,τ(x)\,τ(n(x,λ(xy)))\,), \notag 
\zz
\end{gather} which by applying Lemma~\rf{A816}(\rf{A822}) at $x_o = x$ reduces to $α_x(λ(xy)) = λ′(\,τ(x)\,τ(y)\,)$.

{\em (\rf{A490})}. Take $x⋅∈⋅W$ and $a⋅∈⋅F(x)$.  The definition (\rf{E782}) of $α_x$ implies \begin{gather}
\zz
λ′(\,τ(x)\,τ(n(x,a))\,) = α_x(a).  \label{A824}
\zz
\end{gather} In addition, the definition of $α_x$ implies $α_x(a)⋅∈⋅F′(τ(x))$, which easily implies \linebreak $(τ(x),α_x(a))⋅∈⋅(F′)\gr$, which implies that $(τ(x),α_x(a))$ is in the domain of $n'$.  Thus the definition (\rf{E869}) of $n′$ implies that $n′(τ(x),α_x(a))$ is the unique $y′$ such that \begin{gather}
\zz
λ′(\,τ(x)\,y′\,) = α_x(a). \notag 
\zz
\end{gather} By (\rf{A824}), that $y′$ is $τ(n(x,a))$, and thus the previous sentence implies that $n′(τ(x),α_x(a))$ is $τ(n(x,a))$.
\end{npf}

\begin{lemma}\label{A855} Suppose $Θ$ is a CLT.  Then the action transformation (\rf{E782}) of the tuple $[Θ,Θ,\id_X]$ is $⁅\id_{F(x)}⁆_{x∈W}$.  \end{lemma}

\begin{pf} Use definition (\rf{E782}) to construct the action transformation $⁅α_x⁆_{x∈W}$ of $[Θ,Θ′,τ] = [Θ,Θ,\id_X]$.   Take a decision node $x⋅∈⋅W$.  Then definition (\rf{E782}) implies $α_x{:}F(x)→F′(τ(x))$ which by $τ = \id_X$ and $F′ = F$ reduces to $α_x{:}F(x)→F(x)$.  Thus it remains to show $(∀a∈F(x))$ $α_x(a) = a$.  Toward that end, take a feasible action $a⋅∈⋅F(x)$.  In steps, $α_x(a)$ by definition (\rf{E782}) is $λ′(\,τ(x)\,τ(n(x,a))\,)$, which by $τ = \id_X$ reduces to $λ′(\,x\,n(x,a)\,)$, which by $λ′ = λ$ reduces to $λ(\,x\,n(x,a)\,)$, which by Lemma~\rf{A816}(\rf{A822}) equals $a$.  \end{pf}

\begin{lemma}\label{A826} Suppose $[Θ,Θ′,τ]$ and $[Θ′,Θ″,τ′]$ are morphisms.  Then the action transformation (\rf{E782}) of the tuple $[Θ,Θ″,τ′○τ]$ is $⁅α′_{τ(x)}○α_x⁆_{x∈W}$. \end{lemma}

\begin{pf} First consider the expression $⁅α′_{τ(x)}○α_x⁆_{x∈W}$ in the lemma.  To show that \begin{gather}
\zz
⁅α′_{τ(x)}○α_x{:}F(x)→F″(τ′○τ(x))⁆_{x∈W} \label{F821} 
\zz
\end{gather} is well-defined, consider an individual decision node $x⋅∈⋅W$.  Then $α_x{:}F(x)→F′(τ(x))$ by definition (\rf{E782}) at $x$, and $α′_{τ(x)}{:}F′(τ(x))→F″(τ′○τ(x))$ by the same definition (\rf{E782}) at $x′ = τ(x)$.  Thus (at each $x⋅∈⋅W$) the composition (\rf{F821}) is well-defined.

Second, let \begin{gather}
\zz
⁅α^*_x{:}F(x)→F″(τ′○τ(x))⁆_{x∈W} \label{F822}
\zz
\end{gather} be the action transformation (\rf{E782}) derived directly from the tuple $[Θ,Θ″,τ′○τ]$.  Edge preservation (\rf{cE}) for the morphism $[Θ,Θ′,τ]$ implies $(∀xy∈E)$ $τ(x)τ(y)⋅∈⋅E′$, which by edge preservation (\rf{cE}) for the morphism $[Θ′,Θ″,τ′]$ implies $(∀xy∈E)$ $τ′○τ(x)˙τ′○τ(y)⋅∈⋅E″$.  Hence Proposition~\rf{E871} implies that (\rf{F822}) is well-defined.

Thus it remains to show that $(∀x∈W,a∈F(x))$ $α^*_x(a) = α′_{τ(x)}(α_x(a))$.  Toward that end, take a decision node $x⋅∈⋅W$ and a feasible action $a⋅∈⋅F(x)$.  Then $α^*_x(a)$ by the definition (\rf{E782}) of $α^*_x$ equals \begin{gather}
\zz
λ″(⋅τ′○τ(x)⋅τ′○τ(n(x,a))⋅), \notag
\zz
\end{gather} which by Proposition~\rf{A481}(\rf{A490}) applied to $[Θ,Θ′,τ]$ equals \begin{gather}
\zz
λ″(⋅τ′(τ(x))⋅τ′(n′(τ(x),α_x(a)))⋅), \notag
\notag 
\zz
\end{gather} which by the definition (\rf{E782}) of $α′_{τ(x)}$ equals $α′_{τ(x)}(α_x(a))$. \end{pf}

\begin{npf}[{{\bf for Proposition~\rf{A453}}}]\label{A453p} Since the identity laws and associativity hold by inspection, Claims \rf{F235}(a) and \rf{A759}(a) imply that \ct{CLT} is a well-defined category.  This proves the proposition's part (\rf{F232}).  Its part~(\rf{F233}) follows from Claim~\rf{F235}(b), and its part~(\rf{F234}) follows from Claim~\rf{A759}(b). \begin{cllist}

\yl{F235} {\em Suppose $Θ$ is a CLT.  Then (a) $\id_Θ$ is a morphism and (b) $\id_Θ$'s action transformation is $⁅\id_{F(x)}⁆_{x∈W}$.} Definition (\rf{E790}) sets $\id_Θ$ equal to the tuple $[Θ,Θ,\id_X]$.  Thus Lemma~\rf{A855} shows that $\id_Θ$'s action transformation is $⁅\id_{F(x)}⁆_{x∈W}$.  This proves~(b).  Hence it remains to show that $[Θ,Θ,\id_X]$ satisfies (\rf{cE}), (\rf{cH}), and (\rf{cL}) of Definition~\rf{E773}.  The first two hold by inspection.  For (\rf{cL}), note that the action transformation is $⁅\id_{F(x)}⁆_{x∈W}$ by (b), and that this varies continuously over $W$ because $⁅F(x)⁆_{x∈W}$ varies continuously over $W$ by (\rf{C4}) of Definition~\rf{E772}. 

\yl{A759} {\em Suppose $θ = [Θ,Θ′,τ]$ and $θ′ = [Θ′,Θ″,τ′]$ are morphisms.  Then (a) $θ′○θ$ is a morphism and (b) $θ′○θ$'s action transformation is $⁅α′_{τ(x)}○α_x⁆_{x∈W}$.} Definition~(\rf{E790}) sets $θ′○θ$ equal to the tuple $[Θ,Θ″,τ′○τ]$.  Thus Lemma~\rf{A826} shows that $θ′○θ$'s action transformation is $⁅α′_{τ(x)}○α_x⁆_{x∈W}$.  This proves~(b).  Hence it remains to show that the tuple $[Θ,Θ″,τ′○τ]$ satisfies (\rf{cE}), (\rf{cH}), and (\rf{cL}) from Definition~\rf{E773}.  In this context, (\rf{cE}) for $[Θ,Θ″,τ′○τ]$ is \begin{gather}
\zz
(∀xy∈E)⋅τ′○τ(x)\,τ′○τ(y)⋅∈⋅E″. \notag  
\zz
\end{gather} To see this, take an edge $xy⋅∈⋅E$.  Then (\rf{cE}) for $θ$ implies $τ(x)τ(y)⋅∈⋅E′$, which by (\rf{cE}) for $θ′$ implies $τ′○τ(x)\,τ′○τ(y)⋅∈⋅E″$.  Further, (\rf{cH}) for $[Θ,Θ″,τ′○τ]$ is \begin{gather}
\zz
(τ′○τ)|_{W,W″}⋅\text{is continuous}. \notag
\zz
\end{gather} To see this, note that $τ|_{W,W′}$ is continuous by (\rf{cH}) for $θ$, and that $τ′|_{W′,W″}$ is continuous by (\rf{cH}) for $θ′$.  Thus $(τ′○τ)|_{W,W″} = τ′|_{W′,W″}\,○\,τ|_{W,W′}$ is continuous.  Finally, this claim's part (b) implies that (\rf{cL}) for $[Θ,Θ″,τ′○τ]$ is \begin{gather}
\zz
⁅α′_{τ(x)}○α_x⁆_{x∈W}⋅\text{is continuous from}⋅W. \notag
\zz
\end{gather} To see this, suppose \ilc{F823} $x_1$ and $x_2$ are in the same information set in $\HH$.  By (\rf{cL}) for~$θ$, \rf{F823} implies \li{A761} $α_{x_1} = α_{x_2}$.  By Proposition~\rf{B063}, \rf{F823} also implies $τ(x_1)$ and $τ(x_2)$ are in the same information set in $\HH′$, which by (\rf{cL}) for $θ′$ implies \li{A762} $α′_{τ(x_1)} = α′_{τ(x_2)}$.  Finally, \rf{A761} and \rf{A762} imply $α′_{τ(x_1)}○α_{x_1} = α′_{τ(x_2)}○α_{x_2}$.
\end{cllist} \unskipcl \nichts{yyy} \end{npf}

\nichts{}
\vspace{1mm}\section{For Definition of \ct{Gm}}\label{F223}
\markb{\sc Appendix \rf{F223}. For Definition of \ct{Gm}}

\begin{npf}[{{\bf for Proposition~\rf{E791}}}]\label{E791p} Since $κ$ is a function from $W$ by (\rf{G2}), and since $I$ is the codomain of $κ$ by (\rf{F218}), \begin{gather}
\zz
κ{:}W→I. \label{F801}
\zz
\end{gather} By the same reasoning applied to $Γ′$, \begin{gather}
\zz
κ′{:}W′→I′. \label{F802}
\zz
\end{gather}  

{\em (a)}. For each source decision node $x⋅∈⋅W$, $κ(x)⋅∈⋅I$ by (\rf{F801}) and $κ′(τ(x))⋅∈⋅I′$ by Proposition~\rf{A481}(\rf{A482}) and (\rf{F802}).  Thus the set $ι\gr$ (\rf{E785}) is a well-defined subset of $I×I′$.  Further, the projection of $ι\gr$ on its first coordinate is $I$ because $κ$ is surjective by~(\rf{G2}).  Thus it remains to show that each $i⋅∈⋅I$ is paired by $ι\gr$ with exactly one $i′⋅∈⋅I′$.  Toward that end, suppose $(i,i′_1)$ and $(i,i′_2)$ were two members of $ι\gr$.  Then the definition (\rf{E785}) of $ι\gr$ implies there are $x_1⋅∈⋅W$ and $x_2⋅∈⋅W$ such that \begin{gather}
\zz
(i,i′_1) = (κ(x_1),κ′(τ(x_1)))⋅\text{and} \nt
(i,i′_2) = (κ(x_2),κ′(τ(x_2))). \notag 
\zz
\end{gather}The first components of these two equalities imply that $κ(x_1) = κ(x_2)$, which by control preservation (\rf{E788}) implies $κ′○τ(x_1) = κ′○τ(x_2)$, which by the second components of the equalities implies that $i′_1 = i′_2$.  

{\em (b)}. By inspection, the definition (\rf{E785}) of $ι$ implies that $(∀x∈W)$ $ι(κ(x)) = κ′(τ(x))$.  The composition $ι○κ$ is well-defined because $I$ is both the codomain of $κ$ by (\rf{F801}), and the domain of $ι$ by definition (\rf{E785}).  The composition $κ′○τ|_{W,W′}$ is well-defined because $W′$ is both the codomain of $τ|_{W,W′}$ by inspection, and the domain of $κ′$ by (\rf{F802}).  Further, $W$ is the domain of both $ι○κ$ by (\rf{F801}), and $κ′○τ|_{W,W′}$ by inspection.  Finally, $I′$ is the codomain of both $ι○κ$ by definition (\rf{E785}), and $κ′○τ|_{W,W′}$ by (\rf{F802}).   \end{npf}

\begin{npf}[{{\bf for Proposition~\rf{A458}}}]\label{A458p}  Take any run $Z⋅∈⋅\ZZ$.  By the definition (\rf{E786}) of $ζ$, it suffices to show that $P′○τ(r)⋃\dtau(Z)⋅∈⋅\ZZ′$.  Before beginning, note that the definition of $P′$ implies \begin{gather}
\zz
\text{$P′○τ(r)⋃⎨τ(r)⎬$ is the walk from $r′$ to $τ(r)$ in $(X′,E′)$}. \label{A595} 
\zz
\end{gather} 

On the one hand, suppose $Z⋅∈⋅\ZZf$.  Then there is a \lic{A596} $y⋅∈⋅X⧷W$ such that \il{A597} $Z$~is the walk from $r$ to $y$ in $(X,E)$.  Note \rf{A596} and end-preservation (\rf{E787}) imply \il{A593} $τ(y)⋅∈⋅X′⧷W′$.  Meanwhile, \rf{A597} and edge preservation (\rf{cE}) imply that $\dtau(Z)$ is the walk from $τ(r)$ to $τ(y)$ in $(X′,E′)$, which by (\rf{A595}) implies that the concatenation $P′○τ(r)⋃\dtau(Z)$ is the walk from $r′$ to $τ(y)$ in $(X′,E′)$.  Thus \rf{A593} and the definition of $\ZZf$ (above (\rf{E813})) implies $P′○τ(r)⋃\dtau(Z)⋅∈⋅\ZZf′$.

On the other hand, take $Z⋅∈⋅\ZZi$.  Then $Z$ is an infinite walk from $r$ in $(X,E)$.  Thus edge preservation (\rf{cE}) implies $\dtau(Z)$ is an infinite walk from $τ(r)$ in $(X′,E′)$, which by (\rf{A595}) implies that the concatenation $P′○τ(r)⋃\dtau(Z)$ is an infinite walk from $r′$ in $(X',E′)$.  Thus the definition of $\ZZi$ (above (\rf{E813})) implies $P′○τ(r)⋃\dtau(Z)⋅∈⋅\ZZi′$.  \mbox{\hspace{5cm}} \nichts{yyy} \end{npf}

\begin{lemma}\label{A854} Suppose $Γ$ is a game and consider the tuple $[Γ,Γ,\id_X]$.  Then (a) its player transformation (\rf{E785}) is $\id_I$, and (b) its run transformation (\rf{E786}) is $\id_{\ZZ}$. \end{lemma}

\begin{pf} {\em (a)}. Use definition (\rf{E785}) to construct the player transformation $ι$ of $[Γ,Γ′,τ] = [Γ,Γ,\id_X]$.  Then definition (\rf{E785}) implies $ι{:}I→I′$, which by $I′ = I$ reduces to $ι{:}I→I$.  Thus it remains to show $(∀i∈I)⋅ι(i) = i$.  Toward that end, take a player $i⋅∈⋅I$.  By the definition (\rf{F218}) of $I$, there is $x⋅∈⋅W$ such that \lic{F803} $κ(x) = i$.  In steps, $ι(i)$ by \rf{F803} is $ι(κ(x))$, which by the definition (\rf{E785}) of $ι$ is $κ′(τ(x))$, which by $κ′ = κ$ and $τ = \id_X$ reduces to $κ(x)$, which by \rf{F803} is $i$.

{\em (b)}. Use definition (\rf{E786}) to construct the run transformation $ζ$ of $[Γ,Γ′,τ] = [Γ,Γ,\id_X]$.  Then definition (\rf{E786}) implies $ζ{:}\ZZ→\ZZ′$, which by $\ZZ = \ZZ′$ reduces to $ζ{:}\ZZ→\ZZ$.  Thus it remains show $(∀Z∈\ZZ)$ $ζ(Z) = Z$.  Toward that end, take a run $Z⋅∈⋅\ZZ$.  Then $ζ(Z)$ by definition (\rf{E786}) is $P′○τ(r)⋃\bar{τ}(Z)$ which by $τ = \id_X$ reduces to $P′(r)⋃Z$, which by $P′ = P$ reduces to $P(r)⋃Z$, which by $P(r) = ∅$ is~$Z$.  \end{pf}

\begin{lemma}\label{F239} Suppose that $Γ$, $Γ′$, and $Γ″$ are games, that $[Θ,Θ′,τ]$ and $[Θ′,Θ″,τ′]$ are CLT morphisms, and that the tuples $[Γ,Γ′,τ]$ and $[Γ′,Γ″,τ′]$ are control-preserving (\rf{E788}).  Then (a) $[Γ,Γ″,τ′○τ]$ is control-preserving and (b) its player transformation (\rf{E785}) is $ι′○ι$. \end{lemma}

\begin{pf} {\em (a)}.  By the definition (\rf{E788}) of control preservation, it suffices to show \begin{gather}
\zz
(∀x_1∈W,x_2∈W)⋅κ(x_1) = κ(x_2)⋅\text{implies}⋅κ″(τ′○τ(x_1)) = κ″(τ′○τ(x_2)). \notag 
\zz
\end{gather} Toward that end, take decision nodes $x_1⋅∈⋅W$ and $x_2⋅∈⋅W$ such that $κ(x_1) = κ(x_2)$.  Then $[Γ,Γ′,τ]$ preserving control implies $κ′(τ(x_1)) = κ′(τ(x_2))$.  Further, Proposition~\rf{A481}(\rf{A482}) implies that $τ(x_1)$ and $τ(x_2)$ are decision nodes in $W′$, so the previous sentence and $[Γ′,Γ″,τ′]$ preserving control imply $κ″(τ′○τ(x_1)) = κ″(τ′○τ(x_2))$.  

{\em (b)}. Let $ι^*$ be the player transformation of $[Γ,Γ″,τ′○τ]$.  Note $ι^*$ is well-defined by part~(a) and Proposition~\rf{E791}(a).  Thus by definition~(\rf{E785}), $ι^*{:}I→I″$ and  \begin{gather}
\zz
(ι^*)\gr = ⎨⋅(κ(x),κ″(τ′○τ(x)))⋅|⋅x∈W⋅⎬. \label{F804}
\zz
\end{gather} Now consider $ι′○ι$.  Note $ι′○ι{:}I→I″$, because $ι{:}I→I′$ and $ι′{:}I′→I″$ by two applications of definition (\rf{E785}).  Thus it suffices to show that $(∀i∈I)$⋅$ι^*(i) = ι′○ι(i)$.  Toward that end, take $i⋅∈⋅I$.  By the definition (\rf{F218}) of~$I$, there is $x⋅∈⋅W$ such that $κ(x) = i$.  Thus $ι^*(i)$ equals $ι^*(κ(x))$, which by (\rf{F804}) is $κ″(τ′○τ(x))$, which by $x⋅∈⋅W$ and Proposition~\rf{A481}(\rf{A482}) is $κ″○τ′|_{W′,W″}○τ|_{W,W′}(x)$, which by Proposition~\rf{E791}(b) is $ι′○κ′○τ|_{W,W′}(x)$, which by Proposition~\rf{E791}(b) again is $ι′○ι○κ(x)$, which by the definition of $x$ is $ι′○ι(i)$.  \end{pf}

\begin{lemma}\label{A604} Suppose $[Θ,Θ′,τ]$ and $[Θ′,Θ″,τ′]$ are end-preserving (\rf{E787}) morphisms.  Then (a) $[Θ,Θ″,τ′○τ]$ is an end-preserving morphism, and (b) its run transformation (\rf{E786}) is $ζ′○ζ$. \end{lemma}

\begin{pf}  Parts (a) and (b) follow from Claims~\rf{B182} and \rf{B183}, respectively. \begin{cllist}

\yl{B182} {\em $[Θ,Θ″,τ′○τ]$ is an end-preserving morphism.}  Proposition~\rf{A453}(a) (the well-definition of \ct{CLT}) implies $[Θ,Θ″,τ′○τ]$ is a morphism.  For end preservation (\rf{E787}), take $x⋅∈⋅X⧷W$.  Since $[Θ,Θ′,τ]$ is end-preserving, $τ(x)⋅∈⋅X′⧷W′$.  Thus since $[Θ′,Θ″,τ′]$ is end-preserving, $τ′○τ(x)⋅∈⋅X″⧷W″$.

\yl{A606} {\em $P″○τ′(r′)⋅⋃⋅\overline{τ′}(P′○τ(r)⋃⎨τ(r)⎬) = P″○τ′○τ(r)⋃⎨τ′○τ(r)⎬$.} 
\par\vspace{1mm}First, the definition of $P″$ implies \begin{gather}
\zz
P″○τ′(r′)⋃⎨τ′(r′)⎬⋅\text{is the walk in $(X″,E″)$ from}⋅r″⋅\text{to}⋅τ′(r′),  \notag
\zz
\end{gather} Second, the definition of $P′$ implies $P′○τ(r)⋃⎨τ(r)⎬$ is the walk in $(X′,E′)$ from $r′$ to $τ(r)$, which by edge preservation (\rf{cE}) for $[Θ′,Θ″,τ′]$ implies \begin{gather}
\zz
\overline{τ′}(P′○τ(r)⋃⎨τ(r)⎬)⋅\text{is the walk in $(X″,E″)$ from}⋅τ′(r′)⋅\text{to}⋅τ′○τ(r). \notag
\zz
\end{gather} Note $τ′(r′)$ is in the second walk since it is the walk's start.  Thus by concatenation,  \begin{gather}
\zz
\text{$P″○τ′(r′)⋅⋃⋅\overline{τ′}(P′○τ(r)⋃⎨τ(r)⎬)$ is the walk in $(X″,E″)$ from $r″$ to $τ′○τ(r)$}.\notag 
\zz
\end{gather} Yet by the definition of $P″$, we know $P″○τ′○τ(r)⋃⎨τ′○τ(r)⎬$ is the walk from $r″$ to $τ′○τ(r)$.  Thus the two walks are equal.\nichts{}

\yl{A605} $(∀Z∈\ZZ)⋅P″○τ′(r′)\,⋃\,\overline{τ′}(P′○τ(r)⋃\dtau(Z)) 
=P″○τ′○τ(r)\,⋃\,\overline{τ′○τ}(Z).$ 
\par\vspace{1mm}Consider a run $Z⋅∈⋅\ZZ$.  Because the root $r$ is in $Z$, the left-hand side equals \begin{gather}
\zz
P″○τ′(r′)⋅⋃⋅\overline{τ′}(⋅P′○τ(r)⋃⎨τ(r)⎬⋅⋃⋅\dtau(Z)⋅), \notag 
\zz
\end{gather} which by manipulation equals \begin{gather}
\zz
P″○τ′(r′)⋅⋃⋅\overline{τ′}(P′○τ(r)⋃⎨τ(r)⎬)⋅⋃⋅\overline{τ′○τ}(Z), \notag 
\zz
\end{gather} which by Claim~\rf{A606} equals  \begin{align}
\zz
P″○τ′○τ(r)⋃⎨τ′○τ(r)⎬⋅⋃⋅\overline{τ′○τ}(Z), \notag
\zz
\end{align} which by $r⋅∈⋅Z$ equals the right-hand side.  

\yl{B183} {\em The run transformation of $[Θ,Θ″,τ′○τ]$ is $ζ′○ζ$.}  Since the morphisms $[Θ,Θ′,τ]$ and $[Θ′,Θ″,τ′]$ are end-preserving by assumption, Proposition~\rf{A458} implies $ζ{:}\ZZ→\ZZ′$ and $ζ′{:}\ZZ′→\ZZ″$ are well-defined.   Thus $ζ′○ζ{:}\ZZ→\ZZ″$ is well-defined.  Now let $ζ^*{:}\ZZ→\ZZ″$ denote the run transformation of $[Θ,Θ″,τ′○τ]$.  Claim~\rf{B182} and Proposition~\rf{A458} imply $ζ^*{:}\ZZ→\ZZ″$ is well-defined.  So it remains to show that $(∀Z∈\ZZ)$ $ζ′○ζ(Z) = ζ^*(Z)$.  Towards this end, consider a run $Z⋅∈⋅\ZZ$.  Then Claim~\rf{A605} and the definition (\rf{E786}) of $ζ$ imply \begin{gather}
\zz
P″○τ′(r′)\,⋃\,\overline{τ′}(⋅ζ(Z)⋅) = P″○τ′○τ(r)\,⋃\,\overline{τ′○τ}(Z), \notag 
\zz
\end{gather} which by applying the definition (\rf{E786}) of $ζ′$ to the left-hand side, and by applying the definition (\rf{E786}) of $ζ^*$ to the right-hand side, yields $ζ′○ζ(Z) = ζ^*(Z)$.  \end{cllist} \unskipcl \nichts{yyy} \end{pf}

\begin{npf}[{{\bf for Proposition~\rf{A454}}}]\label{A454p}  Since the identity laws and associativity hold by inspection, Claims \rf{F240}(a) and \rf{A828}(a) imply that \ct{Gm} is a well-defined category.  This proves the proposition's part (\rf{F236}).  Its part~(\rf{F237}) follows from Claim~\rf{F240}(b,c), and its part~(\rf{F238}) follows from Claim~\rf{A828}(b,c). \begin{cllist}

\yl{F240} {\em Suppose $Γ$ is a game.  Then (a) $\id_Γ$ is a morphism, (b) its player transformation is $\id_I$, and (c) its run transformation is $\id_{\ZZ}$.} Definition (\rf{E789}) sets $\id_Γ$ equal to the tuple $[Γ,Γ,\id_X]$.  Thus Lemma~\rf{A854} shows that its player transformation is $\id_I$, and its run transformation is $\id_{\ZZ}$.  This proves (b) and (c).  Hence it remains to show that $[Γ,Γ,\id_X]$ satisfies (\rf{gZ}), (\rf{gK}), and (\rf{gU}) of Definition~\rf{E774}.  In this context, (\rf{gZ}) is \begin{gather}
\zz
[Θ,Θ,\id_x]⋅\text{is an end-preserving (\rf{E787}) CLT morphism}. \notag 
\zz
\end{gather} This holds because $[Θ,Θ,\id_X]$ is the CLT identity $\id_Θ$, and because this identity obviously preserves ends.  Further, (\rf{gK}) reduces to \begin{gather}
\zz
[Γ,Γ,\id_X]⋅\text{is control-preserving (\rf{E788})}, \notag 
\zz
\end{gather} which holds vacuously.  Finally, this claim's parts (b) and (c) reduce (\rf{gU}) to \begin{gather}
\zz
(∀i∈I)⋅\id_{\ZZ}⋅\text{is monotone from}⋅(\ZZ,\lesssim_i)⋅\text{to}⋅(\ZZ,\lesssim_i), \notag 
\zz
\end{gather} which holds vacuously.  

\yl{A828} {\em Suppose $γ = [Γ,Γ′,τ]$ and $γ′ = [Γ′,Γ″,τ′]$ are morphisms.  Then (a) $γ′○γ$ is a morphism, (b) its player transformation is $ι′○ι$, and (c) its run transformation is $ζ′○ζ$.}  First consider $γ = [Γ,Γ′,τ]$ and $γ′ = [Γ′,Γ″,τ′]$ without composing them.  Since both are game morphisms, Definition~\rf{E774} implies that both $γ$ and $γ′$ are control-preserving, and that both $[Θ,Θ′,τ]$ and $[Θ′,Θ″,τ′]$ are end-preserving morphisms.  Thus all the assumptions of Lemmas~\rf{F239} and \rf{A604} are met.

The definition of $○$ (\rf{E789}) sets $γ′○γ$ equal to $[Γ,Γ″,τ′○τ]$.  Thus Lemma~\rf{F239}(b) shows that its player transformation is $ι′○ι$, and Lemma~\rf{A604}(b) shows that its run transformation is $ζ′○ζ$.  This proves (b) and (c).  Hence it remains to show that $[Γ,Γ″,τ′○τ]$ satisfies conditions (\rf{gZ}), (\rf{gK}), and (\rf{gU}) of Definition~\rf{E774}.  Lemma~\rf{A604}(a) implies $[Γ,Γ″,τ′○τ]$ satisfies (\rf{gZ}).  Lemma~\rf{F239}(a) implies $[Γ,Γ″,τ′○τ]$ satisfies (\rf{gK}).  Finally, (\rf{gU}) for $γ$ and (\rf{gU}) for $γ′$ imply \begin{gather}
\zz
(∀i∈I)⋅ζ′○ζ⋅\text{is monotone from}⋅(\ZZ,\lesssim_i)⋅\text{to}⋅(\ZZ″,\lesssim″_{ι′○ι(i)}). \notag 
\zz
\end{gather} Hence this claim's parts (b) and (c) imply that $[Γ,Γ″,τ′○τ]$ satisfies (\rf{gU}).\end{cllist}\unskipcl\nichts{yyy}\end{npf}

\begin{npf}[{{\bf for Proposition~\rf{A455}}}]\label{A455p} By condition (\rf{G1}) of game Definition~\rf{E775}, $\FO$ maps any game to a CLT.  By condition (\rf{gZ}) of game-morphism Definition~\rf{E774}, $\FA$ maps any game morphism to a CLT morphism.  By inspection, $\FB$ preserves source and target.  Thus the proposition follows from Claims~\rf{A763} and \rf{A764}.\nichts{} 
\begin{cllist}

\yl{A763} {\em $\FB$ preserves identity.}  By the definition of $\id$ in \ct{Gm} (\rf{E789}), the image $\FA(\id_Γ)$ is equal to \begin{gather}
\zz
\FA([Γ,Γ,\id_X]), \notag 
\zz
\end{gather} which by the definition of $\FA$ in the proposition is equal to \begin{gather}
\zz
[\FO(Γ),\FO(Γ),\id_X], \notag 
\zz
\end{gather} which by the definition of $\id$ in \ct{CLT} (\rf{E790}), is equal to the identity $\id_{\FO(Γ)}$. 

\yl{A764} {\em $\FB$ preserves composition.}  By the definition of $○$ in \ct{Gm} (\rf{E789}), the image $\FA([Γ′,Γ″,τ′]○[Γ,Γ′,τ])$ is equal to \begin{gather}
\zz
\FA([Γ,Γ″,τ′○τ]), \notag 
\zz
\end{gather} which by the definition of $\FA$ in the proposition is equal to \begin{gather}
\zz
[\FO(Γ),\FO(Γ″),τ′○τ],\notag 
\zz
\end{gather} which by the definition of $○$ in \ct{CLT} (\rf{E790}) is equal to \nichts{} \begin{gather}
\zz
[\FO(Γ′),\FO(Γ″),τ′]⋅○⋅[\FO(Γ),\FO(Γ′),τ], \notag 
\zz
\end{gather} which by two applications of the proposition's definition of $\FA$ is equal to the composition $\FA([Γ′,Γ″,τ′])⋅○⋅\FA([Γ,Γ′,τ])$. \end{cllist} \unskipcl \nichts{yyy} \end{npf}

\nichts{}
\vspace{1mm}\section{For Monomorphisms}\label{B026}
\markb{\sc Appendix \rf{B026}. For Monomorphisms}

\begin{npf}[{{\bf for Proposition~\rf{B185}}}]\label{B185p}  Consider a morphism $[Θ,Θ′,τ]$.  By inspection, $[Θ,Θ′,τ]$ is monic if $τ$ is injective.  To show the converse, suppose that $τ$ is not injective.  Then there are $x_1⋅∈⋅X$ and $x_2⋅∈⋅X$ such that \ilc{A721} $x_1⋅≠⋅x_2$ and \li{A722} $τ(x_1) = τ(x_2)$. 

In accord with the last sentences of Section~\rf{B033}, let $Θ^* = [X^*,E^*,\HH^*,λ^*]$ be a one-node CLT in which $X^* = ⎨r^*⎬$, $E^* = ∅$, $\HH^* = ∅$, and $λ^*$ is the empty function.  Then construct the tuple $[Θ^*,Θ,τ^1]$ in which $τ^1(r^*) = x_1$.  This tuple is a morphism because (\rf{cE}) holds by $E^* = ∅$, and (\rf{cH}) and (\rf{cL}) hold by $W^* = ∅$.  Similarly construct the morphism $[Θ^*,Θ,τ^2]$ in which $τ^2(r^*) = x_2$.  

The morphisms $[Θ^*,Θ,τ^1]$ and $[Θ^*,Θ,τ^2]$ are distinct because $τ^1(r^*) = x_1$ and $τ^2(r^*) = x_2$ are distinct by \rf{A721}.  Yet the composition $[Θ,Θ′,τ]○[Θ^*,Θ,τ^1]$ equals the composition $[Θ,Θ′,τ]○[Θ^*,Θ,τ^2]$ because $τ○τ^1(r^*) = τ(x_1)$ and $τ○τ^2(r^*) = τ(x_2)$ are equal by \rf{A722}. Thus $[Θ,Θ′,τ]$ is not monic.  \end{npf}

\begin{lemma}\label{A713} Suppose $[Θ,Θ′,τ]$ is a monomorphism, and let $⁅α_x⁆_{x∈W}$ be its action transformation.  Then $(∀x∈W)$ $α_x$ is injective. \end{lemma} 

\begin{pf} By Proposition~\rf{B185}, $τ$ is injective.  Now take a decision node $x⋅∈⋅W$ and feasible actions $a_1⋅∈⋅F(x)$ and $a_2⋅∈⋅F(x)$.  It suffices to show  that $α_x(a_1) = α_x(a_2)$ implies $a_1 = a_2$.  Toward that end, suppose $α_x(a_1) = α_x(a_2)$.  Then $α_x$'s definition (\rf{E782}) implies \begin{gather}
\zz
λ′(\,τ(x)\,τ(n(x,a_1))\,) = λ′(\,τ(x)\,τ(n(x,a_2))\,). \notag 
\zz
\end{gather} Thus $λ′$˙'s local injectivity (\rf{C3}) implies $τ(n(x,a_1)) = τ(n(x,a_x))$.  Thus the injectivity of $τ$ implies $n(x,a_1) = n(x,a_2)$.  Thus Lemma~\rf{A816}(\rf{A822}) at $x_o = x$ implies $a_1 = a_2$. \end{pf}

\begin{lemma}\label{B129} Suppose $γ = [Γ,Γ′,τ]$ is a morphism.  Then $γ$ is monic if its run transformation $ζ$ is injective.  \end{lemma}

\begin{pf} To show the contrapositive, suppose $γ$ is not monic.  Then there are $γ^1$ and $γ^2$ such that \begin{gather}
\zz
γ○γ^1 = γ○γ^2⋅\text{and}⋅γ^1⋅≠⋅γ^2. \label{F805}
\zz
\end{gather} The equality in (\rf{F805}) implies that $γ^1$ and $γ^2$ have a common source, so denote that source $Γ^*$ and let $γ^1 = [Γ^*,Γ,τ^1]$ and $γ^2 = [Γ^*,Γ,τ^2]$.  Then (\rf{F805}) implies \begin{gather}
\zz
τ○τ^1 = τ○τ^2⋅\text{and}⋅τ^1⋅≠⋅τ^2. \label{B106} 
\zz
\end{gather} The inequality in (\rf{B106}) implies there is $x^*⋅∈⋅X^*$ such that  \begin{gather}
\zz
τ^1(x^*)⋅≠⋅τ^2(x^*). \label{B107} 
\zz
\end{gather} Further, since any node is in at least one run, there is $Z^*⋅∈⋅\ZZ^*$ such that \begin{gather}
\zz
x^*⋅∈⋅Z^*. \label{B111} 
\zz
\end{gather} The equality in (\rf{F805}) and two applications of Proposition~\rf{A454}(\rf{F238}) imply $ζ(ζ^1(Z^*)) = ζ(ζ^2(Z^*))$.  Thus Claim~\rf{B110} shows $ζ$ is not injective. \begin{cllist}

\yl{B109} {\em Neither} $τ^1(x^*) \prec τ^2(x^*)$ {\em nor} $τ^1(x^*) \succ τ^2(x^*)$.  By symmetry, it suffices to show the first is impossible.  If $τ^1(x^*)⋅≺⋅τ^2(x^*)$, then Proposition~\rf{A481}(\rf{A484}) for $γ$ would imply $τ○τ^1(x^*)⋅≺⋅τ○τ^2(x^*)$, which would contradict (\rf{B106})'s equality.

\yl{B110} $ζ^1(Z^*)⋅≠⋅ζ^2(Z^*)$.  To see this, suppose $ζ^1(Z^*) = ζ^2(Z^*)$.  Then the definition (\rf{E786}) of $ζ$ implies $P○τ^1(r^*)⋃\overline{τ^1}(Z^*) = P○τ^2(r^*)⋃\overline{τ^2}(Z^*)$, which by (\rf{B111}) implies  \begin{gather}
\zz
τ^1(x^*)⋅∈⋅P○τ^2(r^*)⋅⋃⋅\overline{τ^2}(Z^*). \notag 
\zz
\end{gather} 

On the one hand, suppose $τ^1(x^*)⋅∈⋅P○τ^2(r^*)$.  Then $τ^1(x^*)⋅≺⋅τ^2(r^*)$.  Meanwhile, $r^*⋅{≼^*}⋅x^*$ and Proposition~\rf{A481}(\rf{A483}) for $γ^2$ imply $τ^2(r^*)⋅≼⋅τ^2(x^*)$.  The last two sentences imply $τ^1(x^*)⋅≺⋅τ^2(x^*)$, which contradicts Claim~\rf{B109}.  

On the other hand, suppose $τ^1(x^*)⋅∈⋅\overline{τ^2}(Z^*)$.  Then (\rf{B111}) implies \begin{gather}
\zz
\text{both}⋅τ^1(x^*)⋅\text{and}⋅τ^2(x^*)⋅\text{belong to}⋅\overline{τ^2}(Z^*).  \label{F806}
\zz
\end{gather} Since $Z^*$ is a walk (\rf{E824}) in $(X^*,E^*)$, edge preservation (\rf{cE}) for $γ^2$ implies $\overline{τ^2}(Z^*)$ is a walk in $(X,E)$.  Further, since a walk in an out-tree visits each of this nodes exactly once, $(\overline{τ^2}(Z^*),≼)$ is a chain.  Thus (\rf{F806}) implies $τ^1(x^*)$ and $τ^2(x^*)$ belong to the same chain, which contradicts the combination of (\rf{B107}) and Claim~\rf{B109}. \end{cllist} \unskipcl \nichts{yyy} \end{pf}

\newcommand{\notegriso}{\footnote{\label{F074}A graph isomorphism from one out-tree $(X^A,E^A)$ to another out-tree $(X^B,E^B)$ is a bijection $τ{:}X^A→X^B$ such that $⎨˙τ(x)τ(y)˙|˙xy∈E^A˙⎬ = E^B$.\nichts{}}}
\begin{lemma}[from graph theory]\label{B132} Suppose that $(X,E)$ and $(X′,E′)$ are out-trees, that $τ{:}X→X′$ satisfies $(∀xy∈E)$ $τ(x)τ(y)⋅∈⋅E′$, and that $N$ is a walk in $(X,E)$ (expressed as a node set).  Then (a) $\dtau(N)$ is a walk in $(X′,E′)$ and (b) $τ|_{N,\dtau(N)}$ is a graph isomorphism{\notegriso} from the out-tree $(N,E⋂N^2)$ to the out-tree $(\dtau(N),E′⋂\dtau(N)^2)$.\nichts{} \end{lemma}

\begin{lemma}\label{B130} Suppose $γ = [Γ,Γ′,τ]$ is a morphism.  Then $γ$ is monic only if its run transformation $ζ$ is injective. \end{lemma}

\begin{pf} To prove the contrapositive, suppose $ζ$ is not injective.  Then there are runs $Z^1⋅∈⋅\ZZ$ and $Z^2⋅∈⋅\ZZ$ such that \subi \label{F068} \begin{gather}
\zz
Z^1⋅≠⋅Z^2⋅\text{and} \label{B119} \\
ζ(Z^1) = ζ(Z^2). \label{B120}
\zz
\end{gather} \subo By $\ZZ$'s definition (\rf{E813}), $Z^1$ and $Z^2$ are walks (\rf{E824}) in $(X,E)$ which are expressed as node sets.  Let \begin{gather}
\zz
E^1 = E⋂(Z^1)^2⋅\text{and}⋅E^2 = E⋂(Z^2)^2 \label{F073}
\zz
\end{gather} be their respective edge sets, so that $(Z^1,E^1)$ and $(Z^2,E^2)$ are the same walks expressed as out-trees.

Define the tuple \begin{gather}
\zz
Γ^1 = (Z^1,E^1,\HH^1,λ^1,κ^1,U^1) \label{F067}
\zz
\end{gather} by augmenting the out-tree $(Z^1,E^1)$ as follows.  First partition the decision-node set \mbox{$W^1 = π_1E^1$} by the collection $\HH^1 = ⎨˙⎨x⎬˙|˙x∈W^1˙⎬$.  Second define the labeling function $λ^1{:}E^1→⎨\f{go}⎬$ by $(∀xy∈E^1)$ $λ^1(xy) = \f{go}$ (the idea is let the only action be to ``$\f{go}$'' along with the walk $Z^1$).  Third define the control-assigning function $κ^1 = \id_{W^1}$ (Claim~\rf{B115}(b) will show that this identifies the player set $I^1$ with the decision-node set $W^1$; the idea is to let each decision node have its own player).  Finally define $U^1 = ⁅U^1_{i}{:}⎨Z^1⎬→\eR⁆_{i∈W^1}$ by $(∀i∈W^1)$ $U^1_{i}(Z^1) = 0$ (the idea is to let each $U^1_i$ have a singleton domain). 

Further, define the tuple \begin{gather}
\zz
γ^1 = [Γ^1,Γ,˙\inc_{Z^1,X}], \label{F069}
\zz
\end{gather} which is well-defined since the run $Z^1⋅∈⋅\ZZ$ is a subset of $X$ ($\inc_{Z^1,X}$ is the inclusion function from $Z^1$ to $X$).  Next define $δ{:}Z^1→Z^2$ by \begin{gather}
\zz
δ = (τ|_{Z^2,\dtau(Z^2)})^{-1}⋅○⋅τ|_{Z^1,\dtau(Z^1)}, \label{F070}
\zz
\end{gather} which Claim~\rf{F072} will show is well-defined.  Finally define the tuple \begin{gather}
\zz
γ^2 = [Γ^1,Γ,˙\inc_{Z^2,X}○δ], \label{F071}
\zz
\end{gather} which is well-defined since $δ$ maps $Z^1$ to $Z^2$, and since $Z^2⋅∈⋅\ZZ$ is a subset of~$X$.  

Figure~\rf{F209} depicts three node transformations.  First, the node transformation $τ$ is part of the morphism $γ = [Γ,Γ′,τ]$, which is assumed by (\rf{F068}) to have a non-injective run transformation at $Z^1$ and $Z^2$.  Second, $\inc_{Z^1,X}$ is part of the tuple $γ^1$ in (\rf{F069}).  Third, $\inc_{Z^2,X}○δ$ is part of the tuple $γ^2$ in (\rf{F071}).  

\begin{figure}[t] \newcommand{\hgth}{4.00}
\begin{picture}(0,\hgth) \myoutergrid{\hgth} \renewcommand{\sella}[1]{#1}
  \put(-0.20, 2.14){\makebox(0,0){\scalebox{0.95}{ \begin{pspicture} \end{pspicture} }}} \end{picture}
\caption{\small The three node transformations of Lemma~\rf{B130}'s proof.  By assumption, the distinct runs $Z^1⋅⊆⋅X$ and $Z^2⋅⊆⋅X$ satisfy $ζ(Z^1) = ζ(Z^2)$ where $ζ$ is the run transformation of $γ = [Γ,Γ′,τ]$.  The function $δ = (τ|_{Z^2,\dtau(Z^2)})^{-1}○τ|_{Z^1,\dtau(Z^1)}$ is a bijection from $Z^1$ to $Z^2$.} \label{F209} \end{figure}
 
\begin{cllist}
\yl{B125} {\em (a) $Θ^1 = (Z^1,E^1,\HH^1,λ^1)$ is a CLT. (b) $\ZZ^1 = ⎨Z^1⎬$.} By definition~(\rf{F073}), $(Z^1,E^1)$ is an out-tree whose only run is the walk $Z^1$.  This establishes (\rf{C1}) of Definition~\rf{E772} and part (b) of this Claim~\rf{B125}.  Thus it remains to show (\rf{C2})--(\rf{C4}).  (\rf{C2}) follows immediately from the definition of $\HH^1$ below (\rf{F067}).  For (\rf{C3}), recall $Z^1$ is a walk, which implies that each element of $Z^1$ has exactly one immediate successor, which vacuously implies $λ^1$ is locally injective.  Further, $λ^1$ is surjective by its definition below (\rf{F067}).  Finally, (\rf{C4}) holds vacuously because $\HH^1$ generates the discrete topology.

\yl{B115} {\em (a) $Γ^1$, from definition (\rf{F067}), is a game. (b) $I^1 = W^1$.}  For (b), $I^1$ is by definition (\rf{F218}) equal to the codomain of $κ^1$, which by the definition $κ^1 = \id_{W^1}$ below (\rf{F067}) is equal to $W^1$.  For (a), it suffices to show that the tuple satisfies (\rf{G1})--(\rf{G3}) of Definition~\rf{E775}.  (\rf{G1}) holds by Claim~\rf{B125}.  For (\rf{G2}), $κ^1 = \id_{W^1}$ is surjective by inspection and continuous because $\HH^1$ generates the discrete topology for $W^1$.  Finally for (\rf{G3}), note that $U^1$ by its definition below (\rf{F067}) has the form $⁅U^1_i{:}⎨Z^1⎬→\eR⁆_{i∈W^1}$, which by Claim~\rf{B125}(b) is $⁅U^1_{i}{:}\ZZ→\eR⁆_{i∈W^1}$, which by this claim's part (b) is $⁅U^1_{i}{:}\ZZ→\eR⁆_{i∈I^1}$.

\yl{B126} {\em $[Θ^1,Θ,\inc_{Z^1,X}]$ is a CLT morphism.}  $Θ^1$ is a CLT by Claim~\rf{B125}(a) and $Θ$ is a CLT by assumption.  Thus it suffices to show that the tuple satisfies (\rf{cE}), (\rf{cH}), and (\rf{cL}) of Definition~\rf{E773}.  For edge preservation (\rf{cE}), $E^1$'s definition (\rf{F073}) implies that every edge in $E^1$ is also in $E$, and thus the node transformation $\inc_{Z^1,X}$ preserves edges.  Further, (\rf{cH}) and (\rf{cL}) hold vacuously because $\HH^1$ generates the discrete topology. 

\yl{B113} {\em $γ^1 = [Γ^1,Γ,\inc_{Z^1,X}]$, from definition (\rf{F069}), is a game morphism.} $Γ^1$ is a game by Claim~\rf{B115}(a) and $Γ$ is a game by assumption.  Thus it suffices to show (\rf{gZ}), (\rf{gK}), and (\rf{gU}) of Definition~\rf{E774}.  

For (\rf{gZ}), Claim~\rf{B126} shows $[Θ^1,Θ,\inc_{Z^1,X}]$ is a morphism.  Thus it suffices to show $[Θ^1,Θ,\inc_{Z^1,X}]$ preserves ends (\rf{E787}).  In other words, it suffices to show \begin{gather}
\zz
\overline{\inc_{Z^1,X}}(X^1⧷W^1)⋅⊆⋅X⧷W. \label{F824} 
\zz
\end{gather} Recall from (\rf{F067}) that $Θ^1$'s out-tree \begin{gather}
\zz
\text{$(X^1,E^1) = (Z^1,E^1)$} \label{F825} 
\zz
\end{gather} is the run $Z^1⋅∈⋅\ZZ$ expressed as an out-tree.  Further, $Z^1⋅∈⋅\ZZ$ and definition (\rf{E813}) imply $Z^1⋅∈⋅\ZZi⋃\ZZf$.  On the one hand, suppose $Z^1⋅∈⋅\ZZi$.  Then (\rf{F825}) implies that $Θ^1$'s end-node set $X^1⧷W^1$ is empty and thus (\rf{F824}) holds vacuously.  On the other hand, suppose $Z^1⋅∈⋅\ZZf$.  Then (\rf{F825}) implies that $Θ^1$'s end-node set $X^1⧷W^1$ is the singleton consisting of the last node in the run $Z^1⋅∈⋅\ZZf$, which by the definition of $\ZZf$ is an end node in $X⧷W$. This implies (\rf{F824}). 
 
For (\rf{gK}), control preservation (\rf{E788}) holds if  \begin{gather}
\zz
(∀x_A∈W^1,x_B∈W^1)⋅κ^1(x_A) = κ^1(x_B)⋅\text{implies}⋅κ(\inc_{Z^1,X}(x_A)) = κ(\inc_{Z^1,X}(x_B)).\notag 
\zz
\end{gather} Since $κ^1 = \id_{W^1}$ by its definition below (\rf{F067}), $κ^1(x_A) = κ^1(x_B)$ implies $x_A = x_B$, which vacuously yields the implication's conclusion.

Finally, (\rf{gU}) holds if \begin{gather}
\zz
(∀i∈I^1)⋅ζ^1⋅\text{is monotone from}⋅(\ZZ^1,\lesssim^1_i)⋅\text{to}⋅(\ZZ,\lesssim_{ι^1(i)}), \notag 
\zz
\end{gather} where $I^1$ and $⁅\lesssim^1_i⁆_{i∈I^1}$ is derived from $Γ^1$, and where $ζ^1$ and $ι^1$ are derived from $γ^1$.  This holds vacuously because Claim~\rf{B125}(b) shows $\ZZ^1$ is the singleton $⎨Z^1⎬$.\footnote{Proposition~\rf{E791}(a) shows that the player transformation $ι^1$ is well-defined.  It may be nontrivial, but this is irrelevant to showing (\rf{gU}).}

\yl{B139} {\em (a) $τ|_{Z^1,\dtau(Z^1)}$ is a graph isomorphism (note \rf{F074}) from $(Z^1,E^1)$ to $(\dtau(Z^1),E′⋂\dtau(Z^1)^2)$.  (b) $τ|_{Z^2,\dtau(Z^2)}$ is a graph isomorphism from $(Z^2,E^2)$ to $(\dtau(Z^2),E′⋂\dtau(Z^2)^2)$. }  Lemma~\rf{B132} at $N = Z^1$ implies that $τ|_{Z^1,\dtau(Z^1)}$ is a graph isomorphism from $(Z^1,E⋂(Z^1)^2)$ to $(\dtau(Z^1),E′⋂\dtau(Z^1)^2)$.  This implies part (a) because $E^1 = E⋂(Z^1)^2$ by definition (\rf{F073}).  Part (b) holds similarly. 

\yl{B123} {\em $\dtau(Z^1) = \dtau(Z^2)$.} Equation (\rf{B120}) and $ζ$'s definition (\rf{E786}) imply \begin{gather}
\zz
P′○τ(r)⋅⋃⋅\dtau(Z^1) = P′○τ(r)⋅⋃⋅\dtau(Z^2). \label{B138}
\zz
\end{gather} Since $(∀x∈X)$ $r⋅≼⋅x$, Proposition~\rf{A481}(\rf{A483}) implies $(∀x∈X)⋅τ(r)⋅≼′⋅τ(x)$, which implies $P′○τ(r)$ and $\dtau(X)$ are disjoint.  So $Z^1⋅⊆⋅X$ implies $P′○τ(r)$ and $\dtau(Z^1)$ are disjoint, and $Z^2⋅⊆⋅X$ implies $P′○τ(r)$ and $\dtau(Z^2)$ are disjoint.  So (\rf{B138}) implies the claim. 

\yl{F072} {\em $δ = (τ|_{Z^2,\dtau(Z^2)})^{-1}⋅○⋅τ|_{Z^1,\dtau(Z^1)}$, from definition (\rf{F070}), is a graph isomorphism from $(Z^1,E^1)$ to $(Z^2,E^2)$.}  Claims~\rf{B139}(b) implies that \begin{gather}
\zz
τ|_{Z^2,\dtau(Z^2)}⋅\text{is a graph isomorphism from}⋅(Z^2,E^2)⋅\text{to}⋅(\dtau(Z^2),E′⋂\dtau(Z^2)^2).\notag 
\zz
\end{gather} Thus Claim~\rf{B123} implies that \begin{gather}
\zz
τ|_{Z^2,\dtau(Z^2)}⋅\text{is a graph isomorphism from}⋅(Z^2,E^2)⋅\text{to}⋅(\dtau(Z^1),E′⋂\dtau(Z^1)^2).\notag 
\zz
\end{gather} Hence its inverse is a graph isomorphism from $(\dtau(Z^1),E′⋂\dtau(Z^1)^2)$ to $(Z^2,E^2)$.  The result then follows from Claim~\rf{B139}(a).

\yl{B127} {\em $[Θ^1,Θ,\inc_{Z^2,X}○δ]$ is a CLT morphism.}  $Θ^1$ is a CLT by Claim~\rf{B125}, and $Θ$ is a CLT by assumption.  Further, Claim~\rf{F072} implies that the composition $\inc_{Z^2,X}○δ$ maps from $Z^1$, which is the node set of $Θ^1$ by definition (\rf{F067}), to $X$, which is the node set of $Θ$. Thus it suffices to show (\rf{cE}), (\rf{cH}), and (\rf{cL}).  For (\rf{cE}), take $xy⋅∈⋅E^1$.  Then Claim~\rf{F072} implies $δ(x)δ(y)⋅∈⋅E^2$, which by $E^2$'s definition (\rf{F073}) implies $δ(x)δ(y)⋅∈⋅E$.  Finally, (\rf{cH}) and (\rf{cL}) hold vacuously because $\HH^1$ (\rf{F067}) generates the discrete topology.  

\yl{B114} {\em $γ^2 = [Γ^1,Γ,˙\inc_{Z^2,X}○δ]$, from definition (\rf{F071}), is a game morphism.} \linebreak $Γ^1$ is a game by Claim~\rf{B115}(a), and $Γ$ is a game by assumption.  Thus is suffices to show (\rf{gZ}), (\rf{gK}), and (\rf{gU}). 

For (\rf{gZ}), Claim~\rf{B127} shows $[Θ^1,Θ,\inc_{Z^2,X}○δ]$ is a morphism.  Thus it suffices to show $[Θ^1,Θ,\inc_{Z^2,X}○δ]$ preserves ends (\rf{E787}).  In other words, it suffices to show \begin{gather}
\zz
\overline{\inc_{Z^2,X}○δ}(X^1⧷W^1)⋅⊆⋅X⧷W. \label{F826} 
\zz
\end{gather} Recall from (\rf{F067}) that $Θ^1$'s out-tree \begin{gather}
\zz
\text{$(X^1,E^1) = (Z^1,E^1)$} \label{F827} 
\zz
\end{gather} is the run $Z^1⋅∈⋅\ZZ$ expressed as an out-tree.  Further, $Z^1⋅∈⋅\ZZ$ and definition (\rf{E813}) imply either $Z^1⋅∈⋅\ZZi⋃\ZZf$.  On the one hand, suppose $Z^1⋅∈⋅\ZZi$.  Then (\rf{F827}) implies that $Θ^1$'s end-node set $X^1⧷W^1$ is empty and thus (\rf{F826}) holds vacuously.  On the other hand, suppose $Z^1⋅∈⋅\ZZf$.  Then (\rf{F827}) implies that $Θ^1$'s end-node set $X^1⧷W^1$ is the singleton consisting of the last node in the finite walk $Z^1$, which by Claim~\rf{F072} implies $\bar{δ}(X^1⧷W^1)$ is the singleton consisting of the last node in the finite walk $Z^2$, which by $Z^2⋅∈⋅\ZZ$ (above (\rf{F068})) and the definition of $\ZZf$ (above (\rf{E813})) implies $\bar{δ}(X^1⧷W^1)$ is an end node in $X⧷W$. This implies~(\rf{F826}).  

For (\rf{gK}), control preservation (\rf{E788}) holds if  \begin{gather}
\zz
(∀x_A∈W^1,x_B∈W^1)⋅κ^1(x_A) = κ^1(x_B)⋅\text{implies}⋅κ(\inc_{Z^2,X}○δ(x_A)) = κ(\inc_{Z^2,X}○δ(x_B)).\notag 
\zz
\end{gather} Since $κ^1 = \id_{W^1}$ by its definition below (\rf{F067}), $κ^1(x_A) = κ^1(x_B)$ implies $x_A = x_B$, which vacuously yields the implication's conclusion.

Finally, (\rf{gU}) holds if \begin{gather}
\zz
(∀i∈I^1)⋅ζ^2⋅\text{is monotone from}⋅(\ZZ^1,\lesssim^1_i)⋅\text{to}⋅(\ZZ,\lesssim_{ι^2(i)}), \notag 
\zz
\end{gather} where $I^1$ and $⁅\lesssim^1_i⁆_{i∈I^1}$ are derived from $Γ^1$, and where $ζ^2$ and $ι^2$ are derived from $γ^2$.  This holds vacuously because Claim~\rf{B125}(b) shows $\ZZ^1$ is the singleton $⎨Z^1⎬$.

\yl{B116} {\em $γ^1⋅≠⋅γ^2$.}  It suffices to show $\inc_{Z^1,X}⋅≠⋅\inc_{Z^2,X}○δ$.  Thus it suffices to show there is a node $x^1⋅∈⋅Z^1$ such that $δ(x^1)⋅≠⋅x^1$.  Toward that end, we know from~(\rf{B119}) that $Z^1⧷Z^2$ or $Z^2⧷Z^1$ is nonempty.  In the first contingency, take $x^1⋅∈⋅Z^1⧷Z^2$.  Then $x^1⋅∈⋅Z^1$ and Claim~\rf{F072} imply $δ(x^1)⋅∈⋅Z^2$, which by $x^1⋅∉⋅Z^2$ implies $δ(x^1)⋅≠⋅x^1$.  In the second contingency, take $x^2⋅∈⋅Z^2⧷Z^1$.  First, $x^2⋅∈⋅Z^2$ and Claim~\rf{F072} imply there is \ilc{B147} $x^1⋅∈⋅Z^1$ such that \il{B148} $δ(x^1) = x^2$.  Second, $x^2⋅∉⋅Z^1$ and \rf{B148} imply $δ(x^1)⋅∉⋅Z^1$, which by \rf{B147} implies $δ(x^1)⋅≠⋅x^1$.

\yl{B117} {\em $γ○γ^1 = γ○γ^2$.} Since $γ = [Γ,Γ′,τ]$ is a morphism by assumption, and since $γ^1 = [Γ^1,Γ,˙\inc_{Z^1,X}]$ and $γ^2 = [Γ^1,Γ,˙\inc_{Z^2,X}○δ]$ are morphisms by Claims~\rf{B113} and~\rf{B114}, it suffices to show  \begin{gather}
\zz
τ⋅○⋅\inc_{Z^1,X} = τ⋅○⋅\inc_{Z^2,X}⋅○⋅δ. \notag 
\zz
\end{gather} Since $τ{:}X→X′$, this is equivalent to \begin{gather}
\zz
\inc_{\dtau(Z^1),X′}⋅○⋅τ|_{Z^1,\dtau(Z^1)} = \inc_{\dtau(Z^2),X′}⋅○⋅τ|_{Z^2,\dtau(Z^2)}⋅○⋅δ, \notag 
\zz
\end{gather} which by the definition of $δ$ (\rf{F070}) is equivalent to  \begin{gather}
\zz
\inc_{\dtau(Z^1),X′}⋅○⋅τ|_{Z^1,\dtau(Z^1)} = \inc_{\dtau(Z^2),X′}⋅○⋅τ|_{Z^2,\dtau(Z^2)}⋅○⋅(τ|_{Z^2,\dtau(Z^2)})^{-1}⋅○⋅τ|_{Z^1,\dtau(Z^1)}, \notag 
\zz
\end{gather} which by cancellation is equivalent to  \begin{gather}
\zz
\inc_{\dtau(Z^1),X′}⋅○⋅τ|_{Z^1,\dtau(Z^1)} = \inc_{\dtau(Z^2),X′}⋅○⋅τ|_{Z^1,\dtau(Z^1)}, \notag 
\zz
\end{gather} which holds after applying Claim~\rf{B123}'s equality to the domains of the inclusion operators. \end{cllist}

{\em Conclusion.} Claims~\rf{B116} and \rf{B117} imply that $γ$ is not monic. \end{pf}

\begin{npf}[{{\bf for Proposition~\rf{B178}}}]{\label{B178p}}  Lemma~\rf{B129} shows that the injectivity of $ζ$ is sufficient. Lemma~\rf{B130} shows it is necessary.  \end{npf}

\nichts{}
\vspace{1mm}\section{For Isomorphisms}\label{F224}
\markb{\sc Appendix \rf{F224}.  For Isomorphisms}

\begin{lemma}\label{A537} Suppose that $(X,E)$ and $(X′,E′)$ are out-trees, that $τ{:}X→X′$ preserves edges (\rf{E870}), and that $τ$ is a bijection.  Then the following hold. \begin{tlist}
\yl{A546x} $E⋅∋⋅xy \mapsto τ(x)τ(y)⋅∈⋅E′$ is a bijection.
\yl{A501x} $τ|_{W,W′}$ is a bijection.
\yl{A502x} $τ|_{X⧷W,X′⧷W′}$ is a bijection.
\yl{A492x} $(∀x∈X,y∈X)⋅x⋅≼⋅y⋅⟺⋅τ(x)⋅≼′⋅τ(y)$. 
\yl{A493x} $(∀x∈X,y∈X)⋅x⋅≺⋅y⋅⟺⋅τ(x)⋅≺′⋅τ(y)$. 
\yl{A533x} $τ(r) = r′$.
\yl{A494x} $(∀y∈X)⋅\dtau(P(y)) = P′(τ(y))$.
\yl{A498x} $\dtau|_{\ZZ,\ZZ′}$ is a bijection, and its inverse is $\overline{τ^{-1}}|_{\ZZ′,\ZZ}$.
\end{tlist}\end{lemma}

\begin{pf} {\em (\rf{A546x})}. Let $E^τ = ⎨⋅τ(x)τ(y)⋅|⋅xy∈E⋅⎬$.  Edge preservation (\rf{E870}) implies \begin{gather}
\zz
E^τ⋅⊆⋅E′. \label{F828}
\zz
\end{gather} Further, since $τ$ is injective, $E⋅∋⋅xy \mapsto τ(x)τ(y)⋅∈⋅E^τ$ is a bijection.  Thus it remains to show that $E^τ⋅⊇⋅E′$.  

Before beginning, we make two observations.  First, since $(X,E)$ is an out-tree and since $τ{:}X→X′$ is a bijection, \begin{gather}
\zz
\text{$(X′,E^τ)$ is an out-tree with root $τ(r)$}. \label{F807}
\zz
\end{gather} Second, we show \begin{gather}
\zz
τ(r) = r′, \label{F188} 
\zz
\end{gather} where $r′$ is the root of the out-tree $(X′,E′)$.  To see this, suppose  $τ(r)⋅≠⋅r′$.  Then $(X′,E′)$ being an out-tree (\rf{E776}) implies there is a nontrivial walk in $(X′,E′)$ from $r′$ to $τ(r)$.  Further, (\rf{F807}) implies there is a nontrivial walk in $(X′,E^τ)$ from $τ(r)$ to~$r′$, and thus (\rf{F828}) implies there is a nontrivial walk in $(X′,E′)$ from $τ(r)$ to $r′$.  The concatenation of these two walks is a nontrivial walk in $(X′,E′)$ from $r′$ to itself, which contradicts $(X′,E′)$ being an out-tree.

To show $E^τ⋅⊇⋅E′$, suppose \begin{gather}
\zz
x′y′⋅∈⋅E′⧷E^τ. \label{F189}
\zz
\end{gather} By (\rf{F807}), there are walks in $(X′,E^τ)$ from $τ(r)$ to $x′$, and from $τ(r)$ to~$y′$.  By (\rf{F828}), these are walks in $(X′,E′)$.  Further, by (\rf{F188}), these are walks in $(X′,E′)$ from $r′$ to $x′$, and from $r′$ to $y′$.  Since $x′y′⋅∈⋅E′$ by (\rf{F189}), the first walk can be augmented into a walk in $(X′,E′)$ from $r′$ to $y′$, and this augmented first walk cannot equal the second walk since $x′y′⋅∉⋅E^τ$ by (\rf{F189}) and since the second walk is in $(X′,E^τ)$ by definition. Thus there are two distinct walks in $(X′,E′)$ from $r′$ to $y′$, in contradiction to $(X′,E′)$ being an out-tree.  

{\em (\rf{A501x})--(\rf{A498x})}. Part (\rf{A546x}) implies that the out-trees $(X,E)$ and $(X′,E′)$ are graph-isomorphic (note \rf{F074}).\nichts{}  This immediately implies parts (\rf{A501x})--(\rf{A498x}).
\end{pf}

\begin{lemma}\label{F226} Suppose $Θ$ and $Θ′$ are CLTs and $τ{:}X→X′$ is a bijection.  Then the following are equivalent. \begin{tlist}
\smallskip\yl{F229} $τ|_{W,W′}$ is a homeomorphism.
\smallskip\yl{F808} $(∀H∈\HH)$ $\dtau(H)⋅∈⋅\HH′$ and $(∀H′∈\HH′)$ $\overline{τ^{-1}}(H′)⋅∈⋅\HH$.
\smallskip\yl{F231} $\dtau|_{\HH,\HH′}$ is a bijection and its inverse is $\overline{τ^{-1}}|_{\HH′,\HH}$.
\smallskip \yl{F813} $(∀H∈\HH)⋅\dtau(H)⋅∈⋅\HH′$. 
\end{tlist} \end{lemma}

\begin{pf} Before beginning, note that $\HH$ partitions $X$ by (\rf{C2}) for $Θ$, and that $\HH′$ partitions $X′$ by (\rf{C2}) for $Θ′$.

{\em (\rf{F229}$⇒$\rf{F808}).}  Assume (\rf{F229}).

To show the first half of (\rf{F808}), take an information set $H⋅∈⋅\HH$.  Since $τ|_{W,W′}$ is a bijection by (\rf{F229}), we have $\dtau(W)⋅⊆⋅W′$, which implies the assumptions of Proposition~\rf{B063} are met.  Thus the forward direction of the proposition implies there is $H′⋅∈⋅\HH′$ such that $\dtau(H)⋅⊆⋅H′$, which by the bijectivity of $τ$ implies \begin{gather}
\zz
H⋅⊆⋅\overline{τ^{-1}}(H′). \label{F227}
\zz
\end{gather} Conversely, since $τ|_{W,W′}$ is a bijection by (\rf{F229}), we have $\dtau^{-1}(W′)⋅⊆⋅W$, which implies the assumptions of Proposition~\rf{B063} for $[Θ′,Θ,τ^{-1}]$ are met.  Thus, since $H′⋅∈⋅\HH′$, the forward direction of the proposition at $[Θ′,Θ,τ^{-1}]$ implies there is $H^*⋅∈⋅\HH$ such that \begin{gather}
\zz
\overline{τ^{-1}}(H′)⋅⊆⋅H^*. \label{F228}
\zz
\end{gather} But since $H$ and $H^*$ are cells in the partition $\HH$, equations (\rf{F227}) and (\rf{F228}) imply $H = H^*$ and $H = \dtau^{-1}(H′)$.  Since $τ$ is bijective, the latter equation implies $\dtau(H) = H′$.  This suffices since $H′⋅∈⋅\HH′$ by construction.  

The second half of (\rf{F808}) holds by a symmetric argument.

{\em (\rf{F229}$⇐$\rf{F808}).} Assume (\rf{F808}).  It suffices to show that both $τ|_{W,W′}$ and $τ^{-1}|_{W′,W}$ are continuous.

To show $τ|_{W,W′}$ is continuous, note that the first half of (\rf{F808}), together with $\HH$ partitioning $W$ and $\HH′$ partitioning $W′$, implies $\dtau(W)⋅⊆⋅W′$.  Thus the assumptions of Proposition~\rf{B063} hold.  Hence the proposition's reverse direction and (\rf{F808})'s first half imply the continuity of $τ|_{W,W′}$.  

The continuity of $τ^{-1}|_{W′,W}$ holds by a symmetric argument.\nichts{}

{\em (\rf{F808}$⇒$\rf{F231}).} Assume (\rf{F808}).  Since $\HH$ partitions $W$ and $\HH′$ partitions $W′$, the first half of (\rf{F808}) implies that $\dtau(W)⋅⊆⋅W′$.  For the same reason, the second half of (\rf{F808}) implies that $\overline{τ^{-1}}(W′)⋅⊆⋅W$.  Thus the bijectivity of $τ{:}X→X′$ implies that $τ|_{W,W′}$ is a bijection.  Thus, $\HH$ partitioning $W$ and $\HH′$ partitioning $W′$ imply (\rf{F808}) implies (\rf{F231}). 

{\em (\rf{F808}$⇐$\rf{F231}).} Assume (\rf{F231}).  The first half of (\rf{F808}) follows from the well-definition of $\dtau|_{\HH,\HH′}$ in (\rf{F231})'s first half.  The second half of (\rf{F808}) follows from the well-definition of $\overline{τ^{-1}}|_{\HH′,\HH}$ in (\rf{F231})'s second half. 

\newcommand{\tihp}{\overline{τ^{-1}}(H′)}
{\em (\rf{F808}$⇔$\rf{F813}).}  Since (\rf{F808})$⇒$(\rf{F813}) is obvious, it suffices to prove the converse.  Toward that end, suppose (\rf{F813}) and take a target information set \lic{F814} $H′⋅∈⋅\HH′$.  Since $H′$ is a cell in a partition, $H′$ is nonempty, which implies that $\tihp$ is nonempty, which implies that there is some source information set \il{F815} $H⋅∈⋅\HH$ which intersects $\tihp$.  This intersection implies that $\dtau(H)$ intersects $\dtau(\tihp)$, which by $\dtau(\tihp) = H′$ (from $τ$'s bijectivity) implies that $\dtau(H)$ intersects $H′$.  This intersection, with $H′⋅∈⋅\HH′$ from~\rf{F814}, and with $\dtau(H)⋅∈⋅\HH′$ by \rf{F815} and (\rf{F813}), implies that $\dtau(H) = H′$.  Thus $τ$'s bijectivity implies $H = \tihp$, which by \rf{F815} implies $\tihp⋅∈⋅\HH$.  \end{pf}

\begin{lemma}\label{A779} Suppose $[Θ,Θ′,τ]$ is a morphism and $τ$ is bijective.  Then the following hold. \begin{tlist}
\yl{B188} $(∀x∈W)$ $α_x{:}F(x)→F′(τ(x))$ is bijective.%
\yl{F190} The action transformation of the tuple $[Θ′,Θ,τ^{-1}]$ is $⁅α^{-1}_{τ^{-1}(x′)}⁆_{x′∈W′}$.
\yl{F193} $[Θ,Θ′,τ]$ is end-preserving.
\yl{B189} $ζ = \dtau|_{\ZZ,\ZZ′}$.
\end{tlist} 
\end{lemma}

\begin{pf} {\em (\rf{B188})}. Take a decision node $x⋅∈⋅W$.  By definition (\rf{E782}), $α_x{:}F(x)→F′(τ(x))$.  To show its injectivity, take feasible actions $a_1⋅∈⋅F(x)$ and $a_2⋅∈⋅F(x)$.  It suffices to show that $α_x(a_1) = α_x(a_2)$ implies $a_1 = a_2$.  Toward that end, suppose $α_x(a_1) = α_x(a_2)$.  Then $α_x$'s definition (\rf{E782}) implies \begin{gather}
\zz
λ′(\,τ(x)\,τ(n(x,a_1))\,) = λ′(\,τ(x)\,τ(n(x,a_2))\,). \notag 
\zz
\end{gather} Thus $λ′$˙'s local injectivity (\rf{C3}) implies $τ(n(x,a_1)) = τ(n(x,a_x))$.  Thus the injectivity of $τ$ implies $n(x,a_1) = n(x,a_2)$.  Thus Lemma~\rf{A816}(\rf{A822}) at $x_o = x$ implies $a_1 = a_2$.

It remains to show that $α_x$ is surjective.  Since $α_x{:}F(x)→F′(τ(x))$ by definition (\rf{E782}), it suffices to show $\overline{α_x}(F(x))⋅⊇⋅F′(τ(x))$.  Toward that end, take $a′⋅∈⋅F′(τ(x))$.  By the definition (\rf{E779}) of $F′$, there is $y′⋅∈⋅X′$ such that $a′ = λ′(\,τ(x)\,y′\,)$.  Thus by the bijectivity of $τ$, there is $y⋅∈⋅X$ such that $a′ = λ′(\,τ(x)\,τ(y)\,)$.  Thus Proposition~\rf{A481}(\rf{A804}) implies \lic{A780} $a′ = α_x(λ(xy))$.  Meanwhile, the definition (\rf{E779}) of $F$ implies $λ(xy)⋅∈⋅F(x)$.  So \rf{A780} implies $a′⋅∈⋅\overline{α_x}(F(x))$.

{\em (\rf{F190})}. Let $α^*$ denote the action transformation of $[Θ′,Θ,τ^{-1}]$.  Then definition (\rf{E782}) characterizes by $α^*$ by [1] $α^* = ⁅α^*_{x′}⁆_{x′∈W′}$ and [2] for each decision node $x′⋅∈⋅W′$  \begin{gather}
\zz
α^*_{x′}{:}F′(x′)→F(τ^{-1}(x′))⋅\text{and} \nt
(∀a′∈F′(x′))⋅α^*_{x′}(a′) = λ(⋅τ^{-1}(x′)˙τ^{-1}(n′(x′,a′))˙). \notag
\zz
\end{gather} Thus it suffices to show, for each decision node $x′⋅∈⋅W′$, that $α^{-1}_{τ^{-1}(x′)}$ is well-defined,  \begin{gather}
\zz
α^{-1}_{τ^{-1}(x′)}{:}F′(x′)→F(τ^{-1}(x′)),⋅\text{and} \nt
(∀a′∈F′(x′))⋅α^{-1}_{τ^{-1}(x′)}(a′) = λ(⋅τ^{-1}(x′)˙τ^{-1}(n′(x′,a′))˙). \notag
\zz
\end{gather} Toward that end, take $x′⋅∈⋅W′$.  Define $x = τ^{-1}(x′)$.  Lemma~\rf{A537}(\rf{A501x}) implies $x⋅∈⋅W$.  So by several substitutions it suffices to show that $α^{-1}_x$ is well-defined, \subi\label{F811} \begin{gather}
\zz
α^{-1}_x{:}F′(τ(x))→F(x),⋅\text{and} \label{F810}\\
(∀a′∈F′(τ(x)))⋅α^{-1}_{x}(a′) = λ(⋅x˙τ^{-1}(n′(τ(x),a′))˙). \label{F809}
\zz
\end{gather} \subo Part (\rf{B188}) implies that $α^{-1}_x$ is well-defined and that (\rf{F810}) holds.  Further, since $α_x{:}F(x)→F′(τ(x))$ is a bijection by part (\rf{B188}), (\rf{F809}) is equivalent to \begin{gather}
\zz
(∀a∈F(x))⋅a = λ(⋅x⋅τ^{-1}(n′(τ(x),α_x(a)))⋅). \notag
\zz
\end{gather} To show this equality, take a feasible action $a⋅∈⋅F(x)$.  Then in steps, $a$ by Lemma~\rf{A816}(\rf{A822}) at $x_o = x$ is equal to $λ(\,x\,n(x,a)\,)$, which by inspection is equal to $λ(\,x\,τ^{-1}(τ(n(x,a)))\,)$, which by Proposition~\rf{A481}(\rf{A490}) is equal to the right-hand side.

{\em (\rf{F193})}. Lemma~\rf{A537}(\rf{A502x}) implies that $\dtau(X⧷W) = X′⧷W′$.  Hence $[Θ,Θ′,τ]$ is end-preserving (\rf{E787}).

{\em (\rf{B189}).} Lemma~\rf{A537}(\rf{A533x}) implies $τ(r) = r′$, which implies $P′○τ(r) = ∅$, which by the definition of $ζ$ (\rf{E786}) implies $ζ = \dtau|_{\ZZ,\ZZ′}$.  \end{pf}

\begin{lemma}\label{A781} Suppose $[Θ,Θ′,τ]$ is a morphism such that $τ$ is a bijection and $τ|_{W,W′}$ is a homeomorphism.  Then $[Θ,Θ′,τ]$ is an isomorphism.  \end{lemma}

\begin{pf} It suffices to show that $[Θ′,Θ,τ^{-1}]$ is a morphism.  Thus by Definition~\rf{E773}, it suffices to show that $[Θ′,Θ,τ^{-1}]$ satisfies (\rf{cE}), (\rf{cH}), and (\rf{cL}).  First, (\rf{cE}) for $[Θ′,Θ,τ^{-1}]$ is $(∀x′y′∈E′)$ $τ^{-1}(x′)τ^{-1}(y′)⋅∈⋅E$, which follows from Lemma~\rf{A537}(\rf{A546x}).   Second, (\rf{cH}) for $[Θ′,Θ,τ^{-1}]$ is the continuity of $τ^{-1}|_{W′,W}$, which follows from $τ|_{W,W′}$ being a homeomorphism.  

Finally, by Lemma~\rf{A779}(\rf{F190}), (\rf{cL}) for $[Θ′,Θ,τ^{-1}]$ is equivalent to the continuity of $⁅α^{-1}_{τ^{-1}(x′)}⁆_{x′∈W′}$ from~$W′$.  Since $τ|_{W,W′}$ is a homeomorphism, this is equivalent to the continuity of $⁅α^{-1}_x⁆_{x∈W}$ from~$W$, which is equivalent (as at (\rf{E781})) to \begin{gather}
\zz
(∀H∈\HH,x_1∈H,x_2∈H)⋅α^{-1}_{x_1} = α^{-1}_{x_1}, \notag 
\zz
\end{gather} which by inspection is equivalent to \begin{gather}
\zz
(∀H∈\HH,x_1∈H,x_2∈H)⋅α_{x_1} = α_{x_1}, \notag 
\zz
\end{gather} which is equivalent (as at (\rf{E781})) to the continuity of $⁅α_x⁆_{x∈W}$ from~$W$, which is (\rf{cL}) for $[Θ,Θ′,τ]$. \end{pf}

\begin{npf}[{{\bf for Proposition \rf{A491}}}]\label{A491p}  Lemma~\rf{A781} shows that $[Θ,Θ′,τ]$ is an isomorphism if it satisfies (\rf{F079}) and (\rf{F080}).  So it remains to show that $[Θ,Θ′,τ]$ is an isomorphism only if it satisfies (\rf{F079})--(\rf{F194}).  Toward that end, suppose $[Θ,Θ′,τ]$ is an isomorphism.  

{\em (\rf{F079},\rf{F082})}.  Since $[Θ,Θ′,τ]$ is an isomorphism, \ct{CLT} definition (\rf{E790}) implies that $τ$ is a bijection and that the inverse of $[Θ,Θ′,τ]$ is $[Θ′,Θ,τ^{-1}]$.  These two conclusions are parts (\rf{F079}) and (\rf{F082}). 

{\em (The assumptions of Lemmas \rf{A537}--\rf{A779})}.  Since $[Θ,Θ′,τ]$ is a morphism, both $(X,E)$ and $(X′,E′)$ are out-trees and $τ$ preserves edges.  This and part (\rf{F079}) implies that the assumptions of Lemmas \rf{A537}--\rf{A779} are met.

{\em (\rf{F080})}.  We assemble three facts: [1] $τ|_{W,W′}$ is a bijection by Lemma~\rf{A537}(\rf{A501x}), [2] $τ|_{W,W′}$ is continuous by (\rf{cH}) for $[Θ,Θ′,τ]$, and [3] $(τ|_{W,W′})^{-1}$ is continuous by (\rf{cH}) for $[Θ′,Θ,τ^{-1}]$.  Thus $τ|_{W,W′}$ is a homeomorphism, which is part (\rf{F080}). 

{\em (\rf{A546}--\rf{A494})}.  Lemma~\rf{A537}(\rf{A546x}) implies (\rf{A546}).  Lemma~\rf{A537}(\rf{A502x})--(\rf{A494x}) implies (\rf{A502})--(\rf{A494}). 

{\em (\rf{F225}--\rf{F186})}.  Part~(\rf{F080}) and Lemma~\rf{F226}(\rf{F229}$⇒$\rf{F231}) imply (\rf{F225}).  Lemma~\rf{A779}(\rf{B188}) implies (\rf{F081}).  Lemma~\rf{A779}(\rf{F193}) implies (\rf{F087}).  Finally, Lemma~\rf{A779}(\rf{B189}) implies $ζ = \dtau|_{\ZZ,\ZZ′}$, which by Lemma~\rf{A537}(\rf{A498x}) implies the remainder of (\rf{F186}).   

{\em (\rf{F083}--\rf{F194})}. Lemma~\rf{A779}(\rf{F190}) implies part (\rf{F083}).  For the remainder, note that parts (\rf{F079}) and (\rf{F082}) imply that the inverse $[Θ′,Θ,τ^{-1}]$ satisfies the assumptions of Lemma~\rf{A779}.  Thus Lemma~\rf{A779}(\rf{F193}) applied to $[Θ′,Θ,τ^{-1}]$ implies~(\rf{F192}).  Also Lemma~\rf{A779}(\rf{B189}) applied to $[Θ′,Θ,τ^{-1}]$ implies $ζ′ = \overline{τ^{-1}}|_{\ZZ′,\ZZ}$.  To use this equality, recall that the proposition's part~(\rf{F186}) showed $ζ^{-1} = \overline{τ^{-1}}|_{\ZZ′\ZZ}$.  Thus $ζ′ = ζ^{-1}$.  This equality is (\rf{F194}).  \end{npf}

\begin{lemma}\label{F200} Let $[Γ,Γ′,τ]$ be a tuple such that $Γ$ and $Γ′$ are games and $[Θ,Θ′,τ]$ is a morphism.  Also let $ι^*$ be a function.  Then the following are equivalent. \begin{tlist}
\yl{F201} $[Γ,Γ′,τ]$ is control-preserving (\rf{E788}) and $ι^*$ is its player transformation (\rf{E785}).
\yl{F202} $ι^*○κ = κ′○τ|_{W,W′}$. \end{tlist} \end{lemma}

\begin{pf} {\em (\rf{F201}$⇒$\rf{F202})}.  Proposition~\rf{E791}(b).

{\em (\rf{F201}$⇐$\rf{F202})}. To show $[Γ,Γ′,τ]$ is control-preserving (\rf{E788}), it must be shown that $(∀x_1∈W,x_2∈W)$ $κ(x_1) = κ(x_2)$ implies $κ(τ(x_1)) = κ(τ(x_2))$.  Toward that end, take two decision nodes $x_1⋅∈⋅W$ and $x_2⋅∈⋅W$ such that $κ(x_1) = κ(x_2)$.  Then $ι^*○κ(x_1) = ι^*○κ(x_2)$, which by (\rf{F202}) implies $κ′○τ(x_1) = κ′○τ(x_2)$. 

To show $ι^*$ is the player transformation (\rf{E785}) of $[Γ,Γ′,τ]$, it must be shown that \lic{F203} $ι^*{:}I→I′$ and that \il{F204} $(ι^*)\gr = ⎨⋅(κ(x),κ′(τ(x)))⋅|⋅x∈W⋅⎬$.  For \rf{F203}, note that $ι^*$ is a function by assumption, that its domain by (\rf{F202}) is $κ$'s codomain which by (\rf{F218}) is $I$, and that its codomain by (\rf{F202}) is $κ′$'s codomain which by (\rf{F218}) is $I′$.  Finally, \rf{F204} follows immediately from (\rf{F202}). \end{pf}

\begin{lemma}\label{A874} Suppose $[Γ,Γ′,τ]$ is a morphism such that (a) $[Θ,Θ′,τ]$ is an isomorphism, (b) $ι$ is bijective, and (c) $(∀i∈I)$ $ζ$ is a preorder isomorphism from $(\ZZ,\lesssim_i)$ to $(\ZZ′,\lesssim′_{ι(i)})$.  Then $[Γ,Γ′,τ]$ is an isomorphism. \end{lemma}

\begin{pf} Since $[Θ,Θ′,τ]$ is an isomorphism by assumption, we have all the conclusions of Proposition~\rf{A491}.   Part (\rf{F079}) of that proposition shows $τ$ is a bijection, so $τ^{-1}$ exists, so the tuple $[Γ′,Γ,τ^{-1}]$ exists.  Thus it suffices for this lemma to show that $[Γ′,Γ,τ^{-1}]$ is a morphism.  In other words, it suffices to show that $[Γ′,Γ,τ^{-1}]$ satisfies conditions (\rf{gZ}), (\rf{gK}), and  (\rf{gU}) of Definition~\rf{E774}.

First, (\rf{gZ}) requires that  $[Θ′,Θ,τ^{-1}]$ be an end-preserving morphism.  It is a morphism by the lemma's assumption (a).  It is end-preserving by Proposition~\rf{A491}(\rf{F192}).  

Second, (\rf{gK}) requires that $[Γ′,Γ,τ^{-1}]$ be control-preserving (\rf{E788}).  To show this and that $ι^{-1}$ is the player transformation of $[Γ′,Γ,τ^{-1}]$, note Proposition~\rf{E791}(b) implies $ι○κ = κ′○τ|_{W,W′}$.  Since $τ|_{W,W′}$ is a bijection by Proposition~\rf{A491}(\rf{F080}), this is equivalent to $ι○κ○τ^{-1}|_{W′,W} = κ′$.  Since $ι$ is a bijection by assumption, this is equivalent to $κ○τ^{-1}|_{W′,W} = ι^{-1}○κ′$.  This is equivalent to Lemma~\rf{F200} condition (\rf{F202}) for the tuple $[Γ′,Γ,τ^{-1}]$ and for $ι^* = ι^{-1}$.  Thus Lemma~\rf{F200}(\rf{F201}$⇐$\rf{F202}) implies that $[Γ′,Γ,τ^{-1}]$ is control-preserving, and that $ι^{-1}$ is its player transformation.

Third, consider (\rf{gU}).  Since $ζ^{-1}$ is the run transformation of $[Γ′,Γ,τ^{-1}]$ by Proposition~\rf{A491}(\rf{F194}), and since $ι^{-1}$ is the player transformation of $[Γ′,Γ,τ^{-1}]$ by the previous paragraph, (\rf{gU}) requires that \begin{gather}
\zz
(∀i′∈I′)⋅ζ^{-1}⋅\text{is monotone from}⋅(\ZZ′,\lesssim′_{i′})⋅\text{to}⋅(\ZZ,\lesssim_{ι^{-1}(i′)}). \notag 
\zz
\end{gather}  Since $ι{:}I→I′$ is a bijection by assumption, this is equivalent to \begin{gather}
\zz
(∀i∈I)⋅ζ^{-1}⋅\text{is monotone from}⋅(\ZZ′,\lesssim′_{ι(i)})⋅\text{to}⋅(\ZZ,\lesssim_{i}). \notag 
\zz
\end{gather} This holds because the lemma assumes that $(∀i∈I)$ $ζ$ is a preorder isomorphism from $(\ZZ,\lesssim_i)$ to $(\ZZ′,\lesssim′_{ι(i)})$. \end{pf}

\begin{npf}[{{\bf for Proposition~\rf{A503}}}]\label{A503p}  Lemma~\rf{A874} shows that $[Γ,Γ′,τ]$ is an isomorphism if it satisfies (\rf{F084}), (\rf{F085}), and (\rf{F086}).  So it remains to show that $[Γ,Γ′,τ]$ is an isomorphism only if it satisfies (\rf{F084})--(\rf{F090}).  Toward that end, suppose $[Γ,Γ′,τ]$ is an isomorphism.  We begin with the easiest of (\rf{F084})--(\rf{F090}) and proceed toward the hardest.  

{\em (\rf{F088})}.  Since $[Γ,Γ′,τ]$ is an isomorphism, $[Γ′,Γ,τ^{-1}]$ is its inverse (this follows from the definition of $○$ in (\rf{E789})).

{\em (\rf{F084})}.  Since $[Γ,Γ′,τ]$ is an isomorphism, and since there is a forgetful functor from \ct{Gm} to \ct{CLT} by Proposition~\rf{A455}, $[Θ,Θ′,τ]$ is an isomorphism in \ct{CLT}.  

{\em (\rf{F085},\rf{F090})}.  Let $ι^*$ be the player transformation of $[Γ′,Γ,τ^{-1}]$.  It suffices for both (\rf{F085}) and (\rf{F090}) to show that $ι^* = ι^{-1}$.  Since $[Θ′,Θ,τ^{-1}]○[Θ,Θ′,τ] = \id_Θ$, Proposition~\rf{A454}(b,c) implies \lic{A872} $ι^*○ι = \id_I$.  Similarly, since $[Θ,Θ′,τ]○[Θ′,Θ,τ^{-1}] = \id_{Θ′}$, Proposition~\rf{A454}(b,c) implies \li{A873} $ι○ι^* = \id_{I′}$.  Results \rf{A872} and \rf{A873} imply $ι^* = ι^{-1}$.

{\em (\rf{F086})}.  Since $[Γ,Γ′,τ]$ is a morphism, condition~(\rf{gU}) of Definition~\rf{E774} implies \begin{gather}
\zz
(∀i∈I)⋅ζ⋅\text{is monotone from}⋅(\ZZ,\lesssim_i)⋅\text{to}⋅(\ZZ′,\lesssim′_{ι(i)}). \label{F205}
\zz
\end{gather} In a different direction, part (\rf{F084}) and Proposition~\rf{A491}(\rf{F194}) imply that the run transformation of $[Γ′,Γ,τ^{-1}]$ is $ζ^{-1}$.  Further, part (\rf{F090}) implies that the player transformation of $[Γ′,Γ,τ^{-1}]$ is $ι^{-1}$.  Therefore, since $[Γ′,Γ,τ^{-1}]$ is a morphism, condition~(\rf{gU}) of Definition~\rf{E774} implies \begin{gather}
\zz
(∀i′∈I′)⋅ζ^{-1}⋅\text{is monotone from}⋅(\ZZ′,\lesssim′_{i′})⋅\text{to}⋅(\ZZ,\lesssim_{ι^{-1}(i′)}). \notag 
\zz
\end{gather} Thus, since $ι{:}I→I′$ is a bijection by part (\rf{F085}), \begin{gather}
\zz
(∀i∈I)⋅ζ^{-1}⋅\text{is monotone from}⋅(\ZZ′,\lesssim′_{ι(i)})⋅\text{to}⋅(\ZZ,\lesssim_i). \label{F206} 
\zz
\end{gather} Finally, the combination of (\rf{F205}) and (\rf{F206}) yields \begin{gather}
\zz
(∀i∈I)⋅ζ⋅\text{is preorder isomorphism from}⋅(\ZZ,\lesssim_i)⋅\text{to}⋅(\ZZ′,\lesssim′_{ι(i)}). 
\notag 
\zz
\end{gather} \end{npf}

\markb{\sc References}
\eput

%% file: Streufert-commands.tex
\listfiles %
\usepackage{ 
  amssymb, %
  amsmath, %
  latexsym, %
  graphpap, %
  graphics, %
  verbatim, %
  color, %
  natbib, 
  har2nat, %
  layout, %
  amsbsy, %
  calc, %
  pstricks, %
  pst-plot, %
  pst-pdf, %
  pstricks-add, %
  xr, %
  mathdots, %
  enotez, %
  lipsum, 
  stmaryrd, %
  textcomp %
  }

\usepackage[bottom]{footmisc} %
 
\usepackage[inline]{enumitem} %

\newcommand{\ifdraft}[1]{} %
\newcommand{\ifdrafttext}[1]{} %
\newcommand{\iffinal}[1]{#1} %
\newcommand{\ifallkeys}[1]{} %
\newcommand{\ifbeamer}[1]{} %
  \newcommand{\eput}{ \ifmain{\showbib \end{document}}  } 

\newcommand{\sella}[1]{} 
\newcommand{\sellb}[1]{} 
\newcommand{\sellc}[1]{} 
\newcommand{\selld}[1]{} 
\newcommand{\selle}[1]{} 
\newcommand{\sellf}[1]{} 
\newcommand{\sellg}[1]{} 
\newcommand{\sellh}[1]{} 
\newcommand{\selli}[1]{} 
\newcommand{\sellj}[1]{} 

\newtheorem{lemma}{\indent Lemma} 

\newtheorem{ndef}[lemma]{\indent Definition} 
 
\newtheorem{prop}[lemma]{\indent Proposition}

\theoremstyle{definition} %
\newtheorem{pfa}[lemma]{\indent Proof} 
\newenvironment{npf} %
  { \begin{pfa} } 
  { \hfill $\Box$ \end{pfa} }

\theoremstyle{remark} %

\newtheorem{absatz}[lemma]{\hspace{-1ex}\tt} %
\newcommand{\alabel}{} 
\newcommand{\amark}{} 

\newcommand{\ai}[2][]{\setcounter{equation}{\thelemma} 
  \begin{absatz}\label{#2}\amark \renewcommand{\alabel}{#2} 
  \markb{\rf{#2}}
  \begin{subequations} {\em#1}} %
\newcommand{\ao}{\end{subequations} \end{absatz}}

\newtheorem*{pfb}{\indent Proof} 
\newenvironment{pf} 
  { \begin{pfb} } 
  { \hfill $\Box$ \end{pfb}  } 

\newtheorem{exa}[lemma]{\indent Question} %
\newtheorem{soln}[lemma]{\indent Answer to Question} 
\newcommand{\exercomment}{} %
 
\newcommand{\ifbothexeransr}[1]{} 
  { \smallskip \begin{exa} \label{#1} \exercomment} %
  { \end{exa} } 
  { \end{soln} } %

\newcounter{blistno} %
  { \begin{list} %
      {}%
      { \setlength{\leftmargin}{5mm} %
        \setlength{\rightmargin}{0mm} %
        \setlength{\itemindent}{-5mm} %
        \setlength{\labelwidth}{0mm} %
        \setlength{\labelsep}{0mm} %
        \setlength{\itemsep}{2mm} %
        \setlength{\topsep}{2mm} %
        \usecounter{blistno} 
        \nopagebreak } } %
  { \end{list} }

\newcounter{clistno} %
  { \begin{list} %
      {  [\Alph{clistno}] } %
      {  \setlength{\leftmargin}{0mm} %
        \setlength{\rightmargin}{0mm} %
        \setlength{\itemindent}{-1.4mm} %
        \setlength{\labelwidth}{0mm} %
        \setlength{\labelsep}{0mm} %
        \setlength{\itemsep}{0mm} %
        \setlength{\topsep}{0mm} %
        \usecounter{clistno} 
        \nopagebreak } }%
  { \end{list} } 
\newcounter{dlistno} %
  { \begin{list} %
      {  \arabic{dlistno}. } %
      {  \setlength{\leftmargin}{0mm} %
        \setlength{\rightmargin}{0mm} %
        \setlength{\itemindent}{2mm} %
        \setlength{\labelwidth}{0mm} %
        \setlength{\labelsep}{0mm} %
        \setlength{\itemsep}{2mm} %
        \setlength{\topsep}{0mm} %
        \usecounter{dlistno} 
        \nopagebreak } }%
  { \end{list} } 
\newcommand{\tlistindent}{-4.20mm} 
\newcounter{tlistno} %
\newenvironment{tlist}[1][] %
  { \begin{list} %
      {\indent(\alph{tlistno}#1) } %
      { \setlength{\leftmargin}{0mm} %
        \setlength{\rightmargin}{0mm} %
        \setlength{\itemindent}{\tlistindent} 
        \setlength{\labelwidth}{0mm} %
        \setlength{\labelsep}{0mm} %
        \setlength{\itemsep}{0mm} %
        \setlength{\topsep}{0mm} %
        \usecounter{tlistno} } } %
  { \end{list} } 
\newcommand{\hglistindent}{0mm} 
\newcounter{hglistno} %
  { \begin{list} %
      {\indent\fbox{\q}\hspace{1ex} } %
      { \setlength{\leftmargin}{0mm} %
        \setlength{\rightmargin}{0mm} %
        \setlength{\itemindent}{\hglistindent} 
        \setlength{\labelwidth}{0mm} %
        \setlength{\labelsep}{0mm} %
        \setlength{\itemsep}{0mm} %
        \setlength{\topsep}{0mm} %
        \usecounter{hglistno} } } %
  { \end{list} } 

\newcounter{cllistno} %
\newenvironment{cllist} %
  { \begin{list} %
      {\emph{Claim \arabic{cllistno}:} } 
      { \setlength{\leftmargin}{0mm} %
        \setlength{\rightmargin}{0mm} %
        \setlength{\itemindent}{\parindent} %
        \setlength{\labelwidth}{0mm} %
        \setlength{\labelsep}{0mm} %
        \setlength{\itemsep}{1.2mm}%
        \setlength{\topsep}{1.2mm}%
        \usecounter{cllistno} } } %
  { \end{list} } 
 
\newcommand{\unskipcl}{\vspace{-5.2mm}} %

\newcommand{\itemi}{ \begin{itemize} } 
\newcommand{\itemo}{ \end{itemize} } 
\newcommand{\yl}[1]{\item\label{#1}} 

\newcommand{\ili}{\begin{enumerate*}[label=$\lbrack$\alph*$\rbrack$]} 
\newcommand{\ilr}{\begin{enumerate*}[resume*]} 
\newcommand{\ilo}{\end{enumerate*}} 
\newcommand{\ilitem}[1]{\item\label{#1}\ilo\hspace{-1mm}\ifallkeys{\hspace{10mm}}} 
\newcommand{\ilc}[1]{\ili\ilitem{#1}} 
\newcommand{\il}[1]{\ilr\ilitem{#1}} 
 
\newcommand{\lii}{\begin{enumerate*}[label=$\lbrack$\arabic*$\rbrack$]} 
\newcommand{\lir}{\begin{enumerate*}[resume*]} 
\newcommand{\lio}{\end{enumerate*}} 
\newcommand{\liitem}[1]{\item\label{#1}\lio\hspace{-1mm}\ifallkeys{\hspace{10mm}}} 
\newcommand{\lic}[1]{\lii\liitem{#1}} 
\newcommand{\li}[1]{\lir\liitem{#1}}

\newcommand{\zz}{} 
\newcommand{\subi}{\begin{subequations}} 
\newcommand{\subo}{\end{subequations}} 
\newcommand{\nt}{\notag \\ }

\newenvironment{maneq}{%
  \setlength{\arraycolsep}{.1ex} 
   
  \begin{array}{lc} \rule{5ex}{0ex} & \rule{79ex}{0ex} \\[-3ex] }{%
  \end{array}} 
\newcommand{\mqi}{\begin{maneq}} 
\newcommand{\mqo}{\end{maneq}}

\newcommand{\ci}{\par\begin{center}} 
\newcommand{\co}{\end{center}\par\noindent}

\newcommand{\mi}{ \left[ \begin{matrix} } 
\newcommand{\mo}{ \end{matrix} \right] } 
\newcommand{\miz}{ \left. \begin{matrix} } 
\newcommand{\moz}{ \end{matrix} \right. } 
\newcommand{\mip}{ \left( { \ } \begin{matrix} } 
\newcommand{\mop}{ \end{matrix} { \ } \right) } 
\newcommand{\mib}{ \left[ { \ } \begin{matrix} } 
\newcommand{\mob}{ \end{matrix} { \ } \right] } 
\newcommand{\smi}{ \left[ \begin{smallmatrix} } %
\newcommand{\smo}{ \end{smallmatrix} \right] } 
\newcommand{\casei}{ \left\{ \begin{matrix} } %
\newcommand{\caseo}{ \end{matrix} \right. } 
 
\newcommand{\mpi}{ \begin{minipage} } %
\newcommand{\mpo}{ \end{minipage} } 
\newcommand{\mpic}[1]{\begin{minipage}{#1}\begin{center}} 
\newcommand{\mpoc}{\end{center}\end{minipage}}

\newcommand{\datestmp}{%
  $ \the\year \backslash \the\month \backslash \the\day \backslash 
    \the\time $ } 
 
\newcommand{\filestamp}[1]{} %

\newcommand{\markb}[1]{\markboth{#1}{#1}} %
\newcommand{\showit}{\ifdraft{\vspace{-.7cm}{⋅}\vspace{.3cm}}} %
   
\newcommand{\nssec}[2]{%
  \vspace{-1mm} 
  \subsection{#1}\label{#2}\hspace{0mm}%
  \par\vspace{1.6mm} 
  \begin{picture}(0,0) 
  \put(-.50,.33){\color{white} \rule{85ex}{4ex}} 
  \put(-.4,.53){\color{black} \sc \rf{#2}. #1} 
  \end{picture} 
  \par\vspace{-4.5mm}} 

 \newcommand{\ifssec}[1]{#1} %
 \providecommand{\ifssec}[1]{} 
\ifssec{ 
  }

\allowdisplaybreaks  %
 
\newcommand{\toenote}[1]{} 
\newcommand{\qx}{\rule{4ex}{0ex}}
\newcommand{\qy}{\rule[-1.3ex]{0ex}{4ex}}
\newcommand{\q}{\qx\qy}

\newcommand{\nichts}[1]{} %
\newcommand{\showbib}{%
  \markb{\sc References} 
  \bibliographystyle{\tpath/ecta/econometrica} 
  \bibliography{\tpath/bibtex/streuf,\tpath/bibtex/others,references}} 
\definecolor{darkred}{rgb}{.7,0,0} 

\definecolor{darkgreen}{rgb}{0,.7,0} 
   
\definecolor{darkblue}{rgb}{0,0,.97}

\definecolor{igray}{gray}{.87}

\newcommand{\rf}[1]{\ref{#1}} %

\newcommand{\sss}{\scriptscriptstyle}

\newcommand{\ct}[1]{\ensuremath{\mathbf{#1}}}%
 
\newcommand{\ex}[1]{\mathsf{#1}} 

\newcommand{\yyout}[1]{}          %

  \renewcommand{\lim}{\text{\normalfont{lim}}}

  \newcommand{\src}{\mathsf{src}} 
  \newcommand{\trg}{\mathsf{trg}} 
  \newcommand{\id}{\mathsf{id}} 
  \newcommand{\gr}{^{\mspace{1mu}\mathsf{gr}}} 
  \newcommand{\inc}{\mathsf{inc}}

  \newcommand{\nnexists}{\raisebox{0.15ex}{/}\hspace{-1.15ex}\exists}

  \renewcommand{\mapsfrom}{\text{\reflectbox{$\mapsto$}}}  %

  \newcommand{\dra}{\rotatebox[origin=c]{180}{$\Lsh$}} %
  \newcommand{\dtau}{\bar{\tau}}

  \newcommand{\HH}{{ \mathcal H}}

  \newcommand{\PP}{{ \mathcal P}}

  \newcommand{\ZZ}{{ \mathcal Z}} 
  \newcommand{\ZZf}{ \mathcal Z_{\mathsf{ft}}} 
  \newcommand{\ZZi}{\mathcal Z_{\mathsf{inft}}}

  \newcommand{\eR}{\bar{\mathbb R}} %

  \newcommand{\f}[1]{\mathsf{#1}} %
    
   \newcommand{\fb}{{\f{b}}} 
   \newcommand{\fc}{{\f{c}}} 
   \newcommand{\fd}{{\f{d}}} 
   \newcommand{\fe}{{\f{e}}} 
   \newcommand{\ff}{{\f{f}}} 
   \newcommand{\fg}{{\f{g}}}

  \newcommand{\FB}{\mathsf{F}} 
  \newcommand{\FO}{\mathsf{F_0}} 
  \newcommand{\FA}{\mathsf{F_1}} 

\usepackage[utf8]{inputenc} 
\usepackage{amssymb} 
\usepackage{newunicodechar} 
\newcommand{\nuc}[2]{\newunicodechar{#1}{#2}} %

  \nuc{α}{{ \alpha}} 
  \nuc{β}{{ \beta}} 
  \nuc{γ}{{ \gamma}} \nuc{Γ}{{ \varGamma}} 
  \nuc{δ}{{ \delta}} \nuc{Δ}{{ \varDelta}} 
  \nuc{ε}{{ \varepsilon}} 
  \nuc{ζ}{{ \zeta}} 
  \nuc{η}{{ \eta}} 
  \nuc{θ}{{ \theta}} \nuc{Θ}{{ \varTheta}} 
  \nuc{ι}{{ \iota}} 
  \nuc{κ}{{ \kappa}} 
  \nuc{λ}{{ \lambda}} \nuc{Λ}{{ \varLambda}} 
  \nuc{μ}{{ \mu}} 
  \nuc{ν}{{ \nu}} 
  \nuc{ξ}{{ \xi}} \nuc{Ξ}{{ \varXi}} 
  \nuc{π}{{ \pi}} \nuc{Π}{{ \varPi}} 
  \nuc{ρ}{{ \rho}} 
  \nuc{σ}{{ \sigma}} \nuc{Σ}{{ \varSigma}} 
  \nuc{τ}{{ \tau}} 
  \nuc{υ}{{ \upsilon}} \nuc{Υ}{{ \varUpsilon}} 
  \nuc{φ}{{ \phi}} \nuc{Φ}{{ \varPhi}} 
  \nuc{χ}{{ \chi}} 
  \nuc{ψ}{{ \psi}} \nuc{Ψ}{{ \varPsi}} 
  \nuc{ω}{{ \omega}} \nuc{Ω}{{ \varOmega}} 

  \nuc{≠}{{ \ne}} 
  \nuc{∾}{{ \sim}} %
  \nuc{≈}{{ \approx}} 
  \nuc{≡}{{ \equiv}} %
  \nuc{∈}{{ \in}} \nuc{∉}{{ \notin}} 
  \nuc{∋}{{ \ni}} \nuc{∌}{{ \not\ni}} 
  \nuc{⊊}{{ \subset}} \nuc{⊆}{{ \subseteq}} %
  \nuc{⊋}{{ \supset}} \nuc{⊇}{{ \supseteq}} %
  \nuc{≤}{{ \leq}} 
  \nuc{≥}{{ \geq}} 
  \nuc{≪}{{ \ll}} 
  \nuc{≫}{{ \gg}} 
  \nuc{≺}{{ \prec}} \nuc{≼}{{ \preccurlyeq}} 
  \nuc{≻}{{ \succ}} \nuc{≽}{{ \succcurlyeq}}
  \nuc{≲}{{ \lesssim}} %
  \nuc{≳}{{ \gtrsim}} %
  \nuc{‖}{{ \parallel}}  %
  \nuc{≊}{{ \approxeq}} %
  \nuc{≅}{\cong} %
  \nuc{≃}{\simeq} %
  \nuc{≐}{\doteq} 

  \nuc{∑}{{ \Sigma}} %
  \nuc{∫}{{ \textstyle \int\nolimits}} 
  \nuc{∏}{{ \Pi}} %
  \nuc{×}{{ \times}} 
  \nuc{÷}{{ \div}} 
  \nuc{○}{{ \circ}} 
  \nuc{⨅}{{\scalebox{1}[1.2]{\ensuremath{\cap}}}} %
  \nuc{⋂}{{\cap}}                                   %
  \nuc{⨆}{{\scalebox{1}[1.2]{\ensuremath{\cup}}}} %
  \nuc{⋃}{{\cup}}                                   %
  \nuc{⋱}{{ \setminus}} %
  \nuc{⧷}{{ \smallsetminus}} %
  \nuc{∧}{{ \wedge}} \nuc{⋀}{{ \textstyle \bigwedge\nolimits}} 
  \nuc{∨}{{ \vee}} \nuc{⋁}{{ \textstyle \bigvee\nolimits}} 
  \nuc{⊗}{{ \otimes}} 
  \nuc{⊙}{{ \odot}} \nuc{⊛}{{ \textstyle \bigodot\nolimits}} 
  \nuc{⊕}{{ \oplus}} 
  \nuc{•}{{ \bullet}} 
  \nuc{∗}{{ \star}} 
  \nuc{⁕}{{ \ast}} %
  \nuc{∙}{{ \cdot}} %

  \nuc{→}{{ \rightarrow}} 
  \nuc{⇉}{{ \rightrightarrows}} 
  \nuc{←}{{ \leftarrow}} 
  \nuc{⟷}{{ \leftrightarrow}} 
  \nuc{⇒}{{ \Rightarrow}} 
  \nuc{⇐}{{ \Leftarrow}} %
  \nuc{⇔}{{ \Leftrightarrow}} %
  \nuc{⟺}{{ \Leftrightarrow}} %
  \nuc{⇡}{{ \uparrow}} 
  \nuc{⇣}{{ \downarrow}} 

  \nuc{ℓ}{{ \ell}} 
  \nuc{∂}{{ \partial}} 
  \nuc{∞}{{ \infty}} 
  \nuc{‴}{^{ \prime\prime\prime}} %
  \nuc{″}{^{ \prime\prime}} %
  \nuc{′}{^{ \prime}} 
  \nuc{¬}{{ \neg}} 
  \nuc{∀}{{ \forall}} 
  \nuc{∃}{{ \exists}} 
  \nuc{∆}{{ \triangle}} %
  \nuc{∄}{{ \nnexists}} %
  \nuc{∅}{{ \varnothing}} 
  \nuc{□}{{ \square}} 

  \nuc{⎨}{ \lbrace } 
  \nuc{⎬}{ \rbrace } 
  \nuc{⁅}{ \langle } 
  \nuc{⁆}{ \rangle } 

  \nuc{Ṛ}{{ \mathbb R}} 
  \nuc{Ẓ}{{ \mathbb Z}} 
  \nuc{⋅}{{ \ }} %
  \nuc{˙}{{\,}} %
  \nuc{˛}{{,}\mspace{.5mu}} %
  \nuc{₵}{\text{\textcent}}%

\setlength{\unitlength}{1cm} 
 
\newrgbcolor{maroon}{.5 0 0} 
\psset{gridcolor=green} 
 
\newcommand{\myoutergrid}[1]{\ifdraft{ 
  \put(-7.5,0){\color{brown} \framebox(7.5,#1){}} 
  \put(-3.75,0){\color{brown} \framebox(7.5,#1){}} 
  \put(0,0){\color{brown} \framebox(7.5,#1){}}} } 
\psset{unit=1cm} %